\documentclass[twocolumn, twocolappendix]{aastex631} 
\usepackage{soul}
\usepackage{graphicx}
\usepackage{anyfontsize} 
\usepackage{tablefootnote}
\usepackage{verbatim}
\usepackage{xcolor}
\usepackage{array,multirow}
\usepackage{xspace} 
\newcommand\mbh{$M_{\rm BH}$\xspace}
\newcommand\mdyn{$M_{\rm dyn}$\xspace}
\newcommand\mbulgedyn{$M_{\rm bulge,dyn}$\xspace}
\newcommand\sig{$\sigma$\xspace}
\newcommand\sigspat{$\sigma_{\rm spat}$\xspace}
\newcommand\sigap{$\sigma_{\rm ap}$\xspace}
\newcommand\sigstar{$\sigma_\star$\xspace}

\newcommand\oiii{[\ion{O}{iii}]\xspace}
\newcommand\CaT{CaT\xspace}
\newcommand\MgIb{\ion{Mg}{i}b\xspace}

\newcommand\f{$f$\xspace}

\DeclareRobustCommand{\ion}[2]{\textup{#1\,\textsc{\lowercase{#2}}}}
\newcommand*\element[1][]{%
  \def\aa@element@tr{#1}%
  \aa@element
}

\begin{document}

    \title{Combining Direct Black Hole Mass Measurements and Spatially Resolved Stellar Kinematics to Calibrate the \mbh-\sigstar Relation of Active Galaxies}
    
%
\author[0000-0001-9428-6238]{Nico~Winkel}
\affiliation{Max-Planck-Institut f\"ur Astronomie, K\"onigstuhl 17, D-69117 Heidelberg, Germany}
\author[0000-0003-2064-0518]{Vardha~N.~Bennert}
\affiliation{Physics Department, California Polytechnic State University, San Luis Obispo, CA 93407, USA}
\author[0000-0002-0164-8795]{Raymond~P.~Remigio}
\affiliation{Department of Physics and Astronomy, 4129 Frederick Reines Hall, University of California, Irvine, CA 92697, USA}
\author[0000-0002-8460-0390]{Tommaso~Treu}
\affiliation{Department of Physics and Astronomy, University of California, Los Angeles, 430 Portola Plaza, Los Angeles, CA 90095, USA}
\author[0000-0003-3804-2137]{Knud~Jahnke}
\affiliation{Max-Planck-Institut f\"ur Astronomie, K\"onigstuhl 17, D-69117 Heidelberg, Germany}
\author[0000-0002-1912-0024]{Vivian~U}
\affiliation{Department of Physics and Astronomy, 4129 Frederick Reines Hall, University of California, Irvine, CA 92697, USA}
\author[0000-0002-3026-0562]{Aaron~J.~Barth}
\affiliation{Department of Physics and Astronomy, 4129 Frederick Reines Hall, University of California, Irvine, CA 92697, USA}
\author[0000-0001-6919-1237]{Matthew~Malkan}
\affiliation{Department of Physics and Astronomy, University of California, Los Angeles, 430 Portola Plaza, Los Angeles, CA 90095, USA}
\author[0000-0003-2901-6842]{Bernd~Husemann}
\affiliation{European Organisation for the Exploitation of Meteorological Satellites (EUMETSAT), Eumetsat Allee 1, 64295 Darmstadt, Germany}
\affiliation{Max-Planck-Institut f\"ur Astronomie, K\"onigstuhl 17, D-69117 Heidelberg, Germany}
\author[0000-0001-8917-2148]{Xuheng~Ding}
\affiliation{School of Physics and Technology, Wuhan University, Wuhan 430072, China}
\author[0000-0003-3195-5507]{Simon~Birrer}
\affiliation{Department of Physics and Astronomy, Stony Brook University, Stony Brook, NY 11794, USA}

\correspondingauthor{Nico Winkel}
\email{winkel@mpia.de}
%


\begin{abstract}
    The origin of the tight scaling relation between the mass of supermassive black holes (SMBHs; \mbh) and their host-galaxy properties remains unclear.
    Active galactic nuclei (AGNs) probe phases of ongoing SMBH growth and offer the only opportunity to measure \mbh beyond the local Universe. However, determining AGN host galaxy stellar velocity dispersion \sigstar, and their galaxy dynamical masses \mdyn, is complicated by AGN contamination, aperture effects and different host galaxy morphologies.
    We select a sample of AGNs for which \mbh has been independently determined to high accuracy by state-of-the-art techniques: dynamical modeling of the reverberation signal and spatially resolving the broad-line region with VLTI/GRAVITY.
    Using IFU observations, we spatially map the host galaxy stellar kinematics across the galaxy and bulge effective radii. 
    We find that that the dynamically hot component of galaxy disks correlates with \mbh; however, the correlations are tightest for aperture-integrated \sigstar measured across the bulge.
    Accounting for the different \mbh distributions, we demonstrate -- for the first time -- that AGNs follow the same \mbh-\sigstar and \mbh-\mbulgedyn relations as quiescent galaxies.
    We confirm that the classical approach of determining the virial factor as sample-average, yielding ${\rm log }f= 0.65 \pm 0.18$, is consistent with the average \f from individual measurements.
    The similarity between the underlying scaling relations of AGNs and quiescent galaxies implies that the current AGN phase is too short to have altered BH masses on a population level. These results strengthen the local calibration of \f for measuring single-epoch \mbh in the distant Universe.
\end{abstract}


%
\section{Introduction}



Supermassive black holes (SMBHs) are located in the hearts of most, if not all, massive galaxies. 
Their masses \mbh form tight correlations with various properties of their host galaxies. 
Prominent examples include the scaling relations between \mbh and bulge stellar mass \citep{Magorrian:1998, Haring&Rix:2004}, bulge luminosity \citep{Kormendy&Richstone:1995, Marconi&Hunt:2003} or bulge stellar velocity dispersion \sigstar \citep{Ferrarese&Merritt:2000, Gebhardt:2000, Merritt&Ferrarese:2001, Tremaine:2002, Treu:2004}.
One way to interpret these scaling relations is a coupling between the growth of SMBHs to that of their host galaxies \citep[e.g.][]{Ferrarese&Merritt:2000, Kormendy_Ho:2013}, implying a causal connection between processes involved.
Among the scaling relations, the \mbh-\sigstar correlation stands out as particularly tight. 
The \mbh-\sigstar relation exhibits a remarkably small intrinsic scatter of $\sim 0.4 \,{\rm dex}$ over many orders \mbh and host-galaxy mass \citep{Gultekin:2009, Saglia:2016, vandenBosch:2016}, providing important insights into SMBH formation scenarios, such as BH seeding models \citep{Volonteri&Natarajan:2009}, and models for SMBH-galaxy coevolution \citep{Robertson:2006, Hopkins:2006, Mo:2024}.
Considering the tightness of the relation, and that \sigstar is a direct tracer for dynamical mass, the \mbh-\sigstar relation is often interpreted as the most direct probe for the formation and coevolution of SMBHs with their host galaxies \citep{Tremaine:2002, Beifiori:2012, Saglia:2016, vandenBosch:2016, deNicola:2019, Graham:2023}.
As such, the \mbh scaling relations for quiescent galaxies are well-established.
However, any kind of evolutionary study of the \mbh scaling relations relies on \mbh measured in active galactic nuclei (AGNs).

The \mbh scaling relations in local AGNs are essential for various reasons.
For one, broad-line (type-1) AGNs (BLAGNs) are the objects targeted in reverberation mapping (RM) studies, a unique way to determine the mass of the SMBH.
In short, RM observes variability in the accretion-disk luminosity and the time-delayed response of ionized gas in the broad-line region (BLR). While the light travel time provides constraints on the size of the BLR, Doppler-broadened emission lines give the velocity of the BLR clouds. The main uncertainties are due to the unknown geometry and kinematics of the BLR, summarized in the "virial" factor \f. A sample-averaged \f-value has traditionally been determined assuming that AGNs follow the same \mbh-\sigstar~relation as quiescent galaxies.
By combining \f~with the empirical relation between BLR radius and luminosity ("$R$-$L$ relation"), a single spectrum becomes sufficient for estimating \mbh for broad-line AGNs. This estimate is commonly referred to as the single-epoch (SE) method which enables \mbh~measurements across the cosmic time.

Second, host galaxies build up their stellar mass, traced by \sigstar, through secular processes on $>$$10^7\,{\rm yr}$ timescales, much longer than the duration of single AGN episodes of $10^4$-$10^6 \,{\rm yr}$ \citep{Hickox:2014, Schawinski:2015}.
Despite their relatively short lifetimes, the bulk of cosmic SMBH mass growth occurs during luminous AGN phases \citep{Merloni:2004, Schulze:2015}.
This implies that \mbh in AGNs is growing rapidly compared to the host galaxy, so that AGNs might probe a special state during the evolution of the \mbh scaling relations.

Third, AGNs are considered crucial for shaping the \mbh scaling relations. 
The energy from the central accretion disk can significantly affect the host galaxy by either heating the interstellar medium (ISM) or expelling cold gas, which suppresses star formation and limits the build-up of stellar mass in the bulge \citep{DiMatteo:2005, Croton:2006, Somerville:2008, Dubois:2013, Harrison:2017}.
These processes, collectively known as AGN feedback, can also regulate SMBH growth
\citep{Dubois:2012, Massonneau:2023}.
Although the exact timing and mechanisms of AGN feedback are still debated, these effectst are expected to influence the \mbh-\sigstar relation: \cite{Silk&Rees:1998} predict a slope of $\beta \approx 4$ for momentum-driven feedback, while energy-driven feedback should yield $\beta = 5$ \citep{King:2003}. 

As an alternative to self-regulated SMBH growth, the hierarchical assembly of galaxy mass over cosmic time could create a non-causal link between \mbh and host-galaxy properties, mimicking the observed scaling relations \citep{Peng:2007, Hirschmann:2010, Jahnke:2011}. 
To achieve high stellar mass by redshift zero, a galaxy must have experienced multiple mergers, during which the central BHs also merged. 
If \mbh and host-galaxy stellar masses are randomly sampled during each merger, the central limit theorem predicts a correlation between them after several mergers. 
This scenario suggests that AGN feedback is not necessary for the formation of the \mbh-host-galaxy scaling relations over cosmic time.\\

%
%
All these open questions have continued to spark a large interest of the community,
in particular whether the \mbh-galaxy scaling relations of active and
inactive galaxies are identical.
A series of studies have reported shallower \mbh-\sigstar relations for RM AGNs, while also highlighting the challenge of extracting host-galaxy kinematics in luminous AGNs and the small dynamic range in \mbh \citep[][]{Woo:2010, Woo:2013, Park:2012, Batiste:2017b}. 
\cite{Woo:2015} explain the initial tension by selection effects, which are sufficient to explain this flattening of the AGNs' relation. 
Indeed, several groups reported that \mbh-\sigstar relation of AGNs and quiescent galaxies share similar slopes \citep[e.g.,][]{Caglar:2020, Shankar:2019}. 
While larger samples allow comparing the relative slopes, the offset between the \mbh-\sigstar relation AGNs and quiescent galaxies remains unconstrained, because it is used to calibrate the sample-average virial factor \f.

Recent advancements have enabled more robust and independent methods for measuring \mbh in AGNs.
Compared to classical RM, velocity-resolved BLR lags from high signal-to-noise (S/N) and high-cadence spectroscopic data allow to resolve the BLR gas-flow structure \citep[e.g.,][]{Blandford&McKee:1982, Horne:2004}.
For the datasets with the highest S/N, it is possible to extract more detailed properties of the BLR.
However, the information is convolved with the BLR signal through the so-called transfer function, which describes the intrinsic time-delay distribution of the broad emission line \citep{Peterson:1993, Skielboe:2015}.
To overcome degeneracies arising from similar BLR geometries, \cite{Pancoast:2011} have introduced the Bayesian Code for AGN Reverberation and Modeling of Emission Lines (\texttt{CARAMEL}).
\texttt{CARAMEL} provides a phenomenological description of the BLR dynamics, and thereby the inference of the BLR parameters and associated uncertainties in RM datasets.
This method yielded precise and independent \mbh measurements, for a statistically meaningful sample of 30 objects \citep[e.g.,][ for a recent compilation see \citealt{Shen:2024}]{Brewer:2011, Li:2013, Pancoast:2014, Pancoast:2018, Li:2013, Williams:2018, Williams:2022, Villafana:2022, Bentz:2022}.

A novel third method involves spatially resolving the BLR, allowing for independent measurements of \mbh. What was first deemed impossible due to the small angular size of the BLR ($\sim 10^{-4}$ arcsec) has become technically feasible with GRAVITY, the second-generation NIR beam combiner at the Very Large Telescope Interferometer (VLTI). 
The differential phase measures how the photo-center shifts at different wavelengths of the broad line emission compared to the continuum. Fitting the full differential phase spectra (rather than the time-resolved RM data) with a BLR model allows constraining the BLR structure and kinematics.
Based the same BLR model parameterization as for \texttt{CARAMEL} \citep{Pancoast:2014}, so far six objects have robust \mbh from this technique \citep{GravityCollaboration:2018, GravityCollaboration:2020, GravityCollaboration:2024}.
As this approach provides another independent method to constrain \mbh, this sample is complementary to the AGNs modeled with \texttt{CARAMEL}.

In terms of host galaxy's contribution in shaping the \mbh-\sigstar relation, previous studies suggested a dependence on host morphology. 
Specifically, galaxies with structures like bars and pseudobulges deviate from the elliptical-only relation seen in quiescent galaxies \cite[e.g.,][]{Graham:2008, Hu:2008,Gultekin:2009}. 
This morphological dependence is particularly relevant for AGNs, as \sigstar measurements are typically based on single-aperture spectra, in which bars and pseudobulges are often unresolved \citep[e.g.,][]{Graham:2011, Woo:2013}.
Aperture-integrated kinematics often are the only diagnostic available when covering a large dynamic range in \mbh.
Consequently, inclination \citep{Xiao:2011, Bellovary:2014}, substructures \citep{Hartmann:2014} and rotational broadening from the disk contribution are likely impacting various recent calibrations of the AGN \mbh-\sigstar relation, such as e.g., \cite{Woo:2015, Caglar:2020, Caglar:2023}.
Long-slit spectroscopy partially addresses this challenge by resolving the host galaxy along the slit axis. 
Using this technique, \cite{Bennert:2015} demonstrated that \sigstar measurements can vary by up to 40\% on average across different definitions.
Nevertheless, slit orientation relative to substructures, such as bars, can still dramatically impact \sigstar \citep{Batiste:2017a}.
\cite{Batiste:2017b} find a 10\% shallower slope for the \mbh-\sigstar relation when accounting for rotational broadening in spatially resolved AGN host galaxies. However, their re-calibration is indistinguishable from that of previous studies due to a small sample of only 10 RM AGNs.
Likewise, many previous studies suffered from a combination of lacking spatial resolution, poorly constrained \mbh and/or limited dynamic range in \mbh and \sigstar. 
In a recent study, \cite{Molina:2024} used spatially resolved kinematics in luminous AGNs from the Close AGN Reference Survey \citep[CARS,][]{Husemann:2022} and the Palomar-Green Bright Quasar Survey (PG). While they report no difference between the \mbh-\sig relation active and inactive galaxies, their calibration is still based on single-epoch BH mass estimates and did not consider biases from selecting the brightest type 1 AGNs.
These limitations have hindered an direct calibration of the AGN \mbh-\sigstar relation, quantifying its intrinsic scatter and identifying trends with AGN parameters or host-galaxy properties.\\

In this work, we use deep high-spectral-resolution integral-field spectroscopic (IFU) observations to spatially resolve \sigstar across various host-galaxy components in a robust local AGN sample. 
High angular resolution imaging from Hubble Space Telescope (HST) will be used in a companion paper to decompose the host galaxy into its morphological components \citep[][in prep.]{Bennert:2024}. 
We match the apertures for stellar kinematics extraction to the radii determined from imaging, addressing aperture effects to account for differences in galaxy morphologies, AGN luminosities, and distances. 
This approach ensures a consistent framework for calibrating the \mbh-\sigstar across a wide range of AGN properties.

This paper is organized as follows.
Sect.~\ref{Sec:Sample_Selection} covers sample selection, while Sect.~\ref{Sec:Observations_and_Data_Reduction} details the IFU observations and data reduction, and Sect.~\ref{Sec:Analysis} the data analysis.
In Sect.~\ref{Sec:Results&Discussion} we present and discuss the \mbh-\sigstar relation in the context of previous work. 
Sect.~\ref{Sec:Summary} provides a summary. The appendix includes details on fitting procedures, comparisons of different IFU datasets, and the impact of the AGN subtraction method in our 3D spectroscopic data.
Throughout this work, we have adopted $ H_0 = 67.8 \, {\rm km/s/Mpc}$, $\Omega_{\rm m}$ = 0.308, and $\Omega_{\rm vac} = 0.692$ \citep{Planck:2016}.
In the following, we refer to the stellar velocity dispersion, commonly denoted as \sigstar, as \sig.


\section{Sample Selection}
\label{Sec:Sample_Selection}
\subsection{AGN Sample}
\label{SubSec:AGN_sample_selection}
\begin{deluxetable*}{lccccccccc}
\tabletypesize{\small} 
\tablecaption{ \emph{Black hole masses.} \label{tbl:bh_parameters}} 
\vspace{-10pt} 
\tablehead{ 
 \colhead{AGN Name} &  \colhead{Sample}  &  \colhead{$ \tau_{\rm cen}^{\rm rms}$} &  \colhead{$ \tau $} &  \colhead{$ \sigma_{\rm line}^{\rm rms}$} &  \colhead{$v$-ind.} &  \colhead{$ {\rm VP}$} &  \colhead{$M_{\rm BH}$}  & \colhead{$ M_{\rm BH}$} &  \colhead{${\rm log} f_{\rm dyn}$} 
 \vspace{-6pt}  \\  
 & \colhead{}  &  \colhead{$[{\rm d}] $} &   \colhead{$\,{\rm Ref.}$} &  \colhead{$[{\rm km\,s}^{-1}]$} &   \colhead{$\,{\rm Ref.}$} &  \colhead{$\,[{\rm log\,M}_\odot]$} & \colhead{$\,[{\rm log\,M}_\odot]$} & \colhead{${\rm Ref.}$ }   
 \vspace{-6pt}   
          } 
\colnumbers  
\startdata 
NGC\,3227          &  CARAMEL            &  $4.03 ^{+ 0.85}_{- 0.94}$   &  B23a               &  $1682 \pm {39}$  &  B23a               &  $6.35 ^{+ 0.11}_{- 0.13}$  &  $7.04 ^{+ 0.11}_{- 0.11}$  &  B23a               &  $0.72 ^{+ 0.15}_{- 0.17}$ \\  
 NGC\,6814          &  CARAMEL            &  $6.64 ^{+ 0.87}_{- 0.90}$   &  B09b               &  $1610 \pm {108}$  &  B09b               &  $6.52 ^{+ 0.11}_{- 0.12}$  &  $6.42 ^{+ 0.24}_{- 0.16}$  &  P14                &  $-0.14 ^{+ 0.26}_{- 0.20}$ \\  
 NGC\,4593          &  CARAMEL            &  $3.54 ^{+ 0.76}_{- 0.82}$   &  W18                &  $1601 \pm {40}$  &  B15                &  $6.25 ^{+ 0.11}_{- 0.13}$  &  $6.65 ^{+ 0.27}_{- 0.15}$  &  W18                &  $0.41 ^{+ 0.29}_{- 0.20}$ \\  
 NGC\,3783          &  CARAMEL            &  $9.60 ^{+ 0.65}_{- 0.72}$   &  B21a               &  $1619 \pm {137}$  &  B21a               &  $6.69 ^{+ 0.10}_{- 0.11}$  &  $7.45 ^{+ 0.07}_{- 0.10}$  &  B21b               &  $0.82 ^{+ 0.12}_{- 0.15}$ \\  
                    &  GRAVITY            &  -  &  -  &  -  &  -  &  -  &  $7.40 ^{+ 0.13}_{- 0.14}$  &  G21                &  - \\  
 NGC\,2617          &  cRM                &  $6.38 ^{+ 0.44}_{- 0.50}$   &  F17                &  $2424 \pm {91}$  &  F17                &  $6.86 ^{+ 0.06}_{- 0.07}$  &  $7.51 ^{+ 0.47}_{- 0.47}$  &  this work          &  *\,0.65 \\  
 IC\,4329\,A        &  CARAMEL            &  $16.33 ^{+ 2.59}_{- 2.28}$   &  B23b               &  $2112 \pm {93}$  &  B23b               &  $7.15 ^{+ 0.10}_{- 0.11}$  &  $7.64 ^{+ 0.53}_{- 0.25}$  &  B23b               &  $0.49 ^{+ 0.54}_{- 0.27}$ \\  
                    &  GRAVITY            &  -  &  -  &  -  &  -  &  -  &  $7.15 ^{+ 0.38}_{- 0.26}$  &  G24                &  - \\  
 Mrk\,1044          &  cRM                &  $10.50 ^{+ 3.30}_{- 2.70}$   &  D15                &  $^{\dagger}$$831 \pm {43}$  &  D15                &  $6.15 ^{+ 0.15}_{- 0.19}$  &  $6.45 ^{+ 0.12}_{- 0.10}$  &  this work          &  *\,0.65 \\  
 NGC\,5548          &  CARAMEL            &  $4.17 ^{+ 0.36}_{- 0.36}$   &  P17                &  $4115 \pm {513}$  &  P17                &  $7.14 ^{+ 0.14}_{- 0.15}$  &  $7.64 ^{+ 0.21}_{- 0.18}$  &  W20                &  $0.37 ^{+ 0.25}_{- 0.24}$ \\  
 NGC\,7469          &  cRM                &  $8.00 ^{+ 0.80}_{- 1.50}$   &  L21                &  $1485 \pm {34}$  &  L21                &  $6.53 ^{+ 0.08}_{- 0.09}$  &  $7.18 ^{+ 0.05}_{- 0.09}$  &  this work          &  *\,0.65 \\  
 Mrk\,1310          &  CARAMEL            &  $3.66 ^{+ 0.59}_{- 0.61}$   &  B09b               &  $755 \pm {138}$  &  B09b               &  $5.61 ^{+ 0.21}_{- 0.25}$  &  $7.42 ^{+ 0.26}_{- 0.27}$  &  P14                &  $1.63 ^{+ 0.34}_{- 0.37}$ \\  
 Mrk\,1239          &  GRAVITY            &  -  &  -  &  -  &  -  &  -  &  $7.47 ^{+ 0.15}_{- 0.92}$  &  G24                &  - \\  
 Arp\,151           &  CARAMEL            &  $3.99 ^{+ 0.49}_{- 0.68}$   &  B09b               &  $1252 \pm {46}$  &  B09b               &  $6.08 ^{+ 0.09}_{- 0.10}$  &  $6.62 ^{+ 0.10}_{- 0.13}$  &  P14                &  $0.51 ^{+ 0.14}_{- 0.16}$ \\  
 Mrk\,50            &  CARAMEL            &  $8.66 ^{+ 1.63}_{- 1.51}$   &  W18                &  $2020 \pm {103}$  &  B15                &  $6.84 ^{+ 0.12}_{- 0.13}$  &  $7.51 ^{+ 0.06}_{- 0.07}$  &  W18                &  $0.72 ^{+ 0.13}_{- 0.15}$ \\  
 Mrk\,335           &  CARAMEL            &  $18.86 ^{+ 1.81}_{- 2.34}$   &  G17                &  $1239 \pm {78}$  &  G17                &  $6.75 ^{+ 0.10}_{- 0.11}$  &  $7.25 ^{+ 0.10}_{- 0.10}$  &  G17                &  $0.59 ^{+ 0.14}_{- 0.15}$ \\  
 Mrk\,590           &  cRM                &  $20.50 ^{+ 4.50}_{- 3.00}$   &  P98                &  -  &  -  &  -  &  $7.58 ^{+ 0.07}_{- 0.07}$  &  this work          &  *\,0.65 \\  
 SBS\,1116+583A     &  CARAMEL            &  $2.31 ^{+ 0.62}_{- 0.49}$   &  B09b               &  $1528 \pm {184}$  &  B09b               &  $6.02 ^{+ 0.19}_{- 0.23}$  &  $6.99 ^{+ 0.32}_{- 0.25}$  &  P14                &  $0.96 ^{+ 0.37}_{- 0.34}$ \\  
 Zw\,229-015        &  CARAMEL            &  $3.86 ^{+ 0.69}_{- 0.90}$   &  B11                &  $1590 \pm {47}$  &  B11                &  $6.28 ^{+ 0.11}_{- 0.13}$  &  $6.94 ^{+ 0.14}_{- 0.14}$  &  W18                &  $0.66 ^{+ 0.18}_{- 0.19}$ \\  
 Mrk\,279           &  CARAMEL            &  $16.00 ^{+ 5.40}_{- 5.80}$   &  W18                &  $1778 \pm {7}$  &  B15                &  $6.99 ^{+ 0.13}_{- 0.19}$  &  $7.58 ^{+ 0.08}_{- 0.08}$  &  W18                &  $0.78 ^{+ 0.16}_{- 0.21}$ \\  
 Ark\,120           &  CARAMEL            &  $18.70 ^{+ 5.90}_{- 4.50}$   &  U22                &  $1882 \pm {42}$  &  U22                &  $7.11 ^{+ 0.13}_{- 0.16}$  &  $8.26 ^{+ 0.12}_{- 0.17}$  &  V22                &  $1.15 ^{+ 0.17}_{- 0.23}$ \\  
 3C\,120            &  CARAMEL            &  $25.90 ^{+ 2.30}_{- 2.30}$   &  G12                &  $1514 \pm {65}$  &  G12                &  $7.06 ^{+ 0.07}_{- 0.08}$  &  $7.84 ^{+ 0.14}_{- 0.19}$  &  G17                &  $0.75 ^{+ 0.16}_{- 0.21}$ \\  
 MCG\,+04-22-042    &  CARAMEL            &  $13.30 ^{+ 2.40}_{- 1.80}$   &  U22                &  $977 \pm {29}$  &  U22                &  $6.39 ^{+ 0.09}_{- 0.10}$  &  $7.59 ^{+ 0.42}_{- 0.28}$  &  V22                &  $1.06 ^{+ 0.43}_{- 0.30}$ \\  
 Mrk\,1511          &  CARAMEL            &  $5.44 ^{+ 0.67}_{- 3.00}$   &  W18                &  $1506 \pm {42}$  &  B15                &  $6.38 ^{+ 0.15}_{- 0.20}$  &  $7.11 ^{+ 0.20}_{- 0.17}$  &  W18                &  $0.63 ^{+ 0.25}_{- 0.26}$ \\  
 PG\,1310-108       &  CARAMEL            &  $7.20 ^{+ 2.41}_{- 3.11}$   &  W18                &  $^{\dagger}$$1978 \pm {104}$  &  B15                &  $6.74 ^{+ 0.19}_{- 0.26}$  &  $6.48 ^{+ 0.21}_{- 0.18}$  &  W18                &  $-0.26 ^{+ 0.28}_{- 0.31}$ \\  
 Mrk\,509           &  GRAVITY            &  -  &  -  &  -  &  -  &  -  &  $8.00 ^{+ 0.06}_{- 0.23}$  &  G24                &  - \\  
 Mrk\,110           &  CARAMEL            &  $27.80 ^{+ 4.30}_{- 5.10}$   &  U22                &  $1314 \pm {69}$  &  U22                &  $6.97 ^{+ 0.11}_{- 0.13}$  &  $7.17 ^{+ 0.67}_{- 0.26}$  &  V22                &  $0.20 ^{+ 0.68}_{- 0.29}$ \\  
 Mrk\,1392          &  CARAMEL            &  $26.70 ^{+ 3.50}_{- 3.90}$   &  U22                &  $1501 \pm {38}$  &  U22                &  $7.07 ^{+ 0.08}_{- 0.09}$  &  $8.16 ^{+ 0.11}_{- 0.13}$  &  V22                &  $1.01 ^{+ 0.13}_{- 0.16}$ \\  
 Mrk\,841           &  CARAMEL            &  $11.20 ^{+ 4.80}_{- 3.50}$   &  U22                &  $2278 \pm {96}$  &  U22                &  $7.05 ^{+ 0.17}_{- 0.24}$  &  $7.62 ^{+ 0.50}_{- 0.30}$  &  V22                &  $0.60 ^{+ 0.53}_{- 0.38}$ \\  
 Zw\,535-012        &  cRM                &  $20.30 ^{+ 8.10}_{- 4.60}$   &  U22                &  $1259 \pm {112}$  &  U22                &  $6.80 ^{+ 0.19}_{- 0.24}$  &  $7.57 ^{+ 0.15}_{- 0.10}$  &  this work          &  *\,0.65 \\  
 Mrk\,141           &  CARAMEL            &  $5.63 ^{+ 2.64}_{- 2.98}$   &  W18                &  $^{\dagger}$$2473 \pm {125}$  &  B15                &  $6.83 ^{+ 0.22}_{- 0.35}$  &  $7.46 ^{+ 0.15}_{- 0.21}$  &  W18                &  $0.70 ^{+ 0.27}_{- 0.40}$ \\  
 RBS\,1303          &  CARAMEL            &  $18.70 ^{+ 3.40}_{- 4.30}$   &  U22                &  $1292 \pm {156}$  &  U22                &  $6.78 ^{+ 0.18}_{- 0.21}$  &  $6.79 ^{+ 0.19}_{- 0.11}$  &  V22                &  $0.04 ^{+ 0.26}_{- 0.24}$ \\  
 Mrk\,1048          &  CARAMEL            &  $7.40 ^{+ 9.70}_{- 9.40}$   &  U22                &  $1726 \pm {76}$  &  U22                &  -  &  $7.79 ^{+ 0.44}_{- 0.48}$  &  V22                &  - \\  
 Mrk\,142           &  CARAMEL            &  $2.74 ^{+ 0.73}_{- 0.83}$   &  B09b               &  $859 \pm {102}$  &  B09b               &  $5.59 ^{+ 0.21}_{- 0.26}$  &  $6.23 ^{+ 0.30}_{- 0.30}$  &  L18                &  $0.74 ^{+ 0.36}_{- 0.39}$ \\  
 RX\,J2044.0+2833   &  CARAMEL            &  $14.40 ^{+ 1.60}_{- 1.90}$   &  U22                &  $870 \pm {50}$  &  U22                &  $6.33 ^{+ 0.10}_{- 0.11}$  &  $7.09 ^{+ 0.17}_{- 0.17}$  &  V22                &  $0.66 ^{+ 0.20}_{- 0.20}$ \\  
 IRAS\,09149-6206   &  GRAVITY            &  -  &  -  &  -  &  -  &  -  &  $8.00 ^{+ 0.30}_{- 0.40}$  &  G20                &  - \\  
 PG\,2130+099       &  CARAMEL            &  $9.60 ^{+ 1.20}_{- 1.20}$   &  G12                &  $1825 \pm {65}$  &  G12                &  $6.79 ^{+ 0.08}_{- 0.09}$  &  $6.92 ^{+ 0.24}_{- 0.23}$  &  G17                &  $0.00 ^{+ 0.25}_{- 0.25}$ \\  
 NPM\,1G+27.0587    &  CARAMEL            &  $8.00 ^{+ 4.70}_{- 4.50}$   &  U22                &  $1735 \pm {136}$  &  U22                &  $6.67 ^{+ 0.26}_{- 0.44}$  &  $7.64 ^{+ 0.40}_{- 0.36}$  &  V22                &  $0.93 ^{+ 0.48}_{- 0.57}$ \\  
 RBS\,1917          &  CARAMEL            &  $11.90 ^{+ 4.30}_{- 3.90}$   &  U22                &  $851 \pm {154}$  &  U22                &  $6.22 ^{+ 0.27}_{- 0.36}$  &  $7.04 ^{+ 0.23}_{- 0.35}$  &  V22                &  $0.54 ^{+ 0.36}_{- 0.50}$ \\  
 PG\,2209+184       &  CARAMEL            &  $13.70 ^{+ 2.80}_{- 2.90}$   &  U22                &  $1353 \pm {64}$  &  U22                &  $6.69 ^{+ 0.12}_{- 0.14}$  &  $7.53 ^{+ 0.19}_{- 0.20}$  &  V22                &  $0.72 ^{+ 0.23}_{- 0.25}$ \\  
 PG\,1211+143       &  cRM                &  $103.00 ^{+ 32.00}_{- 44.00}$   &  K00                &  $^{\dagger}$$981 \pm {120}$  &  K00                &  $7.28 ^{+ 0.24}_{- 0.31}$  &  $8.07 ^{+ 0.11}_{- 0.15}$  &  this work          &  *\,0.65 \\  
 PG\,1426+015       &  cRM                &  $115.00 ^{+ 49.00}_{- 68.00}$   &  K00                &  $^{\dagger}$$3345 \pm {471}$  &  K00                &  $8.40 ^{+ 0.29}_{- 0.44}$  &  $9.02 ^{+ 0.11}_{- 0.15}$  &  this work          &  *\,0.65 \\  
 Mrk\,1501          &  CARAMEL            &  $15.50 ^{+ 2.20}_{- 1.80}$   &  G12                &  $3321 \pm {107}$  &  G12                &  $7.52 ^{+ 0.08}_{- 0.09}$  &  $7.86 ^{+ 0.20}_{- 0.17}$  &  G17                &  $0.34 ^{+ 0.22}_{- 0.19}$ \\  
 PG\,1617+175       &  cRM                &  $34.30 ^{+ 6.80}_{- 3.80}$   &  H21                &  $1288 \pm {347}$  &  H21                &  $7.04 ^{+ 0.27}_{- 0.35}$  &  $7.69 ^{+ 0.21}_{- 0.38}$  &  this work          &  *\,0.65 \\  
 PG\,0026+129       &  cRM                &  $126.80 ^{+ 37.50}_{- 32.50}$   &  P04                &  $1719 \pm {495}$  &  P04                &  $7.86 ^{+ 0.33}_{- 0.44}$  &  $8.50 ^{+ 0.07}_{- 0.11}$  &  this work          &  *\,0.65 \\  
 3C\,273            &  GRAVITY            &  $170.00 ^{+ 9.60}_{- 14.00}$   &  Z19                &  $1099 \pm {40}$  &  Z19                &  $7.60 ^{+ 0.06}_{- 0.06}$  &  $9.06 ^{+ 0.21}_{- 0.27}$  &  L22                &  $1.52 ^{+ 0.22}_{- 0.28}$ \\  
 \enddata 
\tablecomments{
AGNs are listed in order of increasing redshift. 
(1) {Most common identifier.} 
(2) {Sample based on \mbh-measurement.}
(3) {Cross-correlation H$\beta$ emission line lag.} 
(4) {Reference for H$\beta$ lag.} 
(5) {Velocity indicator. Values marked with ($^{\dagger}$) are estimated from $\sigma_{\rm line}^{\rm mean}$ or $\textrm{FWHM}_{\rm line}^{\rm mean}$} 
(6) {Reference for velocity indicator.} 
(7) {Virial Product as calculated from eq.~\ref{eq:virial_product}.} 
(8) {Black hole mass \mbh.} 
(9) {Reference for \mbh. "this work" indicates that we have standardized the $f$-factor.}
(10) {Independent $f$-factor inferred from dynamical modelling. (*) indicates the sample-average for cRM.}
Reference keys are 
                    P98:\,\cite{Peterson:1998}, 
                    K00:\,\cite{Kaspi:2000}, 
                    P04:\,\cite{Peterson:2004}, 
                    B09b:\,\cite{Bentz:2009b}, 
                    B11:\,\cite{Barth:2011}, 
                    G12:\,\cite{Grier:2012}, 
                    P14:\,\cite{Pancoast:2014}, 
                    B15:\,\cite{Barth:2015}, 
                    D15:\,\cite{Du:2015},
                    F17:\,\cite{Fausnaugh:2017}, 
                    G17:\,\cite{Grier:2017}, 
                    P17:\,\cite{Pei:2017}, 
                    L18:\,\cite{Li:2018}, 
                    W18:\,\cite{Williams:2018}, 
                    Z19:\,\cite{Zhang:2019}, 
                    G20:\,\cite{GravityCollaboration:2020}, 
                    W20:\,\cite{Williams:2020}, 
                    B21a:\,\cite{Bentz:2021a}, 
                    B21b:\,\cite{Bentz:2021b}, 
                    G21:\,\cite{GravityCollaboration:2021}.
                    H21:\,\cite{Hu:2021}, 
                    L21:\,\cite{Lu:2021}, 
                    L22:\,\cite{Li:2022}, 
                    V22:\,\cite{Villafana:2022}, 
                    U22:\,\cite{U:2022}, 
                    B23a:\,\cite{Bentz:2023a}, 
                    B23b:\,\cite{Bentz:2023b}, 
                    G24:\,\cite{GravityCollaboration:2024}.
                } 
\end{deluxetable*}

\vspace*{-2\baselineskip} 
The core sample for this work are AGNs with velocity-resolved BLR lags that have been modeled with \texttt{CARAMEL}.
Since this technique constrains the virial factor \f individually ($f_{\rm dyn}$), a major source of systematic uncertainty is eliminated compared to \mbh from classical RM (cRM). 
In other words, AGNs with dynamically modeled \mbh provide the most pristine sample for inferring the underlying scaling relations.
Furthermore, dynamical modeling reduces the statistical uncertainties of individual measurements from $\sim 0.4$\,dex (SE) and $\sim 0.3\,{\rm dex}$ (cRM), to typically $0.2\,{\rm dex}$ \citep{Pancoast:2014, Villafana:2022}.
Thanks to a number of recent campaigns, the sample of CARAMEL AGNs has grown to 30 objects (for a recent compilation see \citealt{Shen:2024}), covering a large range in BH masses and AGN luminosities (${\rm log}(M_{\rm BH}/M_\odot) \sim 6.4-8.3 \,M_\odot; 0.01\leq z\leq 0.16$).

In addition, we complement the sample by AGNs whose \mbh has been measured from spatially resolving the BLR with VLTI/GRAVITY.
This has been achieved for a total of seven objects so far \citep{GravityCollaboration:2018, GravityCollaboration:2020, GravityCollaboration:2021, GravityCollaboration:2024}, of which NGC\,3783 and IC\,4329A overlap with the CARAMEL AGN sample. 
Of the remaining five, we include the four that have deep optical IFU observations plus broad-band HST imaging publicly available: Mrk\,1239, Mrk\,509, IRAS\,09149-6206, 3C\,273. 
In the following, we refer to those six objects as GRAVITY AGNs.

To further increase the range of AGN luminosities, \mbh and host morphologies, but without sacrificing data quality, we additionally include the complete set of unobscured AGNs that have {\it (i)} \mbh determined from cRM, {\it (ii)} existing deep optical 3D spectroscopy and {\it (iii)} archival broad-band imaging at high angular resolution from HST. 
In the following, we refer to these 10 objects as cRM AGNs.
In total, our extended sample consists of 44 objects: 30 CARAMEL AGNs, 6 GRAVITY AGNs, and 10 cRM AGNs.

\subsubsection{Black Hole Masses}
\label{SubSubSec:Black_Hole_Masses}

The black hole masses \mbh for the entire sample are listed in Table~\ref{tbl:bh_parameters}, with column (2) indicating the technique used for \mbh determination. 
For CARAMEL and GRAVITY AGNs, \mbh was determined independently, without assuming the virial factor \f, avoiding assumptions about BLR geometry. NGC\,3783 and IC\,4329A, present in both samples, have \mbh values consistent between both techniques. 
The \mbh from CARAMEL is used for the analysis unless stated otherwise. 
NGC\,3227 is the only AGN with \mbh measured using a third technique, stellar dynamical modeling, suitable for nearby galaxies where the BH sphere of influence is spatially resolved \citep{Davies:2006}. 
The value from this method ${\rm log}(M_{\rm BH}/M_\odot)= 7.0 \pm 0.3$ agrees with the ${\rm log}(M_{\rm BH}/M_\odot)= 7.04 \pm 0.11$ from \texttt{CARAMEL} modeling \citep{Bentz:2023a}, with the latter adopted for analysis.

The cRM AGNs require assuming an \f-factor to determine \mbh.
Previous studies have used different calibrations of $\langle f \rangle$ for deriving \mbh, e.g. 5.5 \citep{Onken:2004}, 5.2 \citep{Woo:2010}, 2.8 \citep{Graham:2011}, 5.1 \citep{Park:2012}, 4.3 \citep{Grier:2013}, or 4.8 \citep{Batiste:2017b}.
For consistency, we standardize the virial product VP by computing it from the broad H$\beta$ emission line time lag, $\tau_{\rm cen}^{\rm rms}$, and the line dispersion $\sigma_{\rm line}^{\rm rms}$ via
\begin{equation}
\label{eq:virial_product}
    {\rm VP} = c\tau_{\rm cen}^{\rm rms} \sigma_{\rm line}^{\rm rms} {}^2 /G .
\end{equation}
If $\sigma_{\rm line}^{\rm rms}$ is unavailable, we estimate it using the relation with $\sigma_{\rm line}^{\rm mean}$, or, if both are not available, ${\rm FWHM}_{\rm line}^{\rm mean}$ \citep[][their table~3]{DallaBonta:2020}.
We then adopt the virial factor of ${\rm log}\,f=0.65$ ($f =4.47$) from \cite{Woo:2015}, consistent with the average of the individual values ${\rm log}\,f_{\rm dyn} = 0.66 \pm 0.07$ determined here \citep[see also][]{Villafana:2023} to derive the BH masses via
\begin{equation}
\label{eq:MBH_f_factor}
M_{\rm BH} = f \frac{\langle \Delta v\rangle^2 \,R_{\rm BLR}}{G}
\end{equation}
where $G$ is the gravitational constant. 
A summary of \mbh, H$\beta$ time lags, line widths, and virial products is provided in Table~\ref{tbl:bh_parameters}. 

\subsection{Quiescent Galaxy Sample}
\label{SubSec:Quiescent_Galaxy_Sample}

To compare the AGN scaling relations between \mbh and \sig to those of quiescent galaxies, we adopt the sample from \citeauthor{Kormendy_Ho:2013} (\citeyear{Kormendy_Ho:2013}; KH13 in the following).
This sample includes 8 local galaxies with \mbh measurements based on dynamical modeling of spatially resolved stellar kinematics. 
Of 86 galaxies in total, we include 44 elliptical galaxies, 20 spiral and S0 galaxies with classical bulges, and 21 spiral and S0 galaxies with pseudobulges. 
%
While more recent compilations extend to lower galaxy masses, definition of host galaxy parameters in the KH13 sample is closest to our properties used in the following analysis. In particular the bulge dynamical mass derived from the spheroid effective radius, allowing for a consistent comparison.
We have tested that changing the quiescent sample to those from \cite{McConnell_Ma:2013} or \cite{vandenBosch:2016} does qualitatively not affect the conclusions.
%

\section{Observations and Data Reduction}
\label{Sec:Observations_and_Data_Reduction}

\subsection{IFU Observations} 
\label{SubSec:IFU_observations}
\begin{deluxetable*}{lccccccc}
\tabletypesize{\small} 
\tablecaption{ \emph{Observational parameters for the IFU data.} \label{tbl:obs_parameters}} 
\tablehead{ 
 \colhead{AGN Name} &  \colhead{$\alpha$ (J2000)} &  \colhead{$ \delta$ (J2000)} &  \colhead{Instrument} &  \colhead{UT Date} &  \colhead{$ t_{\rm exp}\, [{\rm s}]$} &  \colhead{$\theta_{\rm FWHM}$} &  \colhead{Prog. ID}  
 \vspace{-6pt}  
          } 
\colnumbers  
\startdata 
NGC\,3227          &  10:23:30.57  &  $+$19:51:54.28  &  VLT/MUSE    &  2022-03-31  &  2660    &  0.96    &  0108.B-0838(A)   \\ 
NGC\,6814          &  19:42:40.64  &  $-$10:19:24.60  &  Keck/KCWI   &  2023-10-17  &  1650    &  1.06    &  2023B\_U114      \\ 
NGC\,4593          &  12:37:04.67  &  $-$05:04:10.79  &  VLT/MUSE    &  2019-04-28  &  4750    &  0.62    &  099.B-0242(B)    \\ 
NGC\,3783          &  11:39:01.70  &  $-$37:44:19.01  &  VLT/MUSE    &  2015-04-19  &  3600    &  0.90    &  095.B-0532(A)    \\ 
NGC\,2617          &  08:35:38.80  &  $-$04:05:18.00  &  VLT/MUSE    &  2020-12-23  &  2300    &  1.04    &  0106.B-0996(B)   \\ 
IC\,4329\,A        &  13:49:19.26  &  $-$30:18:34.21  &  VLT/MUSE    &  2022-04-01  &  2200    &  0.81    &  60.A-9100(A)     \\ 
Mrk\,1044          &  02:30:05.52  &  $-$08:59:53.20  &  VLT/MUSE    &  2019-08-24  &  1200    &  1.20    &  094.B-0345(A)    \\ 
NGC\,5548          &  14:17:59.54  &  $+$25:08:12.60  &  Keck/KCWI   &  2024-04-29  &  3305    &  0.83    &  2024A\_U118      \\ 
NGC\,7469          &  23:03:15.67  &  $+$08:52:25.28  &  VLT/MUSE    &  2014-08-19  &  2400    &  0.84    &  60.A-9339(A)     \\ 
Mrk\,1310          &  12:01:14.36  &  $-$03:40:41.10  &  Keck/KCWI   &  2024-04-29  &  3840    &  1.02    &  2024A\_U118      \\ 
Mrk\,1239          &  09:52:19.16  &  $-$01:36:44.10  &  VLT/MUSE    &  2021-01-27  &  4600    &  1.14    &  0106.B-0996(B)   \\ 
Arp\,151           &  11:25:36.17  &  $+$54:22:57.00  &  Keck/KCWI   &  2024-01-04  &  1890    &  1.22    &  2023B\_U114      \\ 
Mrk\,50            &  12:20:50.69  &  $+$02:57:21.99  &  Keck/KCWI   &  2018-02-08  &  900     &  1.62    &  2018B\_U171      \\ 
Mrk\,335           &  00:06:19.52  &  $+$20:12:10.50  &  Keck/KCWI   &  2023-10-17  &  2570    &  0.69    &  2023B\_U114      \\ 
Mrk\,590           &  02:14:33.56  &  $-$00:46:00.18  &  VLT/MUSE    &  2017-10-28  &  9900    &  0.76    &  099.B-0294(A)    \\ 
SBS\,1116+583A     &  11:18:57.69  &  $+$58:03:23.70  &  Keck/KCWI   &  2024-01-04  &  2840    &  1.22    &  2023B\_U114      \\ 
Zw\,229-015        &  19:03:50.79  &  $+$42:23:00.82  &  Keck/KCWI   &  2018-08-15  &  3600    &  1.01    &  2018B\_U012      \\ 
Mrk\,279           &  13:53:03.45  &  $+$69:18:29.60  &  Keck/KCWI   &  2024-04-30  &  5400    &  0.84    &  2024A\_U118      \\ 
Ark\,120           &  05:13:37.87  &  $-$00:12:15.11  &  Keck/KCWI   &  2018-02-08  &  4800    &  1.75    &  2018B\_U171      \\ 
3C\,120            &  04:33:11.09  &  $+$05:21:15.61  &  Keck/KCWI   &  2024-01-04  &  2760    &  1.12    &  2023B\_U114      \\ 
MCG\,+04-22-042    &  09:23:43.00  &  $+$22:54:32.64  &  Keck/KCWI   &  2018-02-08  &  5400    &  1.87    &  2018B\_U171      \\ 
Mrk\,1511          &  15:31:18.07  &  $+$07:27:27.90  &  Keck/KCWI   &  2024-04-29  &  5910    &  0.84    &  2024A\_U118      \\ 
PG\,1310-108       &  13:13:05.79  &  $-$11:07:42.40  &  Keck/KCWI   &  2024-04-29  &  5810    &  1.03    &  2024A\_U118      \\ 
Mrk\,509           &  20:44:09.75  &  $-$10:43:24.70  &  Keck/KCWI   &  2024-04-29  &  4830    &  1.18    &  2024A\_U118      \\ 
Mrk\,110           &  09:21:44.37  &  $+$52:30:07.63  &  Keck/KCWI   &  2018-02-08  &  5400    &  2.09    &  2018B\_U171      \\ 
Mrk\,1392          &  15:05:56.55  &  $+$03:42:26.33  &  Keck/KCWI   &  2018-02-08  &  4200    &  1.71    &  2018B\_U171      \\ 
Mrk\,841           &  15:01:36.31  &  $+$10:37:55.65  &  Keck/KCWI   &  2018-02-08  &  5400    &  2.01    &  2018B\_U171      \\ 
Zw\,535-012        &  00:36:20.98  &  $+$45:39:54.08  &  Keck/KCWI   &  2018-10-03  &  4500    &  1.13    &  2018B\_U012      \\ 
Mrk\,141           &  10:19:12.56  &  $+$63:58:02.80  &  Keck/KCWI   &  2024-01-04  &  3770    &  1.25    &  2023B\_U114      \\ 
RBS\,1303          &  13:41:12.88  &  $-$14:38:40.24  &  VLT/VIMOS   &  2009-04-27  &  2000    &  1.19    &  083.B-0801(A)    \\ 
Mrk\,1048          &  02:34:37.88  &  $-$08:47:17.02  &  VLT/MUSE    &  2015-01-12  &  1200    &  1.21    &  094.B-0345(A)    \\ 
Mrk\,142           &  10:25:31.28  &  $+$51:40:34.90  &  Keck/KCWI   &  2024-04-30  &  6690    &  0.76    &  2024A\_U118      \\ 
RX\,J2044.0+2833   &  20:44:04.50  &  $+$28:33:12.10  &  Keck/KCWI   &  2018-08-07  &  5400    &  0.85    &  2018B\_U012      \\ 
IRAS\,09149-6206   &  09:16:09.36  &  $-$62:19:29.56  &  VLT/MUSE    &  2024-05-08  &  1600    &  1.01    &  113.26SK.001(B)  \\ 
PG\,2130+099       &  21:30:01.18  &  $+$09:55:00.84  &  VLT/MUSE    &  2019-06-09  &  2440    &  0.53    &  0103.B-0496(B)   \\ 
NPM\,1G+27.0587    &  18:53:03.87  &  $+$27:50:27.70  &  Keck/KCWI   &  2023-10-20  &  6000    &  0.96    &  2023B\_U114      \\ 
RBS\,1917          &  22:56:36.50  &  $+$05:25:17.20  &  Keck/KCWI   &  2023-10-17  &  5550    &  0.83    &  2023B\_U114      \\ 
PG\,2209+184       &  22:11:53.89  &  $+$18:41:49.90  &  Keck/KCWI   &  2023-10-20  &  6960    &  0.78    &  2023B\_U114      \\ 
PG\,1211+143       &  12:14:17.67  &  $+$14:03:13.18  &  VLT/MUSE    &  2016-04-01  &  2800    &  0.66    &  097.B-0080(A)    \\ 
PG\,1426+015       &  14:29:06.57  &  $+$01:17:06.15  &  VLT/MUSE    &  2016-04-04  &  2800    &  0.45    &  097.B-0080(A)    \\ 
Mrk\,1501          &  00:10:31.01  &  $+$10:58:29.00  &  Keck/KCWI   &  2023-11-03  &  4050    &  0.82    &  2023B\_U114      \\ 
PG\,1617+175       &  16:20:11.27  &  $+$17:24:27.51  &  VLT/MUSE    &  2016-04-04  &  2800    &  0.52    &  097.B-0080(A)    \\ 
PG\,0026+129       &  00:29:13.70  &  $+$13:16:03.94  &  VLT/MUSE    &  2016-07-31  &  2250    &  0.62    &  097.B-0080(A)    \\ 
3C\,273            &  12:29:06.69  &  $+$02:03:08.59  &  VLT/MUSE    &  2016-03-31  &  4750    &  0.47    &  097.B-0080(A)    \\ 
\enddata 
\tablecomments{
AGNs are listed in order of increasing redshift (as in Table~\ref{tbl:bh_parameters}). 
(1) {AGN name.}  
(2) {Right ascension.}  
(3) {Declination.}  
(4) {IFU instrument used to conduct the observations.}  
(5) {Observing date.}  
(6) {Total on source exposure time combined for the final cube after rejecting low-quality individual exposures.}  
(7) {Seeing in the final combined cubes inferred from 2D Moffat modeling of the broad H$\beta$ intensity maps.}  
(8) {Proposal ID of the data set under which the program was executed. Our team has carried out the observations with VLT/MUSE and Keck/KCWI of under the Prog. IDs 097.B-0080(A) and U114, U118, U171 respectively. For approximately a quarter of the sample we collected archival data.}  
} 
\end{deluxetable*}

\vspace*{-2\baselineskip} 
Our team has carried out IFU observations for 33/44 of the AGNs in our sample.
For the remaining objects, archival IFU observations are available from public repositories. The details of the observations are provided in Table~\ref{tbl:obs_parameters}. In the following, we describe data acquisition and reduction.
\subsubsection{Keck/KCWI Observations}
\label{SubSubSec:Observations_KCWI}
Many of the AGNs were initially monitored in the LAMP2011 and LAMP16 RM campaigns to study BLR dynamics and measure \citep{Barth:2015, U:2022}.
We followed up with 3D spectroscopy of their host galaxies using the Keck Cosmic Web Imager \citep[KCWI,][]{Morrissey:2018}  on Keck II under several programs.
Key diagnostic features were the stellar absorption lines, in particular the \ion{Mg}{i}b$\lambda\lambda\lambda 5167,\,5173,\,5184$ triplet (hereafter \MgIb), and the \ion{Fe}{i}+\ion{Fe}{ii} complex.
KCWI was configured with the medium IFU slicer and medium-resolution blue grating, providing a 16\farcs5$\times$20\farcs4 FoV and 0\farcs69 spatial sampling, covering the 4700–5700\,\AA\ range optimized for H$\beta$, \oiii, and \MgIb+Fe lines.

Our observing programs followed the same general strategy: 
Given the rectangular shape of the KCWI FoV we chose its position angle (PA) such that the FoV major axis matches that of the galaxy as estimated from archival images.
For each object, we first took a short exposure (60-120\,s, depending on redshift and AGN luminosity) guaranteeing that at least one exposure is available, for which the AGN emission lines in the center are not saturated. We used this exposure to scale up the exposure time of the following frames such that the continuum in the center is close to saturation. For most objects, except nearby bright AGNs, this resulted in 600\,s or 990\,s science exposures, which we dither-offset by 1\arcsec along the FoV major axis in between adjacent exposures.
In between every other science exposure, we took sky frames by nodding away from the target (T) to obtain external sky exposures (S) e.g., sequence TSTTSTTST.
We chose sky pointings carefully such that they are at least 1\,arcmin away from the AGN, in blank patches of the sky as verified by SDSS, DSS and 2MASS images. 

The pilot program 2018B\_U171 began on Feb 8 2018, with observations under photometric conditions and 1.5-2\arcsec.
We observed during three more nights on Aug 7, Aug 15 and Oct 3 2018 under Prog.ID 2018B\_U012.
In total, our observations during 2018B\_U171 and 2018B\_U012 yielded data of eight AGNs from the LAMP2016 campaign (4200-5400$\,$s on-source times) and for Mrk\,50 from the the LAMP2011 campaign (900\,s on-source).
%
During program 2023B\_U114, conducted on four nights between October 2023 and January 2024, the setup of the BM grating was maintained while using the novel KCWI red arm.
We observed 10 AGNs from the LAMP16 campaign under mostly clear conditions, with total integration times from 1800\,s to 7200\,s.
%
Under program 2024A\_U118, we conducted two consecutive runs, observing the last seven objects from the LAMP2016 campaign with total integration times from 1800\,s to 7200\,s.
In addition, we collecting some more integration on RXJ\,2044.0+2833 and NPM1G\,+27.0587 to improve S/N.
Although observations since 2023 with the Keck Cosmic Reionization Mapper cover the \ion{Ca}{ii}$\lambda\lambda\lambda8498,\,8542,\,8662$ (hereafter \CaT), temporal variation in strong sky emission lines, made their accurate subtraction difficult. 
We tried methods like CubePCA and other approaches based on principal component analysis like the one from \cite{Gannon:2020}, but these were hindered by the absence of empty sky regions in the science exposure, or strong spatial variation of the science spectra.
As a result, we decided to rely solely on KCWI blue spectra for consistent analysis across the AGN sample.

We reduced the data using the Python KCWI Data Reduction Pipeline, including bias subtraction, flat field correction, and flux calibration. Additionally, we aligned science frames, replaced saturated pixels, and coadded reduced data cubes as described in our companion paper \cite[][in prep.]{Remigio:2024}. 
The [\ion{O}{i}$]\lambda$5577 sky emission line indicates an instrumental resolution of ${\rm FWHM}=0.95$\AA\ ($\sim 32$\,${\rm km/s}$), with a common wavelength coverage of $\sim$4700-5600\,\AA\ and 0.28$\,$\AA/pix sampling.

\subsubsection{VLT/MUSE Observations}
\label{SubSubSec:Oobservations_MUSE}
\begin{figure*}
   \centering
   \includegraphics[]{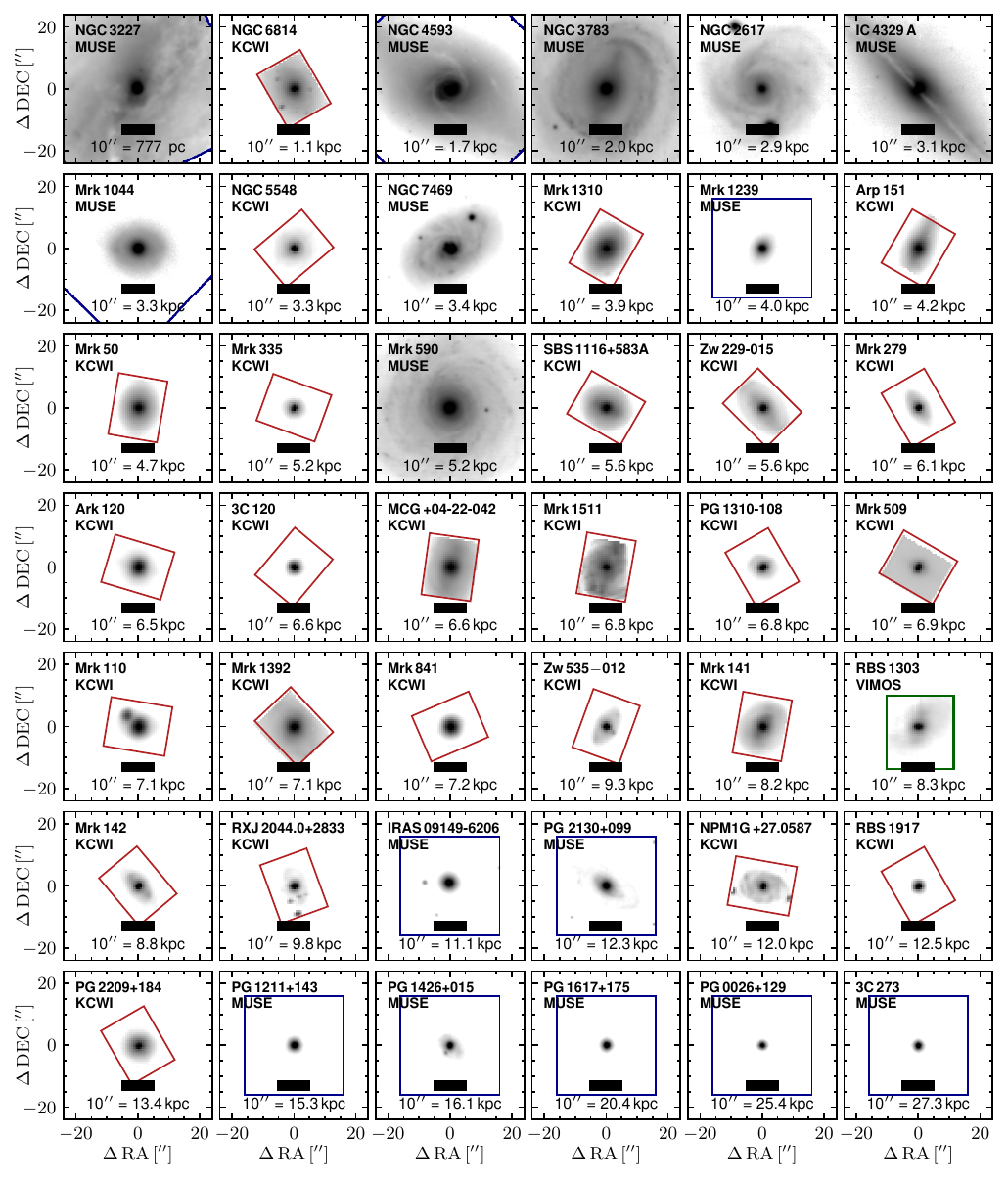} 
   \caption{
   \emph{Gallery of $V$-band images.}
   Images were reconstructed from the IFU data cubes (Table~\ref{tbl:obs_parameters}) with frames centered on the AGN position. North is up and east is to the left. Rectangles indicate the FoV covered by the IFU instrument, VLT/MUSE (blue), Keck/KCWI (red), VLT/VIMOS (green). For more distant AGNs observed with MUSE, cutouts are shown. AGN luminosity, host-galaxy sizes, IFU field coverage and depth of the observations vary substantially between the datasets.}
              \label{fig:gallery_cont}%
\end{figure*}
We acquired IFU observations for 8/44 AGNs using the Multi-Unit Spectroscopic Explorer (MUSE) at the Very Large Telescope (VLT). All observations were taken in MUSE wide field mode (WFM), covering a 1\arcmin $\times$1\arcmin FoV at 0\farcs2 sampling, and 4750–9300\,\AA\ spectral coverage at a spectral resolution of $R \sim 2500$. 
Observations were conducted across various programs with consistent strategies. Mrk\,1044 and Mrk\,1048 had already been observed as part of CARS, while five luminous cRM AGNs were observed under Prog.ID 097.B-0080(A) with integration times between 2800\,s and 4500\,s, employing standard dither-offset strategies. 
Observations were conducted in March, April, and July 2016 under gray moon and clear conditions with seeing of 0\farcs4-1\farcs0.
In addition, IRAS\,09149-6206 was observed under Prog.ID 113.26SK.001(B), with 260\,s exposures split into three observing blocks.
Observations on May 4 and 8, and June 8, 2024, achieved a total integration time of 3360\,s. 
For another nine CARAMEL AGNs and two cRM AGNs, we retrieved phase 3 archival data from the ESO archive.

We processed the data using MUSE pipeline v2.8.3-1 with ESO Reflex v2.11.0, following standard reduction procedures including bias frames, continuum lamp frames, arc lamp frames for wavelength calibration, standard star frames for flux calibration, and twilight flats. 
For AGN host galaxies covering only a small part of the FoV, we created a mean sky spectrum from the lowest 20\% flux in white light images and subtracted it from the cube. 
When the host galaxy filled the FoV, we used dedicated sky exposures from the archive. 
Telluric absorption bands were corrected by dividing the spectra by normalized transmission from standard star exposures taken close in time. 
Residuals in spectra arose from sky-line subtraction issues due to the timing of standard stars and spatial variations in the line spread function. 
To address these, we used \texttt{CubePCA}. This tool identifies the principal components (PCs) in the sky line residuals by fitting orthogonal eigenspectra to the individual spectra, and then subtracts the PCs.

\subsubsection{VLT/VIMOS Observations}
\label{SubSubSec:Observations_VIMOS}
Three of the 30 CARAMEL AGNs (RBS\,1303, PG\,1310-108, and NGC\,5548) were observed with the VIsible Multi-Object Spectrograph IFU (VIMOS) \citep{LeFevre:2003}.
The VIMOS blue and orange cubes cover wavelengths of 3700-5222\,\AA\ and 5250-7400\,\AA, respectively, with a 27$\arcsec \times$27$\arcsec$ FoV and 0\farcs6 pixel sampling. While PG\,1310-108 and NGC\,5548 have higher-resolution, deeper data from Keck/KCWI, RBS\,1303 was only observed with VIMOS. We used reduced data cubes from the Close AGN Reference Survey DR1 \citep{Husemann:2022}, which were initially processed with the \texttt{Py3D} package and included standard reduction steps. 
For specific details on data reduction, including exposure alignment and drizzling, see \cite{Husemann:2022}. 
Our analysis focuses on the blue cubes, as they cover the essential \ion{Mg}{I}b and \ion{Fe}{I} absorption lines for measuring stellar kinematics.

\subsubsection{AGN - Host Galaxy Deblending}
\label{SubSubSec:AGN-host_deblending}

The AGN featureless continuum and broad emission lines (in the wavelength range covered H$\beta$ and \ion{He}{i} and \ion{Fe}{ii}) can easily outshine the underlying host-galaxy spectrum.
It is therefore essential, to subtract the unresolved AGN emission before analyzing the faint host-galaxy emission.
For this task, we use the approach outlined by \cite{Husemann:2022}: 
(1) We first estimate the empirical point spread function (PSF) at and from the broad wavelengths available in each dataset, using \texttt{QDeblend\textsuperscript{3D}} \citep{Husemann:2013, Husemann:2014}.
(2) We model the PSFs with a 2D Moffat profile to suppress noise at large distances from the center. 
(3) If multiple broad lines are available, we interpolate the PSF as a function of wavelength.
(4) We reconstruct the intrinsic host-galaxy surface brightness profile from 2D image modeling.
(5) Finally, we iteratively subtract the point-like AGN emission from the extended host-galaxy emission combining the wavelength-dependent PSF with the host-galaxy surface brightness profile.
For more detailed description of the method and illustration of the deblending, we refer to \cite{Husemann:2022} and \cite{Winkel:2022}.

Deblending is crucial for accurately extracting host-galaxy stellar kinematics, as shown in Appendix~\ref{Appendix:Comparison_AGN-host_Deblending}.
Without deblending, the stellar velocity dispersion \sig can be overestimated by up to a factor of two, particularly near the AGN, which severely biases the luminosity-weighted mean \sig due to poorly fitted spaxels. 
An alternative is to fit the AGN spectrum simultaneously with the host-galaxy emission, as used for a subset of the LAMP AGNs by \cite[][in prep.]{Remigio:2024}.
This method, compared in Appendix~\ref{Appendix:Comparison_Spectral_Fitting_Codes}, generally provides results consistent with our deblending approach within the nominal uncertainties.

\subsection{HST Imaging}
\label{SubSec:HST_observations}
Considering the large range of AGN parameters in our sample, the host galaxies are also likely to cover a large range in stellar masses, sizes and morphologies.
To enable a consistent calibration of the scaling relations, we need a consistent measurement of the host-galaxy kinematics. 
This can be achieved by measuring the kinematics of different host-galaxy morphological components and separating their contributions to the galaxy-integrated kinematics.
We characterize the host-galaxy morphologies from high-resolution images obtained with HST. 
For 33/44 of AGNs in the sample, archival wide-field imaging data exist which were acquired with either WFC3/UVIS, ACS/HRC or WFPC2/PC1 in optical broad or medium bands.
The program HST-GO 17103 (PI: Bennert) acquired broad-band imaging from WFC3/UVIS for the remaining 11 objects of the CARAMEL AGN sample.
A detailed description of the data acquisition, data reduction, PSF subtraction, host-galaxy decomposition, 2D surface photometry and derived host-galaxy parameters will be presented in our companion paper \cite[][in prep.]{Bennert:2024}.

\begin{deluxetable*}{lcccccccccc}
\tabletypesize{\small} 
\tablecaption{ \emph{Host galaxy parameters.} \label{tbl:host_parameters}} 
\tablehead{ 
 \colhead{AGN Name } &  \colhead{Alt. Name } &  \colhead{$z$ }&  \colhead{scale } &  \colhead{Morph. } &  \colhead{Morph. } &  \colhead{Model } &  \colhead{$R^{\rm gal}_{\rm eff}  $ } &   \colhead{$R^{\rm bulge}_{\rm eff} $ } &  \colhead{$i$} &  \colhead{Comment}  
 \vspace{-6pt}  \\  
 \colhead{} & \colhead{} & \colhead{} & \colhead{[kpc/$\arcsec$]} &  &  \colhead{Ref.} &    &  \colhead{$[^{\prime\prime}]$} &  \colhead{$[^{\prime\prime}]$} & \colhead{$[^\circ]$}  
 \vspace{-6pt}   
          } 
\colnumbers  
\startdata 
NGC\,3227          &                     &  0.004  &  0.08  &  SAc                &  this work          &  sd  &  28.6  &  1.7  &  65  &                 \\  
 NGC\,6814          &                     &  0.005  &  0.11  &  SABc               &  S11                &  sdb  &  45.1  &  1.3  &  22  &                 \\  
 NGC\,4593          &  Mrk\,1330          &  0.008  &  0.17  &  SABb               &  S11                &  sdb  &  16.9  &  6.2  &  46  &                 \\  
 NGC\,3783          &                     &  0.010  &  0.20  &  SABb               &  V91                &  sdb  &  14.0  &  2.2  &  33  &                 \\  
 NGC\,2617          &  LEDA\,24141        &  0.014  &  0.29  &  SAa                &  this work          &  sd  &  12.1  &  1.2  &  18  &                 \\  
 IC\,4329\,A        &  RBS\,1319          &  0.015  &  0.31  &  SA                 &  V91                &  sd  &  $^{\wedge}$35.2  &  2.9  &  20  &                 \\  
 Mrk\,1044          &  HE\,0227-0913      &  0.016  &  0.33  &  SABc               &  this work          &  sdb  &  6.0  &  0.8  &  28  &                 \\  
 NGC\,5548          &  Mrk\,1509          &  0.016  &  0.33  &  SAa                &  this work          &  sd  &  11.3  &  8.4  &  28  &  asym. morph.      \\  
 NGC\,7469          &  Mrk\,1514          &  0.017  &  0.34  &  SABc               &  this work          &  sd  &  9.3  &  8.3  &  65  &                 \\  
 Mrk\,1310          &  RBS\,1058          &  0.019  &  0.39  &  SAc                &  B19                &  sd  &  4.1  &  4.2  &  43  &                 \\  
 Mrk\,1239          &  LEDA\,28438        &  0.020  &  0.40  &  S0A                &  this work          &  s  &  3.2  &  3.2  &  41  &                 \\  
 Arp\,151           &  Mrk\,40            &  0.021  &  0.42  &  S0                 &  S11                &  s  &  3.2  &  3.2  &  67  &  interacting       \\  
 Mrk\,50            &  RBS\,1105          &  0.023  &  0.47  &  S0A                &  N10                &  s  &  4.0  &  4.0  &  39  &                 \\  
 Mrk\,335           &  PG\,0003+199       &  0.026  &  0.52  &  E                  &  K21                &  s  &  2.6  &  2.6  &  24  &                 \\  
 Mrk\,590           &  NGC\,863\,         &  0.026  &  0.52  &  SAa                &  S11                &  sdb  &  2.0  &  1.4  &  35  &                 \\  
 SBS\,1116+583A     &  Zw\,291-51         &  0.028  &  0.56  &  SABa               &  this work          &  sdb  &  4.1  &  0.6  &  28  &                 \\  
 Zw\,229-015        &                     &  0.028  &  0.56  &  SBd                &  K21                &  sdb  &  7.3  &  0.8  &  49  &                 \\  
 Mrk\,279           &                     &  0.030  &  0.61  &  SAa                &  this work          &  sd  &  4.2  &  2.3  &  50  &  companion         \\  
 Ark\,120           &  Mrk\,1095          &  0.033  &  0.65  &  SAa                &  this work          &  sd  &  5.7  &  2.0  &  30  &  asym. morph.      \\  
 3C\,120            &  Mrk\,1506          &  0.033  &  0.66  &  S0A                &  S11                &  s  &  2.7  &  2.7  &  39  &  tidal tails       \\  
 MCG\,+04-22-042    &  Zw\,121-75         &  0.033  &  0.66  &  SABb               &  this work          &  sdb  &  11.7  &  0.9  &  56  &                 \\  
 Mrk\,1511          &  NGC\,5940          &  0.034  &  0.68  &  SABc               &  B19                &  sd  &  11.6  &  0.5  &  40  &                 \\  
 PG\,1310-108       &  HE\,1310-1051\,    &  0.034  &  0.68  &  SABa               &  this work          &  sdb  &  3.2  &  0.4  &  24  &  tidal tails       \\  
 Mrk\,509           &                     &  0.035  &  0.69  &  *E2                &  B09a               &  s  &  2.4  &  2.4  &  39  &                 \\  
 Mrk\,110           &  PG\,0921+525       &  0.035  &  0.71  &  *S0                &  this work          &  s  &  1.5  &  1.5  &    &                 \\  
 Mrk\,1392          &  Zw\,48-115         &  0.036  &  0.71  &  SBb                &  this work          &  sdb  &  10.4  &  0.7  &  59  &                 \\  
 Mrk\,841           &  PG\,1501+106       &  0.036  &  0.72  &  E                  &  this work          &  s  &  3.6  &  3.6  &  18  &                 \\  
 Zw\,535-012        &  LEDA\,2172         &  0.048  &  0.93  &  SBb                &  this work          &  sdb  &  5.7  &  0.6  &  58  &                 \\  
 Mrk\,141           &  Zw\,313-11         &  0.042  &  0.82  &  SABa               &  B19                &  sdb  &  5.6  &  0.4  &  40  &  companion         \\  
 RBS\,1303          &  HE\,1338-1423      &  0.042  &  0.83  &  SBa                &  this work          &  sdb  &  7.1  &  0.9  &  53  &                 \\  
 Mrk\,1048          &  NGC\,985           &  0.043  &  0.84  &  SBc                &  S02                &  sd  &  11.9  &  2.7  &  46  &  interacting       \\  
 Mrk\,142           &  PG\,1022+519       &  0.045  &  0.88  &  SBa                &  this work          &  sdb  &  5.6  &  0.4  &  34  &                 \\  
 RX\,J2044.0+2833   &                     &  0.050  &  0.98  &  SBd                &  K21                &  sdb  &  4.2  &  0.2  &  46  &                 \\  
 IRAS\,09149-6206   &                     &  0.057  &  1.11  &  S0                 &  this work          &  s  &  5.2  &  5.2  &  49  &                 \\  
 PG\,2130+099       &  Mrk\,1513          &  0.064  &  1.23  &  Sa                 &  B09a               &  sd  &  2.5  &  0.3  &  52  &                 \\  
 NPM\,1G+27.0587    &                     &  0.062  &  1.20  &  SAb                &  this work          &  sd  &  6.5  &  0.6  &  38  &  companion         \\  
 RBS\,1917          &                     &  0.065  &  1.25  &  SB                 &  this work          &  sdb  &  1.7  &  0.1  &  23  &                 \\  
 PG\,2209+184       &                     &  0.070  &  1.34  &  S                  &  this work          &  sd  &  2.9  &  2.9  &  30  &                 \\  
 PG\,1211+143       &                     &  0.081  &  1.53  &  E2                 &  B09a               &  s  &  0.2  &  0.2  &    &                 \\  
 PG\,1426+015       &  Mrk\,1383          &  0.086  &  1.61  &  E2                 &  B09a               &  s  &  2.0  &  2.0  &    &                 \\  
 Mrk\,1501          &  PG\,0007+107       &  0.087  &  1.63  &  *S0                &  S11                &  s  &  5.3  &  5.3  &  52  &  companion         \\  
 PG\,1617+175       &  Mrk\,877           &  0.112  &  2.04  &  E2                 &  B09a               &  s  &  1.2  &  1.2  &    &                 \\  
 PG\,0026+129       &  RBS\,68            &  0.145  &  2.54  &  E1                 &  B09a               &  s  &  $^{*}$2.3  &  $^{*}$2.3  &    &                 \\  
 3C\,273            &  PG\,1226+023       &  0.158  &  2.73  &  E3                 &  B09a               &  s  &  $^{\dagger}$2.3  &  $^{\dagger}$2.3  &    &                 \\  
 \enddata 
\tablecomments{
AGNs are listed in order of increasing redshift (as in Table~\ref{tbl:bh_parameters}). 
(1) {Most common identifier.} 
(2) {Alternative identifier.} 
(3) {Source redshift from NED.} 
(4) {Physical scale of 1$^{\prime\prime}$.} 
(5) {Host Galaxy morphological classification, simplified to the de Vaucouleurs system. Values marked with ($^{*}$) are uncertain due to strong AGN blending.} 
(6) {Reference key for morphological classification.} 
(7) {Adopted parameterization for the host-galaxy morphology (s = Sérsic only, sd = Sérsic + Disk (n=1) fit; sdb = Sérsic + Disk (n=1) + Bar (n=0.5) fit, a detailed presentation will be outlined in \citealt{Bennert:2024}).} 
(8) {Galaxy effective radius from fitting a single Sérsic component.} 
(9) {Bulge effective radius.} 
(10) {Inclination based on disk axis ratio $a/b$ that is retrieved from the best-fitting \texttt{lenstronomy} model.} 
(11) {Additional note regarding host morphology.} 
($^\wedge$)\,{Adopted from NED.} 
($^{*}$)\,{Adopted from \cite{McLeod:2001}.} 
($^\dagger$)\,{Adopted from \cite{Bahcall:1997}.} 
Reference keys are 
                V91:\,\cite{deVaucouleurs:1991}, 
                S02:\,\cite{Salvato:2002}, 
                J04:\,\cite{Jahnke:2004}, 
                N10:\,\cite{Nair:2010}, 
                B09a:\,\cite{Bentz:2009a}, 
                S11:\,\cite{Slavcheva-Mihova:2011}, 
                A15:\,\cite{Ann:2015}, 
                B19:\,\cite{Buta:2019}, 
                K21:\,\cite{Kim:2021}. 
                } 
\end{deluxetable*}

\vspace*{-2\baselineskip} 

\subsubsection*{Special handling of individual objects}
The host galaxy decomposition based on HST/WFC3 images did not yield stable solutions for three objects at the very low- and high-redshift end.
For the nearby galaxy NGC\,3227, the WFC3 FoV covers only a small fraction of its 5.4\arcmin$\times$3.6\arcmin\ size. 
For NGC\,3227's galaxy effective radius, we adopt the scale radius from an exponential fit to the Sloan Digital Sky Survey (SDSS) photometry in the r-band (\emph{exprad\_r}). Although this approach assumes that the PSF has a minimal impact on NGC\,3227's light profile, it provides a quantity closest to the $R_{\rm eff}^{\rm gal}$ definition used for the other objects.
We encountered the same challenge for IC\,4329A, where the highly inclined galaxy extends beyond the HST ACS/HRC FoV.
While structural decomposition allows fitting the bulge, we adopt the galaxy scale length of 25\farcs2 from NED, that was fitted to the $k$-band photometry from 2MASS.

For PG\,0026+129, an extremely bright quasar, the host galaxy parameters recovered in Sect.\ref{SubSec:Surface_photometry} did not converge to stable solutions. Therefore, we adopted the host-galaxy effective radius $R_{\rm eff}^{\rm gal} = 2\farcs6$ from \cite{McLeod:2001}, which was estimated based on HST/NIC2 F160W imaging.
We encountered the same issue with the HST/WFC3 image of the bright quasar 3C\,273. We adopt an effective radius of 2\farcs3 for the host galaxy, as reported by \cite{Bahcall:1997}. 
Their measurement is based on HST/WFPC F606W imaging and is consistent with the 2\farcs3 - 2\farcs6 range reported by \cite{Martel:2003}, measured from coronagraphic imaging with HST/ACS in the $V$ and $I$ bands, respectively.

\section{Analysis}
\label{Sec:Analysis}

\subsection{Surface Photometry}
\label{SubSec:Surface_photometry}

For the purpose of this work, we are exclusively interested in the stellar kinematics of different host-galaxy components for which we adopt effective radii derived by fitting the 2D surface brightness profiles. 
For this task, we used the public code \texttt{lenstronomy} \citep{Birrer&Amara:2018}, as outlined by \cite{Bennert:2021}.
We measure the host-galaxy effective radius $R^{\rm gal}_{\rm eff}$, from the PSF-subtracted host-galaxy surface profile.
A universal parameterization of a single spheroidal component (s), i.e. using a single-Sérsic component as input for \texttt{lenstronomy}, as it is often done for marginally resolved high-redshift galaxies or massive elliptical galaxies.
In reality, however, only a minority of galaxies in our sample are well-described by a spheroidal model. The majority of our AGN hosts are late-type galaxies, with a large morphological diversity including bars, bulges and disks, which can be seen in the reconstructed continuum images in Fig.~\ref{fig:gallery_cont}.
The HST imaging allows us to decompose the host galaxy into its morphological components. 
For many nearby AGNs, morphological classifications are available in the literature, 
Based on the high-quality imaging data collected for this project \cite[][in prep.]{Bennert:2024}, we complemented (or revised) literature classifications, and standardized the nomenclature to the de Vaucouleurs system (see column (5) of Table.~\ref{tbl:host_parameters}).
We use this information as prior for parameterizing the host model, listed in Table~\ref{tbl:host_parameters}. Models include bulge-only (s), bulge+disk (sd) or bulge+disk+bar (sdb) components. The best-fitting effective radii of the entire galaxy and bulge-only, $R_{\rm eff}^{\rm gal}$ and $R_{\rm eff}^{\rm bulge}$, serve as standardized measure as across which stellar kinematics are extracted.
After running a minimum of ten decompositions for each object using different starting parameters, we estimate $0.1\,{\rm dex}$ systematic uncertainty for effective radii, and $0.2\,{\rm dex}$ if strong residuals from the PSF subtraction are present on scales of the spheroid.
More details on the HST imaging data, the fitting process, and the full set of parameters will be presented in our companion paper \cite[][in prep.]{Bennert:2024}.

\subsubsection*{Disk axis ratio as proxy for inclination}
The inclination of a galaxy disk can be estimated from its axis ratio as $i_{b/a} = \arccos(b/a)$.
However, structural decomposition carried out with \texttt{lenstronomy} is sensitive to the parameterization defined by the user. 
While we are careful to check the parameterization, systematic uncertainties from 
limited FoV, prominent dust lanes crossing the galaxy center, and PSF mismatches likely contribute systematic uncertainties to the structural decomposition \citep[][in prep., see also  Sec.~\ref{Subsec:Host_galaxy_morphologies}]{Bennert:2024}.
To test whether the disk axis ratio is a good proxy for the galaxy inclination, we compare $i_{b/a}$ with a visual estimate of the galaxy inclination $i_{\rm vis}$. In general, it is possible to estimate the inclination if the host galaxy can be robustly separated from the PSF, and a disk component is clearly visible. For the majority of the sample, we based our estimate on the original HST images. However, for NGC\,3227 and NGC\,4593, the WFC3 FoV covers only a fraction of the galaxy, so that we used the PanSTARRS $i$-band images.
We were able to estimate $i_{\rm vis}$ for each of the 29 disk galaxies, which are preferentially located at lower-redshift and show a prominent disk component. Depending on how well the galaxy is resolved, and how dominant the PSF is, we estimate that the associated uncertainties of $i_{\rm vis}$ range from approximately 10$^\circ$ to 20$^\circ$.
Overall, the visual estimates agree with the \texttt{lenstronomy} measurements within these uncertainties.
We conclude that $i_{b/a}$, derived from the disk axis ratio, is a suitable indicator for the galaxy inclination. In the following, we adopt $i_{b/a}$ as proxy for the galaxy inclination, and refer to it as $i$ as listed in Table~\ref{tbl:host_parameters}. 
As a side note, the consistent inclination values provide further evidence that the \texttt{lenstronomy} fits have resulted in realistic physical parameters of the host galaxy.

\subsection{AGN Parameters}
\label{SubSec:AGN_parameters}

\begin{figure*}
   \centering
   \includegraphics[]{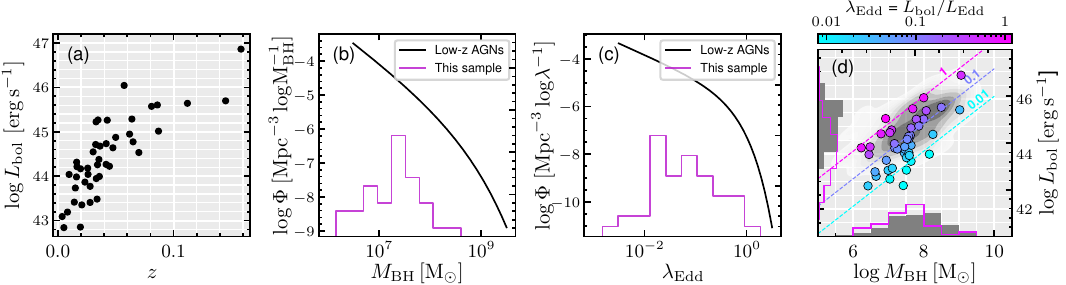} 
   \caption{
   \emph{Properties of our AGN sample.}
   From left to right the panels show the AGN sample in (a) AGN bolometric luminosity as a function of galaxy redshift, (b) distribution of BH masses (purple) the compared to BH mass function (BHMF) obtained from the local AGN population in the Hamburg/ESO (HES) survey and (c) Eddington ratio distribution compared with the global Eddington ratio distribution function (ERDF) from HES. Panels (b) and (c) show that our sample's distribution in \mbh (and $\lambda_{\rm Edd}$) is mostly shaped by the luminosity bias at the low-\mbh end, and the cut-off of the BHMF (ERDF) the at the massive end, respectively.
   Panel (d) shows that our AGNs sample a similar range in \mbh and luminosity compared to HES, characteristic of a flux-limited sample.
   }
              \label{fig:sample_properties}%
\end{figure*}

The AGNs in our sample were exclusively selected based on their spectral properties, more precisely the ability to temporally resolve the broad emission line lags. 
Considering that only a fraction of AGNs show the required variability to monitor them in RM campaigns, we are interested in quantifying to what extent our sample is representative of the overall AGN population. 
Important properties that can be easily compared are the AGN bolometric luminosity $L_{\rm bol}$, the BH mass \mbh and the Eddington ratio $\lambda_{\rm Edd}$.
They can be directly estimated from the unobscured AGN spectra available in the host-subtracted IFU data. 

We constrain the AGN spectral modeling to the H$\beta$-[\ion{O}{iii}] wavelength range, for which various studies have provided calibrations \cite[e.g.][]{, Kaspi:2000, Peterson:2004, Greene&Ho:2005, Vestergaard&Peterson:2006, Bentz:2013, Woo:2015}. 
A detailed description of our fitting methodology is given in Appendix~\ref{Appendix:AGN_spectral_fitting}.
We estimated the bolometric luminosity from the 5100\,\r{A} continuum luminosity using a bolometric correction factor: $L_{\rm bol} = 10 \times L_{5100}$ \citep{Richards:2006}. 
The Eddington ratio is $\lambda_{\rm Edd} = L_{\rm bol}/L_{\rm Edd}$, where $L_{\rm Edd} /{\rm erg\,s}^{-1} = 1.26 \times 10^{38} M_{\rm BH}/M_\odot$ with \mbh taken from Table~\ref{tbl:bh_parameters}.
The AGN parameters are shown in Fig.~\ref{fig:sample_properties}, where we compare our sample with the properties of the overall local AGN population in the flux-limited Hamburg ESO survey \citep{Wisotzki:2000, Schulze&Wisotzki:2010}. 
The unimodal distribution of our AGNs in \mbh (and $\lambda_{\rm Edd}$ analogously) can be explained by the primary sample selection criteria.
At low \mbh, the distribution is incomplete due to the low S/N of the AGN spectral features, whereas at high \mbh the number of AGNs decreases due the cut-off of the SMBH mass function. 
The selection effects are discussed in more detail in Sect.~\ref{SubSec:Controlling_selection_effects}.


\subsection{Spectral Synthesis Modeling}
\label{SubSec:Spectral_Synthesis_Modeling}
To determine the host-galaxy stellar kinematics, We used the first and second moments of the line-of-sight velocity distribution (LOSVD) obtained by fitting the stellar continuum after subtracting AGN emission (see Sect.~\ref{SubSubSec:AGN-host_deblending}).
However, data from Keck/KCWI, VLT/MUSE, and VLT/VIMOS vary in wavelength coverage, field coverage, and resolution. Additionally, the depth of observations and the brightness of the central AGN limit the mapping of stellar kinematics. To ensure a consistent analysis across datasets, we developed a common methodology.

The extraction of stellar kinematics involves several interconnected steps, each affecting the kinematic parameters. 
We tested various approaches to optimize results and maintain general applicability, with details provided in the Appendix.
\begin{itemize}
\item[(1)] We tested stellar kinematics extraction with \texttt{pPXF} \citep{Cappellari:2004,Cappellari:2017}, \texttt{PyParadise} \citep{Husemann:2016a}, and \texttt{BADASS} \citep{Sexton:2021}, all yielding consistent results despite differing methodologies. A detailed comparison is in Appendix~\ref{Appendix:Comparison_Spectral_Fitting_Codes}.

\item[(2)] We tested fitting different wavelength regions ([4750–5300\,\AA], [5150–5200\,\AA], [8450–8650\,\AA]), each containing key diagnostic features for stellar kinematics. A comparison is detailed in Appendix.~\ref{Appendix:Comparison_wavelength_regions}.

\item[(3)] We tested the robustness of our results using various stellar and SSP template libraries: the 2009 Galaxy Spectral Evolution Library \citep[CB09,][]{Bruzual&Charlot:2003}, the high-resolution SSP library from ELODIE \citep[M11,][]{Maraston&Stromback:2011}, the X-shooter Spectral Library \citep[XSL,][]{Verro:2022}, and the Indo-U.S. Library of Coudé Feed Stellar Spectra \citep{Valdes:2004}. A comparison of the impact on stellar kinematics is detailed in Appendix~\ref{Appendix:Comparison_templates}.

\item[(4)] For AGNs observed with multiple IFU instruments (e.g., VLT/MUSE plus Keck/KCWI or VLT/VIMOS), we verified the consistency of our method by analyzing them with the same procedures. Details are provided in Appendix~\ref{Appendix:Comparison_IFU_Datasets}.
\end{itemize}

\begin{figure*}
   \centering
   \includegraphics[width=\textwidth]{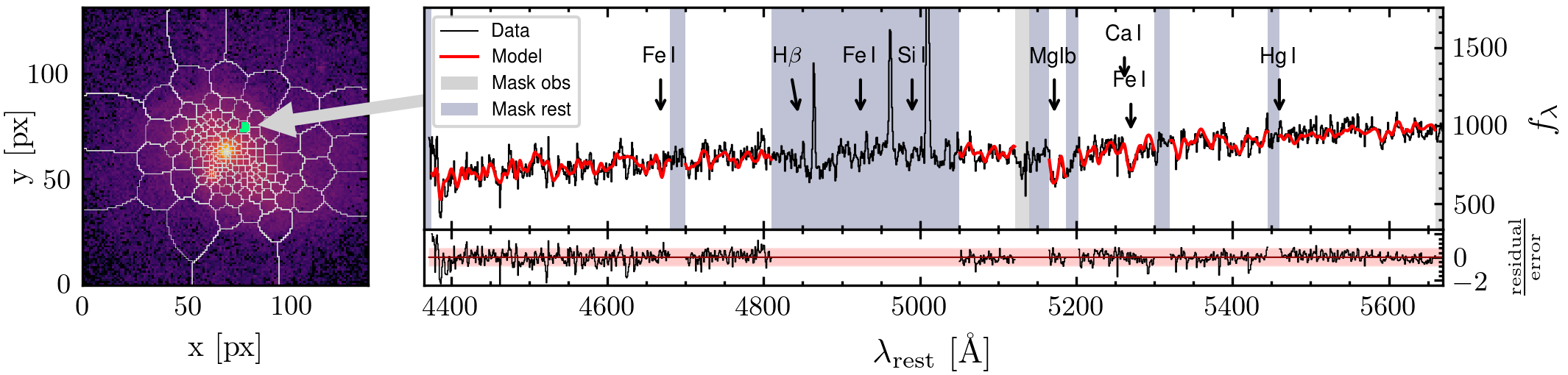} 
   \caption{
   \emph{Spatially resolved spectral synthesis modeling, demonstrated for PG~1426+015.} 
   The left panel shows a continuum image from integrating the wavelength range 5100-5200\,\r{A} of the AGN-subtracted MUSE data cube. The overlaid grid depicts the Voronoi cells, within which spectra are coadded to achieve a minimum S/N of 20. 
   The example spectrum from the arbitrary cell (highlighted green) is shown as a black line in the upper right panel. We constrain the spectral fitting to regions in the rest frame that are free from contamination from broad emission lines and strong narrow emission lines residuals (blue stripes). In the observed frame, we mask sky line residuals (gray stripes). The best-fit stellar continuum model (red line) closely reproduces the continuum emission within the 1$\sigma$ error, as illustrated by the normalized residuals shown in the bottom right panel.}
   \label{fig:Example_Spectral_Fitting}
\end{figure*}

After evaluating the options detailed in the appendices, we summarize our findings:

(1) Template Comparison: \texttt{PyParadise} is superior with large wavelength coverage, e.g., for MUSE spectra, while \texttt{pPXF} offers more robust stellar kinematics extraction for smaller wavelength ranges.

(2) Wavelength Range: A larger wavelength range provides more diagnostic features and better kinematic constraints. However, \CaT cleaned from sky line contamination is covered for objects observed with MUSE.
We adopted the 4750–5200\,\AA~range, which is covered by all datasets and contains key absorption features.

(3) Template Resolution: Higher spectral resolution reduces statistical uncertainties. 
Among higher-resolution templates, XSL and M11 yield consistent results, but XSL's greater number of spectra (130 versus 10) offers more robust absorption line reproduction and better kinematic fits.

(4) Instrumental Comparison: For objects observed with multiple instruments, deep MUSE observations generally provide the highest S/N stellar continuum and superior spatial resolution and field coverage compared to Keck/KCWI and VLT/VIMOS. 
Thus, we prefer VLT/MUSE data for our analysis when available.

For all objects, we adopt the following universal strategy:
After subtracting the point-like AGN emission as described in Sect.~\ref{SubSubSec:AGN-host_deblending}, we increase the S/N of the host-galaxy emission either by taking aperture-integrated spectra (see Sec.~\ref{SubSubSec:two_methods_sigma}), or by binning the cube using Voronoi tessellation to a spectral S/N of 20 in the rest-frame wavelength range 5100-5200\,\AA.
Next, we fit the stellar continuum emission in the 4750–5200\,\AA\ range using the \texttt{pPXF} code \citep{Cappellari:2004,Cappellari:2017}, typically with 5$^{\rm th}$-order polynomials to account for non-physical continuum variations from 3D-PSF subtraction. 
We mask the \ion{Na}{i}$\lambda\lambda$5890,\,5896 sky lines, as well as H$\gamma$, H$\beta$, [\ion{O}{iii}][\ion{O}{iii}]$\lambda\lambda$4960,\,5007 emission lines (hereafter [\ion{O}{iii}]), and the [\ion{O}{i}]$\lambda5577$ night sky line. An example spectrum from a MUSE data cube is shown in Fig.~\ref{fig:Example_Spectral_Fitting}, along with the best-fit stellar continuum model and residuals.

\subsection{Host-Galaxy Stellar Kinematics}
\label{SubSec:Host_Galaxy_Stellar_Kinematics}

Most previous studies investigating the \mbh-\sigstar relation have used aperture-integrated spectra to measure the AGN host-galaxy properties for large datasets \cite[e.g.,][and many more]{Treu:2004, Graham:2011, Grier:2013, Woo:2015, Caglar:2020, Caglar:2023}. 
The statistical power for calibrating scaling relations comes at the cost of larger uncertainties, for example, due to the unknown fraction of the host galaxy covered by the fibers. 
Long-slit spectroscopy in combination with high-resolution imaging has enabled resolving the host-galaxy kinematics along their photometric major axis \citep{Bennert:2015}. 
Thanks to IFU observations, we can now spatially resolve the host-galaxy kinematics in two dimensions, and differentiate them for different host-galaxy morphological components.

\begin{figure*}
   \centering
   \includegraphics[width=\textwidth]{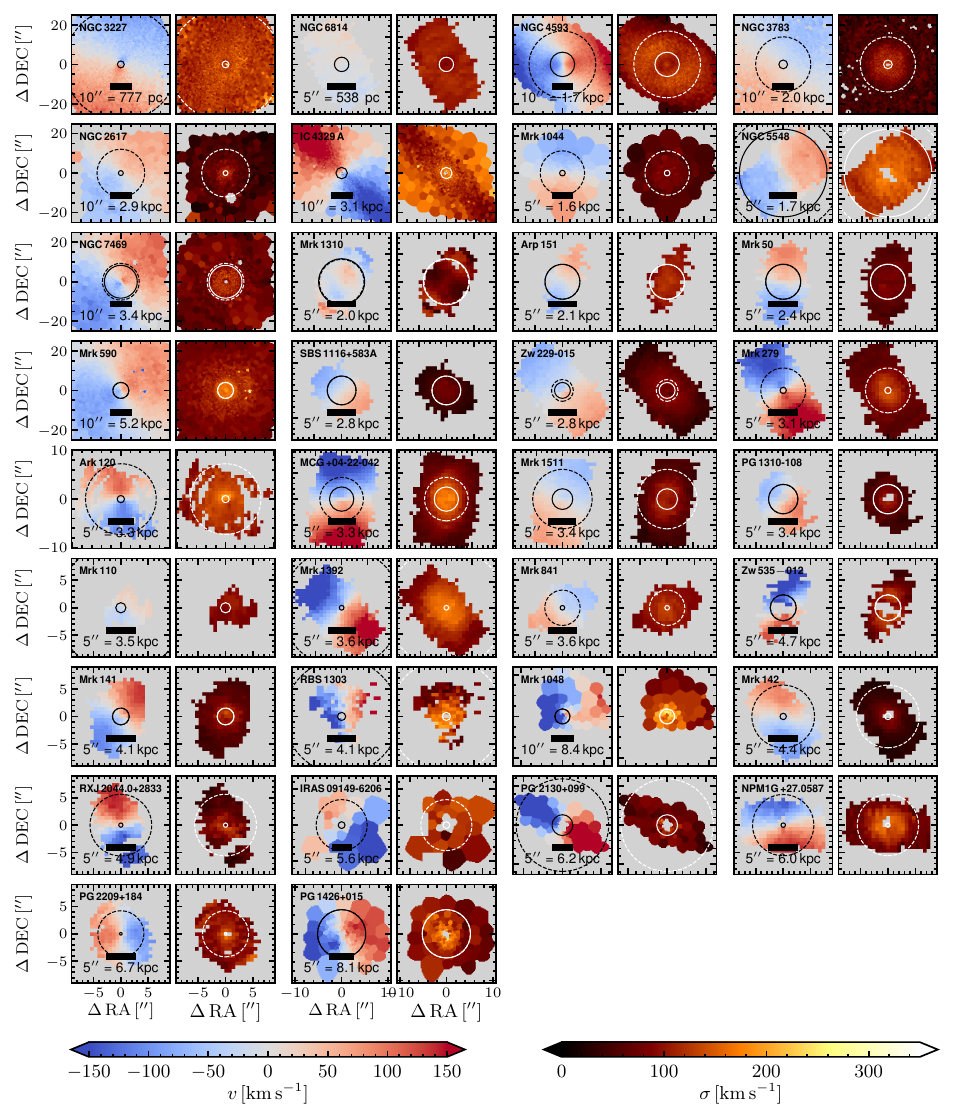} 
   \caption{
   \emph{Spatially resolved stellar kinematic maps of the AGN host galaxies.}
   We show the first moment (line-of-sight-velocity $v$) and second moment (dispersion $\sigma$), measured from 2D continuum modeling of the AGN-subtracted data cubes (see Sect.~\ref{SubSec:Spectral_Synthesis_Modeling}). 
  Two neighboring panels belong to the same objects and the velocity and dispersion colormaps share a common scaling which is indicated by the colorbars at the bottom. 
   Black (left panels) or white (right panels) circles indicate $R_{\rm eff}^{\rm bulge}$ (continuous line) and $R_{\rm eff}^{\rm gal}$ (dashed line), as described in Sect.~\ref{SubSec:Surface_photometry}.
   The AGN host galaxies show a large diversity in their kinematic structures. In some cases, either the kinematics in $R_{\rm eff}^{\rm bulge}$ cannot be resolved, or $R_{\rm eff}^{\rm gal}$ is larger than the FoV covered by the IFU.}
              \label{fig:gallery_kin}%
\end{figure*}

To determine if the kinematics are resolved, we required at least five Voronoi cells with constrained kinematics and centroids within the galaxy's effective radius.
The IFU observations are deep enough to spatially map the host-galaxy stellar kinematics for a 34/44 AGNs.
As illustrated in Fig.~\ref{fig:gallery_kin}, sub-kpc kinematic structures can be resolved in nearby systems. Examples are nuclear disks in NGC\,3227, NGC\,2617, or the counter-rotating disk in Mrk\,1310. 
Such features are commonly identified from photometric decomposition of barred galaxies \citep{Comeron:2010, Gadotti:2020} and have been referred to as "pseudobulges" \citep{Kormendy&Kennicutt:2004}.
Due to lower spatial resolution, kinematic substructures remain unresolved in more distant galaxies. In addition, by selection those distant galaxies tend to host more luminous AGNs. Their blending emission can hampers an accurate mapping of the hots galaxy stellar kinematics, so that for 12/44 galaxies, the galaxy kinematics cannot be spatially mapped. AGNs for which this is the case are typically high specific-accretion-rate AGNs like Mrk~335, Mrk\,1239 3C\,120 or the PG quasars contained in our sample.
Furthermore, we note that accessing the kinematics within $R_{\rm eff}^{\rm bulge}$ and $R_{\rm eff}^{\rm gal}$ can be limited by spatial resolution close to the AGN, or size of the FoV respectively.
Given these limitations, establishing a consistent method for extracting \sig is essential. 
This consistency will enable us to fully leverage the strength of this AGN sample, covering a broad range of \mbh, $L_{\rm bol}$, $z$, and host morphologies.

\subsubsection{Two Methods For Measuring \texorpdfstring{\sig}\ }
\label{SubSubSec:two_methods_sigma}

There is no standard definition for measuring the stellar velocity dispersion \sig from the spatially resolved first and second moments of the LOSVD. 
As a result, it is unclear over what fraction of $R_{\rm eff}^{\rm gal}$ the kinematics should be averaged or how this averaging should be performed. 
The literature presents two different approaches for measuring stellar velocity dispersion.
For measuring the kinematics within the \emph{bulge} effective radius of quiescent galaxies, several studies have favored including rotational broadening by explicitly combining the first and second velocity moments through eq.~\ref{eq:sigma_spat_re} (KH13 refer to this technique as Nuker team practice, e.g., \citealt{Pinkney:2003, Gultekin:2009, Cappellari:2013, vandenBosch:2016}, for AGNs also \citealt{Bennert:2015, Bennert:2021}).
This approach is motivated by the equipartition of energy in the dynamically relaxed bulge, where the combination of $v_{\rm spat}$ and $\sigma_{\rm spat}$ accurately traces the gravitational potential imposed by the stellar mass.
However, the bulge component is often barely resolved in AGN host galaxies, resulting in substantial contributions from disk rotation to dispersion being measured from aperture-integrated spectra. 
When removing rotational broadening through spatially resolving the LOSVD, \cite{Batiste:2017b} reported that \sig\ on average is 13 km/s lower \sig.
They underscore that the difference is strongest for inclined spiral galaxies with significant substructure, highlighting the necessity of maintaining a consistent definition. 
We briefly review the details of both methods for measuring \sig specific to our sample.
\begin{figure}
   \centering
   \includegraphics[]{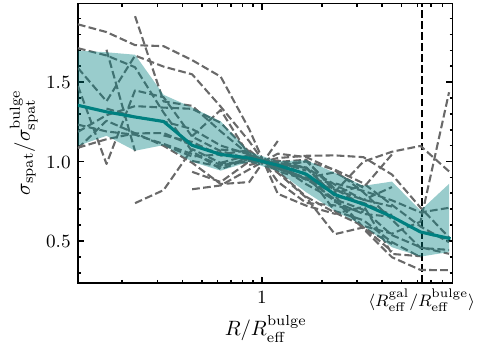} 
   \caption{
   \emph{Radial profile of the spatially resolved stellar velocity dispersion \sigspat across AGN host-galaxy bulges.}
   We measure the spatially resolved stellar dispersion \sigspat in concentric annuli centered on the AGN position. 
   Values of \sigspat are normalized to the value measured at the bulge effective radius $R_{\rm eff}^{\rm bulge}$. 
   Dashed lines show the spatially resolved \sigspat of individual AGN host galaxies, shaded green regions the 16$^{th}$ to 84$^{th}$ percentile range of the stacked profile.
   }
    \label{fig:disp_radial}%
\end{figure}

\subsubsection*{Spatially resolved kinematics}
For the first method, we average the spatially resolved velocity dispersion \sig within a chosen aperture. 
In the following, we refer to this quantity as the \emph{spatially resolved} stellar velocity dispersion \sigspat. 
We note that this quantity is different from the definition used by \cite{Bennert:2015}, who reconstructed the aperture-integrated dispersion from the spatially resolved first and second velocity moment. We have defined a similar quantity $\sigma_{\rm spat}^{\rm recon}$ and explain its behavior relative to \sigap more detail in Appendix~\ref{Appendix:Aperture_vs_Spatially_Resolved}.
In short, the definition from \cite{Bennert:2015} explicitly includes rotational broadening, whereas our \sigspat implicitly removes rotational broadening from kinematic structures down to the spatial scales that are resolved. 
We estimate the uncertainties of \sigspat from the scatter, half of the $16^{th}$ to $84^{th}$ percentile range, divided by the square root of the number of independent \sig measurements.
To account for the systematic uncertainties from limited spectral resolution (see Appendix~\ref{Appendix:Comparison_templates}), we quadratically add the resolution limit to the respective template used to determine the statistical uncertainties. 
Due to the number of individual spectra, the resulting uncertainties of the \sigspat are typically much smaller than what we get from fitting a single aperture-integrated spectrum.

In Fig.~\ref{fig:disp_radial} we show the radial profile of the spatially resolved dispersion component \sigspat, as a function of distance from the center $R$.
While all late-type galaxies (LTGs) in the sample are displayed, measuring \sigspat in early-type galaxies (ETGs) is often not possible due to the bright AGN, or \sigspat only sparsely samples the $R$ range; therefore, these are not included.
The spatially resolved stellar dispersion of LTGs exhibits a steep radial profile.
While on average, the offset between \sigspat measured at $ R_{\rm eff}^{\rm gal}$ and $R_{\rm eff}^{\rm bulge}$ is a factor of $1.9\pm 0.4$, it can be as large as a factor of three for individual galaxies. 
This underscores the importance of considering the aperture size over which \sigspat is measured.

\subsubsection*{Aperture-integrated kinematics}
Another approach is to coadd the spectra in a given aperture, providing a rotationally broadened spectrum, from which the aperture-integrated kinematics can be derived.
We refer to this quantity as the \emph{aperture-integrated} stellar velocity dispersion \sigap.
Since the most luminous AGNs are typically hosted by ETGs, which do not exhibit a detectable disk component, disk rotational broadening is expected to contribute a minor contamination in \sigap.
Varying the aperture size allows us to study the radial behavior of \sigap across different morphological components. More precisely, we trace bulge velocity dispersion $\sigma_{\rm ap}^{\rm bulge}$ or galaxy-wide velocity dispersion $\sigma_{\rm ap}^{\rm gal}$ by aligning the aperture with the bulge's luminosity-weighted centroid and matching its size to $R_{\rm eff}^{\rm bulge}$. 
More details on comparing aperture-integrated with spatially resolved measurements of $\sigma$ are described in Appendix~\ref{Appendix:Aperture_vs_Spatially_Resolved}.
While this approach reduces the spatial resolution of the radial axis, coadding the spectra has the advantage of substantially higher S/N. 
This is particularly beneficial for luminous AGNs, where extracting \sigspat is often hampered by the poor contrast between the AGN continuum and the underlying stellar absorption lines.
Moreover, using aperture-integrated spectra diminishes the contribution from systematic artifacts caused by PSF subtraction, which can be especially severe near the galaxy center.
%
%
%
\begin{figure}
   \centering
   \includegraphics[]{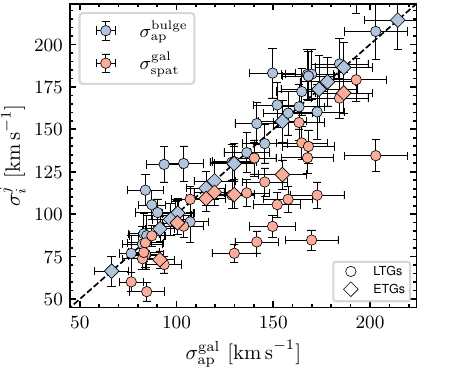} 
   \caption{
   \emph{Comparing methods for measuring stellar velocity dispersion.}
   Aperture-integrated dispersion measured over $R_{\rm eff}^{\rm gal}$ is shown on the x-axis. On the y-axis, we show the aperture integrated dispersion over $R_{\rm eff}^{\rm bulge}$ (blue) and the spatially resolved dispersion over $R_{\rm eff}^{\rm gal}$ (orange), respectively.
   Compared to the one-to-one correlation, denoted by the dash-dotted line, the mean bulge dispersion is on average slightly higher than the dispersion measured across the galaxy. In contrast, spatially resolving the kinematics results in significantly lower dispersion. 
   }
    \label{fig:sig_parameters}%
\end{figure}
%

The results of measuring dispersion using the two methods are summarized in Table~\ref{tbl:sig_parameters}, and illustrated in Fig.~\ref{fig:sig_parameters}.
Overall, the values of $\sigma_{\rm ap}^{\rm bulge}$ tend to be higher than those of $\sigma_{\rm ap}^{\rm gal}$. Averaged over the entire sample, this offset is small (7\,km/s, or 5\%), likely related to $\sigma_{\rm ap}^{\rm gal}$ capturing significant rotational broadening from galaxy disk that flattens any aperture-size dependence if the galaxy disk is viewed at high inclination (see Sect.~\ref{SubSubSec:Correcting_aperture_effects}).
More notably, on galaxy scales $\sigma_{\rm spat}^{\rm gal}$ is smaller than $\sigma_{\rm ap}^{\rm gal}$ by, on average, 25 km/s, or 12\%. 
Comparing the same for the bulge, $\sigma_{\rm spat}^{\rm bulge}$ versus $\sigma_{\rm ap}^{\rm bulge}$, yields similar but less pronounced offset of 9\%, suggesting an increased contribution from rotational broadening when using galaxy-integrated kinematics.
The stellar velocity dispersion measurements reported in the literature often differ substantially from our measurements for individual objects. 
These discrepancies may arise not only from the different diagnostic features used to constrain the stellar kinematics, e.g., \ion{Mg}{i}b\,$\lambda\lambda\lambda 5167,\,5173,\,5184$ \citep{Batiste:2017a, Husemann:2019}, \ion{Ca}{ii}\,$\lambda\lambda\lambda 8498,\,8542,\,8662$ \citep{Onken:2004, Woo:2010, Caglar:2023}, \ion{Ca}{ii}\,H\&K\,$\lambda\lambda 3969,\,3934$ \citep{Bennert:2015}, \ion{Mg}{i}+CO \citep{Watson:2008, Grier:2013}.
For instance, \cite{Harris:2012} report that the average differences are $\langle \sigma_{\rm Mg\,Ib} / \sigma_{\rm CaT} \rangle = -0.02 \pm 0.01$, i.e. a 5\% bias, that depends on aperture size.
Furthermore, varying aperture sizes across which these literature values are reported may introduce additional scatter.
While galaxy morphology is often unexplored in previous studies, our method for measuring stellar velocity dispersion controls for these systematic uncertainties, making our measurements more robust.

\begin{deluxetable*}{lcccccccc}
\tabletypesize{\small} 
\tablecaption{ \emph{Stellar velocity dispersion measurements.} \label{tbl:sig_parameters}} 
\tablehead{ 
 \colhead{AGN Name}  &  \colhead{$\sigma_{\rm ap}^{\rm gal}$}   &  \colhead{$\sigma_{\rm ap}^{\rm bulge}$}  & \colhead{$\sigma_{\rm spat}^{\rm gal}$}  &  \colhead{$\sigma_{\rm spat}^{\rm bulge}$ }  &  \colhead{lit. $\sigma$} &  \colhead{lit. $\sigma$} &  \colhead{$M_{\rm bulge, dyn}$} &  \colhead{$M_{\rm gal, dyn}$}   
 \vspace{-6pt}  \\  
 \colhead{} &  \colhead{$[{\rm km\,s}^{-1}]$} &   \colhead{$[{\rm km\,s}^{-1}]$} &    \colhead{$[{\rm km\,s}^{-1}]$} &   \colhead{$[{\rm km\,s}^{-1}]$} &   \colhead{$[{\rm km\,s}^{-1}]$} &   \colhead{$\,{\rm Ref.}$} &   \colhead{$[{\rm log\,M_\odot }]$} &    \colhead{$[{\rm log\,M_\odot }]$}    
 \vspace{-6pt}   
          } 
\colnumbers  
\startdata 
NGC\,3227          &  140\,$\pm$\,{11}  &  133\,$\pm$\,{11}  &  133\,$\pm$\,{9}  &  118\,$\pm$\,{8}  &  92\,$\pm$\,{6}  &  W13                &  9.23\,$\pm$\,{0.23}  &  10.49\,$\pm$\,{0.23} \\  
 NGC\,6814          &  107\,$\pm$\,{9}  &  95\,$\pm$\,{8}  &  109\,$\pm$\,{8}  &  92\,$\pm$\,{7}  &  69\,$\pm$\,{3}  &  B17                &  8.96\,$\pm$\,{0.25}  &  10.59\,$\pm$\,{0.23} \\  
 NGC\,4593          &  146\,$\pm$\,{12}  &  142\,$\pm$\,{11}  &  119\,$\pm$\,{8}  &  110\,$\pm$\,{8}  &  144\,$\pm$\,{5}  &  B17                &  10.18\,$\pm$\,{0.23}  &  10.64\,$\pm$\,{0.23} \\  
 NGC\,3783          &  104\,$\pm$\,{8}  &  130\,$\pm$\,{10}  &  93\,$\pm$\,{7}  &  122\,$\pm$\,{9}  &  95\,$\pm$\,{10}  &  O04                &  9.72\,$\pm$\,{0.23}  &  10.32\,$\pm$\,{0.23} \\  
 NGC\,2617          &  84\,$\pm$\,{9}  &  114\,$\pm$\,{9}  &  83\,$\pm$\,{6}  &  109\,$\pm$\,{8}  &  128\,$\pm$\,{9}  &  C23                &  9.52\,$\pm$\,{0.23}  &  10.24\,$\pm$\,{0.29} \\  
 IC\,4329\,A        &  165\,$\pm$\,{13}  &  172\,$\pm$\,{14}  &  142\,$\pm$\,{10}  &  166\,$\pm$\,{12}  &     &     &  10.27\,$\pm$\,{0.23}  &  11.32\,$\pm$\,{0.23} \\  
 Mrk\,1044          &  84\,$\pm$\,{7}  &  76\,$\pm$\,{7}  &  76\,$\pm$\,{8}  &  -  &     &     &  9.01\,$\pm$\,{0.24}  &  9.99\,$\pm$\,{0.24} \\  
 NGC\,5548          &  163\,$\pm$\,{13}  &  163\,$\pm$\,{13}  &  154\,$\pm$\,{11}  &  154\,$\pm$\,{11}  &  162\,$\pm$\,{12}  &  B17                &  10.72\,$\pm$\,{0.23}  &  10.85\,$\pm$\,{0.23} \\  
 NGC\,7469          &  129\,$\pm$\,{10}  &  131\,$\pm$\,{10}  &  111\,$\pm$\,{8}  &  113\,$\pm$\,{8}  &  131\,$\pm$\,{5}  &  N04                &  10.53\,$\pm$\,{0.23}  &  10.56\,$\pm$\,{0.23} \\  
 Mrk\,1310          &  82\,$\pm$\,{7}  &  82\,$\pm$\,{7}  &  74\,$\pm$\,{5}  &  74\,$\pm$\,{5}  &  84\,$\pm$\,{5}  &  W10                &  9.90\,$\pm$\,{0.23}  &  9.89\,$\pm$\,{0.23} \\  
 Mrk\,1239          &  99\,$\pm$\,{8}  &  ${^\dagger}$99\,$\pm$\,{8}  &  -  &  -  &     &     &  9.95\,$\pm$\,{0.33}  &  9.95\,$\pm$\,{0.33} \\  
 Arp\,151           &  120\,$\pm$\,{10}  &  ${^\dagger}$120\,$\pm$\,{10}  &  113\,$\pm$\,{8}  &  ${^\dagger}$113\,$\pm$\,{8}  &  118\,$\pm$\,{4}  &  W10                &  10.13\,$\pm$\,{0.33}  &  10.13\,$\pm$\,{0.33} \\  
 Mrk\,50            &  91\,$\pm$\,{10}  &  ${^\dagger}$91\,$\pm$\,{10}  &  73\,$\pm$\,{5}  &  ${^\dagger}$73\,$\pm$\,{5}  &  109\,$\pm$\,{14}  &  B11                &  10.05\,$\pm$\,{0.38}  &  10.05\,$\pm$\,{0.38} \\  
 Mrk\,335           &  66\,$\pm$\,{6}  &  ${^\dagger}$66\,$\pm$\,{6}  &  -  &  -  &     &     &  9.63\,$\pm$\,{0.36}  &  9.63\,$\pm$\,{0.36} \\  
 Mrk\,590           &  184\,$\pm$\,{15}  &  189\,$\pm$\,{15}  &  168\,$\pm$\,{12}  &  178\,$\pm$\,{12}  &  189\,$\pm$\,{6}  &  N04                &  10.28\,$\pm$\,{0.23}  &  10.40\,$\pm$\,{0.23} \\  
 SBS\,1116+583A     &  77\,$\pm$\,{10}  &  77\,$\pm$\,{10}  &  60\,$\pm$\,{5}  &  74\,$\pm$\,{5}  &  92\,$\pm$\,{4}  &  W10                &  9.13\,$\pm$\,{0.33}  &  9.98\,$\pm$\,{0.32} \\  
 Zw\,229-015        &  83\,$\pm$\,{16}  &  88\,$\pm$\,{7}  &  77\,$\pm$\,{5}  &  70\,$\pm$\,{6}  &     &     &  9.37\,$\pm$\,{0.23}  &  10.30\,$\pm$\,{0.45} \\  
 Mrk\,279           &  158\,$\pm$\,{13}  &  160\,$\pm$\,{13}  &  109\,$\pm$\,{8}  &  129\,$\pm$\,{9}  &  156\,$\pm$\,{17}  &  B17                &  10.41\,$\pm$\,{0.23}  &  10.65\,$\pm$\,{0.23} \\  
 Ark\,120           &  168\,$\pm$\,{13}  &  182\,$\pm$\,{15}  &  133\,$\pm$\,{9}  &  160\,$\pm$\,{11}  &  192\,$\pm$\,{8}  &  W13                &  10.48\,$\pm$\,{0.23}  &  10.87\,$\pm$\,{0.23} \\  
 3C\,120            &  178\,$\pm$\,{14}  &  ${^\dagger}$178\,$\pm$\,{14}  &  -  &  -  &  162\,$\pm$\,{20}  &  N95                &  10.59\,$\pm$\,{0.33}  &  10.59\,$\pm$\,{0.33} \\  
 MCG\,+04-22-042    &  170\,$\pm$\,{14}  &  183\,$\pm$\,{15}  &  85\,$\pm$\,{6}  &  173\,$\pm$\,{12}  &     &     &  10.16\,$\pm$\,{0.23}  &  11.20\,$\pm$\,{0.23} \\  
 Mrk\,1511          &  87\,$\pm$\,{7}  &  106\,$\pm$\,{11}  &  87\,$\pm$\,{6}  &  104\,$\pm$\,{7}  &  115\,$\pm$\,{9}  &  C23                &  9.39\,$\pm$\,{0.27}  &  10.62\,$\pm$\,{0.23} \\  
 PG\,1310-108       &  94\,$\pm$\,{8}  &  129\,$\pm$\,{11}  &  70\,$\pm$\,{8}  &  -  &     &     &  9.53\,$\pm$\,{0.24}  &  10.13\,$\pm$\,{0.24} \\  
 Mrk\,509           &  130\,$\pm$\,{10}  &  ${^\dagger}$130\,$\pm$\,{10}  &  -  &  -  &  184\,$\pm$\,{12}  &  G13                &  10.29\,$\pm$\,{0.33}  &  10.29\,$\pm$\,{0.33} \\  
 Mrk\,110           &  100\,$\pm$\,{8}  &  ${^\dagger}$100\,$\pm$\,{8}  &  95\,$\pm$\,{8}  &  ${^\dagger}$95\,$\pm$\,{8}  &  91\,$\pm$\,{9}  &  C23                &  9.89\,$\pm$\,{0.34}  &  9.89\,$\pm$\,{0.34} \\  
 Mrk\,1392          &  168\,$\pm$\,{13}  &  181\,$\pm$\,{15}  &  140\,$\pm$\,{10}  &  -  &  161\,$\pm$\,{9}  &  C23                &  10.09\,$\pm$\,{0.23}  &  11.17\,$\pm$\,{0.23} \\  
 Mrk\,841           &  115\,$\pm$\,{9}  &  ${^\dagger}$115\,$\pm$\,{9}  &  109\,$\pm$\,{8}  &  ${^\dagger}$109\,$\pm$\,{8}  &     &     &  10.39\,$\pm$\,{0.33}  &  10.39\,$\pm$\,{0.33} \\  
 Zw\,535-012        &  152\,$\pm$\,{12}  &  164\,$\pm$\,{13}  &  106\,$\pm$\,{7}  &  -  &     &     &  10.01\,$\pm$\,{0.23}  &  10.94\,$\pm$\,{0.23} \\  
 Mrk\,141           &  130\,$\pm$\,{10}  &  131\,$\pm$\,{12}  &  77\,$\pm$\,{8}  &  -  &  135\,$\pm$\,{5}  &  C23                &  9.59\,$\pm$\,{0.26}  &  10.74\,$\pm$\,{0.23} \\  
 RBS\,1303          &  203\,$\pm$\,{16}  &  208\,$\pm$\,{17}  &  134\,$\pm$\,{9}  &  176\,$\pm$\,{12}  &     &     &  10.37\,$\pm$\,{0.23}  &  11.23\,$\pm$\,{0.23} \\  
 Mrk\,1048          &  193\,$\pm$\,{15}  &  237\,$\pm$\,{19}  &  179\,$\pm$\,{13}  &  223\,$\pm$\,{16}  &     &     &  10.95\,$\pm$\,{0.23}  &  11.42\,$\pm$\,{0.23} \\  
 Mrk\,142           &  85\,$\pm$\,{11}  &  87\,$\pm$\,{13}  &  54\,$\pm$\,{5}  &  -  &     &     &  9.29\,$\pm$\,{0.37}  &  10.39\,$\pm$\,{0.34} \\  
 RX\,J2044.0+2833   &  141\,$\pm$\,{11}  &  153\,$\pm$\,{12}  &  84\,$\pm$\,{7}  &  -  &     &     &  9.59\,$\pm$\,{0.23}  &  10.76\,$\pm$\,{0.23} \\  
 IRAS\,09149-6206   &  155\,$\pm$\,{12}  &  ${^\dagger}$155\,$\pm$\,{12}  &  123\,$\pm$\,{9}  &  ${^\dagger}$123\,$\pm$\,{9}  &     &     &  10.99\,$\pm$\,{0.33}  &  10.99\,$\pm$\,{0.33} \\  
 PG\,2130+099       &  173\,$\pm$\,{14}  &  160\,$\pm$\,{16}  &  111\,$\pm$\,{8}  &  -  &  163\,$\pm$\,{19}  &  G13                &  9.88\,$\pm$\,{0.27}  &  10.80\,$\pm$\,{0.23} \\  
 NPM\,1G+27.0587    &  150\,$\pm$\,{13}  &  183\,$\pm$\,{15}  &  93\,$\pm$\,{6}  &  -  &     &     &  10.24\,$\pm$\,{0.23}  &  11.09\,$\pm$\,{0.25} \\  
 RBS\,1917          &  90\,$\pm$\,{10}  &  101\,$\pm$\,{10}  &  -  &  -  &     &     &  8.99\,$\pm$\,{0.26}  &  10.08\,$\pm$\,{0.29} \\  
 PG\,2209+184       &  136\,$\pm$\,{11}  &  136\,$\pm$\,{11}  &  113\,$\pm$\,{8}  &  113\,$\pm$\,{8}  &     &     &  10.70\,$\pm$\,{0.23}  &  10.70\,$\pm$\,{0.23} \\  
 PG\,1211+143       &  101\,$\pm$\,{11}  &  ${^\dagger}$101\,$\pm$\,{11}  &  -  &  -  &     &     &  9.24\,$\pm$\,{0.39}  &  9.24\,$\pm$\,{0.39} \\  
 PG\,1426+015       &  186\,$\pm$\,{15}  &  ${^\dagger}$186\,$\pm$\,{15}  &  171\,$\pm$\,{12}  &  ${^\dagger}$171\,$\pm$\,{12}  &  217\,$\pm$\,{15}  &  W08                &  10.89\,$\pm$\,{0.33}  &  10.89\,$\pm$\,{0.33} \\  
 Mrk\,1501          &  97\,$\pm$\,{10}  &  ${^\dagger}$97\,$\pm$\,{10}  &  -  &  -  &     &     &  10.76\,$\pm$\,{0.38}  &  10.76\,$\pm$\,{0.38} \\  
 PG\,1617+175       &  174\,$\pm$\,{20}  &  ${^\dagger}$174\,$\pm$\,{20}  &  -  &  -  &  201\,$\pm$\,{37}  &  G13                &  10.72\,$\pm$\,{0.40}  &  10.72\,$\pm$\,{0.40} \\  
 PG\,0026+129       &  233\,$\pm$\,{21}  &  ${^\dagger}$233\,$\pm$\,{21}  &  -  &  -  &     &     &  11.35\,$\pm$\,{0.35}  &  11.35\,$\pm$\,{0.35} \\  
 3C\,273            &  214\,$\pm$\,{17}  &  ${^\dagger}$214\,$\pm$\,{17}  &  -  &  -  &  210\,$\pm$\,{10}  &  H19                &  11.31\,$\pm$\,{0.33}  &  11.31\,$\pm$\,{0.33} \\  
 \enddata 
\tablecomments{
AGNs are listed in order of increasing redshift (as in Table~\ref{tbl:bh_parameters}). 
(1) {AGN Name.} 
(2) {Aperture-integrated \sig over $R_{\rm eff}^{\rm gal}$.} 
(3) {Aperture-integrated \sig over $R_{\rm eff}^{\rm bulge}$. Values marked with ($^\dagger$) are ETGs, for which $R_{\rm eff}^{\rm bulge}= R_{\rm eff}^{\rm gal}$, and thus $\sigma^{\rm bulge} = \sigma^{\rm gal}$ are equal.} 
(4) {Spatially-resolved \sig over $R_{\rm eff}^{\rm gal}$.} 
(5) {Spatially-resolved \sig over $R_{\rm eff}^{\rm bulge}$.} 
(6) {Stellar velocity dispersion reported in the literature.} 
(7) {Reference for the lit. $\sigma$.} 
(8) {Logarithm of the bulge dynamical mass.} 
(8) {Logarithm of the galaxy dynamical mass.} 
Reference keys are 
                N95:\,\cite{Nelson_Whittle:1995}, 
                N04:\,\cite{Nelson:2004}, 
                O04:\,\cite{Onken:2004}, 
                W08:\,\cite{Watson:2008}, 
                W10:\,\cite{Woo:2010}, 
                B11:\,\cite{Barth:2011}, 
                G13:\,\cite{Grier:2013}, 
                W13:\,\cite{Woo:2013}, 
                B17:\,\cite{Batiste:2017a}, 
                H19:\,\cite{Husemann:2019}, 
                C23:\,\cite{Caglar:2023}. 
                } 
\end{deluxetable*}

\vspace*{-2\baselineskip} 

Aperture-integrated measurements can be reconstructed from spatially resolved measurements, as we demonstrate in Appendix~\ref{Appendix:Aperture_vs_Spatially_Resolved}. 
Based on these results, we conclude that across galaxy disks, we can robustly disentangle the contributions of rotation from those of chaotic motions. However, we note that substructures like fast- or counter-rotating disks, which are often observed on scales of several hundred parsecs \citep{Comeron:2010, Gadotti:2020}, below the typical $\sim$arcsec sizes of our bulges, remain unresolved in the majority of AGNs in our sample.

\subsubsection{Systematic Uncertainties for Measuring \texorpdfstring{\sig}\ }
\label{SubSubSec:Systematic_Uncertainties}
To achieve a more accurate calibration of the \mbh host galaxy scaling relations in AGNs, our approach involves the most precise \mbh and \sig measurements available. 
Although the wide dynamic range of AGN parameters is a strength of the sample, it also presents technical challenges in identifying host-galaxy morphological components\citep[see][in prep.]{Bennert:2024}. 
%
At the low-$z$ end, for example, NGC\,3227, NGC\,4593 and NGC\,7469 are cases where plenty of kinematic substructure is resolved, including spiral arms, dust lanes, nuclear rings, nuclear disks, or bulges. In such cases, the simplistic parameterization (s, sd, sdb) is insufficient to describe the morphology accurately (however, the photometry for the main components is adequately recovered even by a simple model).
For the more distant and luminous AGNs in the sample, the PSF subtraction often leaves strong residuals that dominate over the host galaxy on arcsecond scales.
In cases where these residuals coincide with the typical sizes of the bulges, it is impossible to measure accurate bulge sizes.
Also the choice of parameterizing host-galaxy morphology can affect $R_{\rm eff}^{\rm bulge}$ for individual objects. However, for most of the sample, the parameterization is clear, and even in ambiguous cases, adding a component has little impact on the measured sizes.

Another source of systematic uncertainty comes from measuring the kinematics from the IFU data. 
For the nearest AGNs, the FoV of the IFU is smaller than $R_{\rm eff}^{\rm gal}$. In contrast, for the more distant AGNs the lower physical spatial resolution and AGN continuum blending does not allow us to measure \sig within $R_{\rm eff}^{\rm gal}$.
Moreover, beam smearing might contribute to smoothing the radial profiles of \sigspat on small scales, e.g. in Fig.~\ref{fig:disp_radial}. However, this effect cannot be homogeneously controlled without degrading individual datasets.
From the aperture-sizes and methods defined in Sect.~\ref{SubSubSec:two_methods_sigma}, $\sigma_{\rm ap}^{\rm gal}$ provides the measurement that is the least sensitive to systematic effects: Only for 4/44 bright AGNs (PG\,1211+143, PG\,1617+175, PG\,0026+129, 3C\,273), $\sigma_{\rm ap}^{\rm gal}$ is impacted by the PSF subtraction adding systematic uncertainties of $\sim 5$\%. This is caused by a few spaxels that contain signal from the host galaxy heavily blended by AGN emission. 
When excluding these four objects, the slope and intercept of the spatially resolved $\sigma_{\rm spat}^{\rm gal}$ relation is $<$3\%. With such small variation we consider the systematic uncertainty for calibrating the \mbh-\sig relation small.

\subsection{Dynamical Masses}
\label{SubSubSec:Dynamical_Masses}
Based on the kinematics recovered in the previous section, we can derive dynamical masses as
\begin{equation}
    M_{\rm dyn} = c  R_{\rm eff} \sigma_{\rm ap}^2 /G,
\end{equation}
where $c$ is a structural constant that depends on the anisotropy of the system \cite{Courteau:2014}.
While the value of \mdyn for ETGs is best described by coefficient $c=2.5$ \cite{Cappellari:2006}, we adopt $c=3$ for both LTGs and ETGs guaranteeing a consistent comparison with literature (e.g., \citealt{Bennert:2021}).
For LTGs specifically, we adopt $R_{\rm eff}^{\rm bulge}$ and $\sigma_{\rm ap}^{\rm bulge}$ to get the dynamical bulge mass $M_{\rm bulge,dyn}$.
For ETGs, we adopt the parameters that belong to the spheroid, i.e. $R_{\rm eff}^{\rm bulge}$ and $\sigma_{\rm ap}^{\rm bulge}$, and also refer to the derived dynamical mass as $M_{\rm bulge,dyn}$.
With this definition, \mbulgedyn provides a consistent metric for the dynamical mass of the spheroidal component for both LTGs and ETGs.

\subsection{Fitting the \texorpdfstring{\mbh}\ Scaling Relations}
\label{SubSec:M-sigma-relation}
The \mbh scaling relations are parameterized as
\begin{equation}
\label{eq:MBH-sig_relation}
{\rm log} \left( \frac{M_{\rm BH}}{M_\odot}\right) = \alpha + \beta\, \log  X
\end{equation}
where $X$ is the host-galaxy parameter, in our case either $\sigma_\star / 200\,{\rm km\,s^{-1}}$ or $M_{\rm bulge,dyn}  / 10^{11} M_\odot$.%
We fit the relation using the hierarchical Bayesian model \texttt{LINMIX\_ERR} from \cite{Kelly:2007}, which performs a linear regression to observed independent variables $x_i$ and dependent variables $y_i$, accounting for the associated uncertainties of both. 
We re-scale the variables to the mean of their respective distributions to reduce the covariance between the parameters.
Monitoring the convergence to a well-sampled posterior distribution allows us to infer realistic uncertainties of the derived fitting parameters, which also include the intrinsic scatter of the relation $\epsilon$.
Compared to other regression methods that are often used to constrain the \mbh scaling relations, namely \texttt{BCES} \citep{Akritas:1996}, \texttt{FITEXY} \citep{Tremaine:2002} or \texttt{maximum likelihood} \citep{Gultekin:2009, Woo:2010}, \texttt{LINMIX\_ERR} is more general and produces a larger intrinsic scatter \citep{Park:2012}.
For our analysis, we assume that the measurement uncertainties of \mbh and \sig are symmetric in log-space, and symmetrize the measurement uncertainties on \mbh from their upper and lower $1\sigma$ intervals listed in Table~\ref{tbl:bh_parameters}.
We note that the adopted choice of the uncertainties does not significantly impact the results, which has already been reported by \cite{Park:2012}.

\section{Results and Discussion}
\label{Sec:Results&Discussion}

\subsection{Host-galaxy Morphologies}
\label{Subsec:Host_galaxy_morphologies}

If major mergers are responsible for shaping the \mbh scaling relations, only the host-galaxy morphological components bearing the dynamical imprint of merger history should correlate with \mbh, i.e., classical bulges \citep{Cisternas:2011a, Cisternas:2011b, Kormendy_Ho:2013, Bennert:2015}. 
While the dependence of the \mbh scaling relations on host morphology has been extensively studied for quiescent galaxies \citep[e.g.,][]{Gultekin:2009, Greene:2010, McConnell_Ma:2013, Savorgnan&Graham:2015, Sahu:2019, Graham:2023}, they are less well constrained for AGNs due the bright AGN emission \citep[e.g.,][see also Sect.~\ref{SubSubSec:AGN-host_deblending}]{Debattista:2013, Hartmann:2014}, or the narrow dynamic range in \mbh covered \citep[e.g.,][]{Bennert:2021}.

In our sample, 29/44 (66\%) of AGNs are hosted by LTGs. However, disks may remain undetected at high bulge-to-disk ratios, so our estimate should be regarded as an upper limit.
Nevertheless, the fraction is comparable to the fraction of Seyfert hosts with disk-like galaxies among the overall AGN population
(e.g., $\sim$52\% in CANDELS \citep{Kocevski:2012}, or 74\% disk galaxies in CARS \citep{Husemann:2022}), the depth and angular resolution of the HST photometry in our study allow us to identify disk components more robustly than previous studies.
\cite{Bennert:2021}, who used the same methodology as in this work, reported an even higher fraction (95\%) of disk galaxies among local AGNs imaged with HST.

Among the sample of AGNs hosted by disk galaxies, 15/29 show a clear sign of a bar, and are better fitted when including a bar component in the model. 
The intrinsic fraction of bars might be higher, since we used a conservative approach by only including a bar when there is clear signs in the PSF-subtracted images. 
Moreover, a few galaxies have too high disk inclination to identify a bar \citep[for details, see][in prep.]{Bennert:2024}. Typically, the bar fractions of disk-like AGN hosts are reported to be higher \citep[e.g.][]{Cisternas:2011a, Alonso:2013, Husemann:2022}. 
However, we caution against direct comparisons of the bar incidence rate with other surveys, since identification methods, image quality, and intrinsic bar strengths have a significant impact on these numbers, similar to the disk/non-disk classification. In particular, the bar fraction also depends on wavelength range, where higher bar fractions are observed in the infrared compared to identification based on optical photometry \citep[e.g.,][]{Eskridge:2000, Buta:2015, Erwin:2019}.

While 10/44 galaxies have irregular or asymmetric morphologies, only two objects show strong signs of interaction or merger activity (Arp\,151, Mrk\,1048). This corresponds to 5\%, which is consistent with the low fraction of strongly disturbed hosts in the overall AGN population \cite[e.g.,][]{Cisternas:2011a, Schawinski:2012, Mechtley:2016, Marian:2019, Kim:2021}.
As we will demonstrate in Sect.~\ref{subsubsec:Dependence_on_measuring_sigma}, interacting galaxies do not represent the strongest outliers to the \mbh scaling relations and are included in the following analysis.

\subsubsection{Correcting Aperture Effects}
\label{SubSubSec:Correcting_aperture_effects}

\begin{figure}
   \centering
   \includegraphics[]{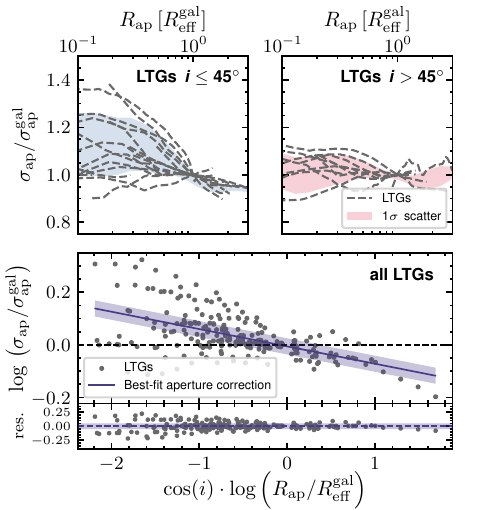} 
   \caption{
   \emph{Inclination-dependent aperture correction for stellar velocity dispersion measurements.}
   (Top) Left and right panels show the behavior of 
   \sigspat as a function of aperture size normalized by the galaxy effective radius, split by disk inclination. Gray dash-dotted lines correspond to individual galaxies, whereas shaded regions denote the scatter (16$^{th}$ to 84$^{th}$ percentile in bins of $R_{\rm ap}$).
   Varying $R_{\rm ap}$ significantly affects \sigap only for the lower-inclination systems.
   (Bottom) Considering all disk galaxies of the AGN sample, we control for inclination by parameterizing the aperture correction formula with eq.~\ref{eq:ap_corr_inclination}.  
   A first-order power law describes the overall trend of decreasing $\sigma_{\rm ap}$ with increasing $R_{\rm ap}$ (top panel), but significant residual structure indicates that galaxy-to-galaxy variation remains likely driven by stellar mass or luminosity.  
   }
    \label{fig:ap_correction}%
\end{figure}

%
As spatially resolved studies will remain unavailable for the majority of distant type 1 AGNs in the Universe, aperture-integrated spectra are often the only means to trace stellar kinematics from bulge to galaxy scales
We therefore investigate the systematic differences induced by the aperture size, depending on host-galaxy morphology.
While differences between $\sigma_{\rm ap}$ and \sigspat for individual AGNs are detailed in Appendix~\ref{Appendix:Aperture_Effects}, we shall here only focus on the sample-integrated behavior and dependencies on morphology.

The spatially resolved kinematics shown in Fig.~\ref{fig:disp_radial} illustrate how galaxy kinematic substructures may impact measurements of \sig: for LTGs with spatially resolved kinematics, the sample-averaged normalized \sigspat exhibits a steep radial profile, underscoring the importance of considering the aperture over which \sigspat is extracted. 
%

Aperture correction recipes are often formulated in the form of a power law:
\begin{equation}
    \label{eq:ap_corr_inclination}
    \frac{\sigma}{\sigma_{\rm eff}} = 
    \left( \frac{R}{R_{\rm eff}} \right) ^{\gamma} 
\end{equation}
For quiescent ETGs, it is established that \sigap typically decreases with increasing aperture size to the center, resulting in $\gamma=-0.04$ \citep{Jorgensen:1995}, $\gamma=-0.06$ \citep{Mehlert:2003}, or $\gamma=-0.066$ \citep{Cappellari:2006}.
The few ETGs in our sample are poorly resolved, so that a statistical analysis of the aperture-size dependence is not possible. 

For quiescent LTGs, recent studies have shown that aperture correction is more complex, due to multiple kinematic components and their anisotropy.
However, compared to galaxy stellar mass and luminosity, we suspect that the galaxy inclination $i$ has the largest effect on measuring \sigap in our AGN sample (Sect.~\ref{SubSubSec:M-sigma-relation_AGNs}).
Galaxy-scale kinematics derived from aperture-integrated spectra of highly inclined disk galaxies are more affected by rotational broadening compared to low-inclination disk galaxies. 
This is reflected in the top panels of Fig.~\ref{fig:ap_correction}, where only disk galaxies viewed at lower inclinations exhibit a trend of \sigap with varying $R_{\rm ap}$, whereas higher-inclination disk galaxies show no significant trend.
This is a result of two opposing trends which cancel each other out at high inclination: stellar-velocity dispersion increases towards the center due to either dynamically hotter bulges or spatially unresolved rotating nuclear disks (see discussion in Sect.~\ref{SubSubSec:Systematic_Uncertainties}), but rotational broadening from the galaxy disk only becomes important at larger distance from the galaxy center.
Although $\sigma_{\rm ap}$ is sometimes measured in elliptical apertures, as for instance in \cite{Falcon-Barroso:2017}, measurements in circular apertures are the default for survey data.
To control for inclination, we included the disk inclination $i$ in the parameterization of the aperture correction:
\begin{equation}
    \label{eq:ap_corr_Falcon-Barroso}
    \frac{\sigma}{\sigma_{\rm eff}} = 
    \left( \frac{R}{R_{\rm eff}} \right) ^{\gamma \cdot {\rm cos} \,(i)}
\end{equation}
Fitting the logarithmic relation with a least-squares minimization provides the best-fitting aperture-correction exponent $\gamma = - 0.063 \pm 0.013$. This value is surprisingly consistent with the aperture correction suggested for ETGs, indicating that when correcting for disk inclination, the \sigap correction of disk galaxies is similar to that of pure spheroidals.
However, significant residual structure of individual galaxies demonstrate that additional parameters must be considered, such as galaxy stellar mass or luminosity \citep{Falcon-Barroso:2017, Zhu:2023}.
For our AGNs, however, the small sample size does not allow us to futher constrain second order dependencies on host-galaxy luminosity or stellar mass.

\subsection{The \texorpdfstring{\mbh}\ Scaling Relations of Quiescent Galaxies}
\label{SubSubSec:M-sigma-relation_quiescent}

The \mbh-\sig relation of the local quiescent galaxy population has been studied across a higher \mbh dynamic range compared to that of AGNs \citep{Gultekin:2009, Kormendy_Ho:2013, McConnell_Ma:2013}.
KH13 compiled \mbh and the "effective dispersion" $\sigma_{\rm eff}$, which they measured within $ R_{\rm eff}^{\rm gal}/2$. Their method involves the intensity-weighted mean of $v^2+\sigma^2$, which close to the definition of our \sigap (see Appendix~\ref{Appendix:Aperture_vs_Spatially_Resolved}).
For a consistent analysis, we have re-fit the \mbh-\sig and \mbh-\mdyn  relation from KH13 with our method (Sect.~\ref{SubSec:M-sigma-relation}).
The results is listed in row (i) of Table \ref{tbl:Results} and reproduce the parameters that have originally been reported.

However, the KH13 sample mainly covers the high-\mbh regime, where RM AGNs are scarce.
For late-type galaxies at the low-\mbh end of the relation, rotational broadening from disk components are non-negligible, and thus the aperture size over which the kinematics are extracted must be considered.
Based on a sample of both LTGs and ETGs, \cite{Batiste:2017a} have compared using an aperture correction factor to estimate $\sigma_{\rm ap}^{\rm gal}$ versus direct measurements of $\sigma_{\rm spat}^{\rm gal}$.
While the former approach has been widely used in the literature, \cite{Batiste:2017a} stress that not only are the effective radii used in the literature uncertain, also the recovered $\sigma_{\rm spat}^{\rm gal}$ are systematically lower by $13\,{\rm km\,s}^{-1}$ compared to $\sigma_{\rm ap}^{\rm gal}$.
As a consequence, \mbh-\sig calibrations using \sigap are offset towards higher intercepts, and tend to result in steeper slopes (e.g., \citealt[][$\beta=5.31$]{Woo:2013}, \citealt[][$\beta=5.04$]{Grier:2013}, \citealt[][$6.34\pm0.8$]{Savorgnan&Graham:2015}).
When using the spatially resolved \sig measurements of LTGs and ETGs, equivalent to our definition of \sigspat, \cite{Batiste:2017a} found $\alpha=8.66 \pm 0.09$ and $\beta=4.76 \pm 0.60$, which are more consistent with the KH13 relation.

\subsection{The \texorpdfstring{\mbh}\ Scaling Relations of AGNs}
\label{SubSubSec:M-sigma-relation_AGNs}

In previous studies, fitting the \mbh-\sig relation of type~1 AGNs required an additional free parameter, the unknown virial factor \f.
To overcome the limited dynamic range when inferring the AGN relation's scatter and intercept (and thereby a sample-average $\langle f \rangle$), previous calibrations often required fixing the slope to that of the quiescent galaxies. This implicitly assumes that AGNs and quiescent galaxies follow the same underlying relations, and selection effects are negligible.
However, so far this assumption does not have any empirical foundation.
In fact, AGNs represent the sites of ongoing SMBH growth, where the present-day SMBH growth may result in different \mbh-host-galaxy scaling relations.
To test this hypothesis, from here on, we will focus on the AGNs in our sample that have independent \mbh measurements.
In contrast to many previous studies, this allows us fitting the AGN \mbh-\sig relation without assumptions on any of the parameters.
Furthermore, we control for host-galaxy morphology by using \sig measured across the bulge or galaxy effective radius, and test how different methods of measuring stellar velocity dispersion impact the \mbh-\sig relation.

\subsubsection{Impact of \texorpdfstring{\sigap}\ versus \texorpdfstring{\sigspat}\ }
\label{subsubsec:Dependence_on_measuring_sigma}

\begin{figure}
   \centering
   \includegraphics[]{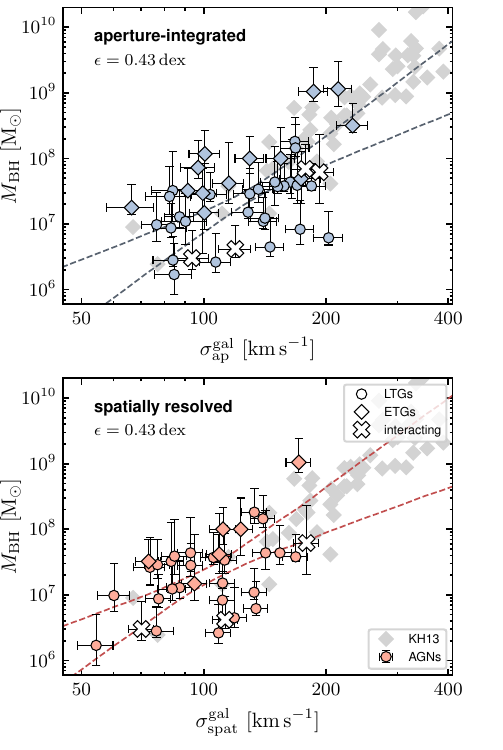} 
   \caption{
   \emph{Effect of using aperture-integrated versus spatially resolved \sig on the \mbh-\sig scaling relation.}
   (Top) Observed relation of AGNs based on $\sigma_{\rm ap}^{\rm gal}$ (row (iii) in Table~\ref{tbl:Results}).
   AGNs show a large scatter, and individual measurements have high uncertainty so that $\alpha$ and $\beta$ are not well constrained. Overall, the AGNs in our sample form the extension of quiescent galaxies towards lower \mbh.
    Interacting galaxies tend to have lower \mbh, but they are not the ones that deviate the most from the relation.
   (Bottom)  Observed relation of AGNs based on  $\sigma_{\rm spat}^{\rm gal}$ (row (iv) in Table~\ref{tbl:Results}). Removing rotational broadening reduces the uncertainty of individual measurements. 
   While the resulting \mbh-\sig correlation is more significant (larger $\beta$) and has a higher intercept, its intrinsic scatter is the same as when using \sigap.
   }
              \label{fig:Mbh-sigma_spat_vs_ap}%
\end{figure}

The majority (29/44) of AGNs in our sample are hosted by LTGs, for which the best-fitting \mbh-\sig relation depends on both the method by which the stellar velocity dispersion is measured (see Sect.~\ref{SubSubSec:two_methods_sigma}), and aperture size.
While the galaxy-wide integrated $\sigma_{\rm ap}^{\rm gal}$ is the closest to the definition used in previous studies \citep[e.g.,][]{Gultekin:2009, Kormendy_Ho:2013, Grier:2013}, \sigspat results in steeper slopes. 
This steepening occurs because \sigspat excludes rotational broadening, 
effectively shifting many LTGs towards lower \sig (top vs. bottom panel of Fig.~\ref{fig:Mbh-sigma_spat_vs_ap}).
This primarily affects high-inclination disk galaxies, whereas \sigap includes this effect (as detailed in Sect.~\ref{SubSubSec:Correcting_aperture_effects}).
The \mbh-\sigspat relation is also offset toward lower dispersion, consistent with findings by \cite{Batiste:2017b}, suggesting that while \mbh \emph{does} correlate with the velocity dispersion of galaxy discs, but the underlying relations are different (see discussion in Sect.~\ref{subsubsec:Dependence_on_host-galaxy_morphology}.
This observation has been predicted by previous studies, which suggested that rotation effects should be corrected for in case of low-mass, disk-dominant galaxies \citep{Bennert:2011, Harris:2012, Woo:2013}.
Despite the differences of how dispersion is extracted either with \sigap or \sigspat, the best-fitting parameters listed in Table~\ref{tbl:sig_parameters} rows (iii) and (iv) indicate that on galaxy scales, both methods result in statistically consistent scaling relations.
On scales of the bulge, many distant galaxies hosting a luminous AGN are dramatically blended by the AGN emission, effectively limiting our ability to resolve \sigspat close to the nucleus.
As a result, the bulge size is smaller than \sigspat for 50\% of the AGNs in our sample. 
For those objects, a robust measurement of $\sigma_{\rm spat}^{\rm bulge}$ is not feasible. The effect of fitting the \mbh-\sig relation of an incomplete sample is discussed in Sect.~\ref{SubSec:Intrinsic_Scatter_M-sigma-relation}.

\begin{deluxetable*}{lccccccccc}
\tabletypesize{\small} 
\tablecaption{\emph{Results of fitting the scaling relations of local AGNs and quiescent galaxies.} \label{tbl:Results}} 
\tablehead{ 
    \colhead{$X$ in relation}  & 
    \colhead{Sample}  &   
    \colhead{\mbh\ distr.}  & 
    \colhead{Aperture} & 
    \colhead{Method}  &
    \colhead{Symbol}  &
    \colhead{$\alpha$}  &  
    \colhead{$\beta$}  & 
    \colhead{$\epsilon$} &
    Row
    }
\colnumbers
\startdata
\multirow{10}{*}{$\sigma  /200\,{\rm km\,s}^{-1}$} & KH13$^\dagger$  & KH13 & galaxy$^\star$ & ap. & $\sigma_{\rm ap}^{\rm gal}$ &
8.53\,$\pm$\,0.05  &  4.53\,$\pm$\,0.32  &  0.34\,$\pm$\,0.04 & (i)
    \\ 
\noalign{\smallskip}
 \cline{2-10} 
\noalign{\smallskip}
  &  AGN   & AGN & bulge  & ap. &  $\sigma_{\rm ap}^{\rm bulge}$   & {\bf 7.90\,$\pm$\,0.16 }  &  {\bf 2.53\,$\pm$\,0.73}  & {\bf 0.47\,$\pm$\,0.08}& (ii) \\ 
  &  AGN   & AGN & galaxy & ap. & $\sigma_{\rm ap}^{\rm gal}$     & 7.93\,$\pm$\,0.16  &  2.48\,$\pm$\,0.72  &  0.47\,$\pm$\,0.08& (iii) \\ 
  &  AGN   & AGN & galaxy & spat. & $\sigma_{\rm spat}^{\rm gal}$   & 8.06\,$\pm$\,0.27  &  2.57\,$\pm$\,0.89  &  0.45\,$\pm$\,0.09& (iv) \\ 
\noalign{\smallskip}
 \cline{3-10} 
\noalign{\smallskip}
   & AGN & KH13 & bulge & ap. & $\sigma_{\rm ap}^{\rm bulge}$     & {\bf 8.71\,$\pm$\,0.13}  &  {\bf 4.10\,$\pm$\,0.93} &  {\bf 0.57\,$\pm$\,0.09}& (v) \\ 
   & AGN & KH13 & galaxy & ap. & $\sigma_{\rm ap}^{\rm gal}$   & 8.80\,$\pm$\,0.13  &  4.51\,$\pm$\,0.88  &  0.53\,$\pm$\,0.08& (vi) \\ 
\noalign{\smallskip}
\cline{2-10} 
\noalign{\smallskip}
  &  AGN LTGs   & AGN LTGs & bulge  & ap. & $\sigma_{\rm ap}^{\rm bulge}$   & 7.72\,$\pm$\,0.16  &  2.80\,$\pm$\,0.80  &  0.27\,$\pm$\,0.11& (vii) \\ 
  &  AGN LTGs   & AGN LTGs & galaxy & ap. & $\sigma_{\rm ap}^{\rm gal}$     & 7.68\,$\pm$\,0.18  &  2.17\,$\pm$\,0.89  &  0.37\,$\pm$\,0.10& (viii) \\ 
  &  AGN LTGs   & AGN LTGs & galaxy & spat. & $\sigma_{\rm spat}^{\rm gal}$   & 7.88\,$\pm$\,0.27  &  2.47\,$\pm$\,0.70  &  0.31\,$\pm$\,0.10& (ix) \\ 
\noalign{\smallskip}
\cline{2-10} 
\noalign{\smallskip}
  &  AGN ETGs   & AGN ETGs & galaxy  & ap. & $\sigma_{\rm ap}^{\rm bulge}$   & 8.44\,$\pm$\,0.30  &  3.00\,$\pm$\,1.28  &  0.54\,$\pm$\,0.18 & (x)\\ 
\noalign{\smallskip}
\cline{1-10} 
\noalign{\smallskip}
\noalign{\smallskip}
%
%
%
\multirow{4}{*}{$M_{\rm bulge,dyn}  / 10^{11} M_\odot$} & KH13 & KH13 & bulge & ap. & $\sigma_{\rm ap}^{\rm bulge}$  & 8.78\,$\pm$\,0.07  &  1.06\,$\pm$\,0.10  &  0.45\,$\pm$\,0.05& (xi) \\
\noalign{\smallskip}
 \cline{2-10} 
\noalign{\smallskip}
    & AGN & AGN & bulge & ap. & $\sigma_{\rm ap}^{\rm bulge}$  & {\bf 8.11\,$\pm$\,0.16}  &  {\bf 0.70\,$\pm$\,0.14}  &  {\bf 0.41\,$\pm$\,0.08} & (xii)\\
    & AGN & AGN & gal & ap. & $\sigma_{\rm ap}^{\rm bulge}$  & 7.80\,$\pm$\,0.19  &  0.83\,$\pm$\,0.52  &  0.54\,$\pm$\,0.10 & (xiii)\\
\noalign{\smallskip}
 \cline{3-10} 
\noalign{\smallskip}
   & AGN & KH13 & bulge & ap. & $\sigma_{\rm ap}^{\rm bulge}$  & {\bf 8.76\,$\pm$\,0.11 } &  {\bf 0.87\,$\pm$\,0.14}  &  {\bf 0.49\,$\pm$\,0.07} & (xiv)\\ 
\noalign{\smallskip}
\hline
\enddata
\tablecomments{
All fits were calculated as part of this paper, including those to quiescent galaxies. Relations that are shown in Fig.~\ref{fig:MBH-scaling-relations} are highlighted in bold-face font.
(1) Scaling relation of the form ${\rm log}(M_{\rm BH}/M_\odot) = \alpha+ \beta \,{\rm log}X$, with $X$ given in the table.
(2) {Sample for which the \mbh-\sig relation was fitted.} 
(3) {\mbh distribution of the sample used for fitting the \mbh-\sig relation. The quiescent galaxy sample from KH13 serves are reference. "AGN" refers to the (sub-)sample of AGNs, specified in the column 2. "KH13" refers to the AGN sample being matched to the KH \mbh distribution, as described in Sect.~\ref{SubSec:Controlling_selection_effects}.}
(4) {Aperture over which the kinematics are evaluated.} 
(5) {Method by which the kinematics are measured. "ap." refers to aperture-integrated, whereas "spat.-res" to spatially resolved kinematics, see Sect.~\ref{SubSubSec:two_methods_sigma}.} 
(6) {Symbol for the stellar dispersion \sig, indicating which aperture size and which method we used to measure it.} 
(7) {Best-fit intercept of the \mbh-\sig relation (eq.~\ref{eq:MBH-sig_relation}).}
(8) {Best-fit slope of the relation.}
(9) {Best-fit intrinsic scatter of the relation.}
(10) {Row number used to refer to the relation.}
($^\dagger$) {KH13 data re-fitted with our method.}
($^\star$) {Galaxy effective radius is poorly constrained from ground-based seeing-limited imaging, as discussed in KH13 and \cite{Batiste:2017b}.}
}                              
\end{deluxetable*}
\vspace*{-2\baselineskip} 

\subsubsection{Impact of Host-Galaxy Morphology}
\label{subsubsec:Dependence_on_host-galaxy_morphology}
The dependence of the \mbh scaling relation on host-galaxy morphology is crucial for understanding its physical drivers \citep[e.g.,][]{Sahu:2019, Graham:2023}. However, studying host galaxies in AGNs, especially at the high-\mbh end, has been challenging. 
With high-quality spectroscopic data, we can now examine the morphology dependence of the \mbh-\sig scaling relation, focusing on the relative behavior of LTGs and ETGs, with best-fitting parameters detailed in rows (vii)-(x) of Table~\ref{tbl:Results}.

\subsubsection*{Late-type AGN hosts}
Only 15/44 AGN host galaxies in our sample are classified as ETGs, whereas the majority 29/44 are hosted by LTGs.
In general, constraining the AGN sample to LTGs significantly lowers the intercept and flattens the observed relation, see row (vii)-(iv) of Table~\ref{tbl:Results}.
While this might partially be caused by quiescent LTGs following a shallower \mbh-\sig relation compared to ETGs \citep{Sahu:2019}, the smaller \mbh dynamic range covered may also contributes to the observed shallower slope (see Sect.\ref{SubSubSec:Correcting_aperture_effects}). 
Of any method and aperture size used for fitting LTGs, the correlation of \mbh with $\sigma_{\rm ap}^{\rm bulge}$ has parameters that are the closest to those of quiescent galaxies. 
This \mbh-$\sigma_{\rm ap}^{\rm bulge}$ relation also shows the least intrinsic scatter of all AGN subsamples. 
However, this might be driven by selection effects: SMBHs are not detected in every LTGs, whereas here we are only selecting those that harbor one. 
We might therefore only be sensitive to the upper envelope of the underlying scaling relation.
As found for the entire AGN sample in Sect.~\ref{subsubsec:Dependence_on_measuring_sigma}, the galaxy-wide stellar velocity dispersion $\sigma^{\rm gal}$ correlates with \mbh of LTG AGNs in a relation that shows small intrinsic scatter.
Compared to $\sigma_{\rm ap}^{\rm gal}$, the correlation is slightly more pronounced for $\sigma_{\rm spat}^{\rm gal}$ which is largely rotation-free and thus traces the older, dynamically hot stellar component of the galaxy disk.
Such a correlation is in contradiction to previous studies \citep[e.g.,][]{Greene:2010, Kormendy:2011, Kormendy_Ho:2013}, that argued that \mbh does not correlate with the properties of the disk. 
These studies suggest that stellar feedback, rather than black hole feedback, plays a more significant role in regulating the growth of galaxy disks.
While this might be true for the dynamically cold component, recently formed inside-out through smooth gas accretion \citep[e.g.,][]{Pichon:2011, El-Badry:2018}, there is no a priori reason to assume that an old disk component should not be affected by early BH feedback, similar to classical bulges.
Indeed, recent observations showed that galaxy disk progenitors had already formed at z$>$3 (e.g., \citealt{Lelli:2021, Roman-Oliveira:2023, Ferreira:2023, Jacobs:2023, Robertson:2023}), well before the peak of cosmic SMBH growth, potentially carrying information about the SMBH-galaxy coevolution.
In this context, the \mbh-$\sigma_{\rm spat}^{\rm gal}$ correlation suggests that SMBHs \emph{do} coevolve with galaxy disks, but this may be limited to early epochs of galaxy disk formation, as traced by the dynamically hot disk component.

\subsubsection*{Early-type AGN hosts}
Among the ETGs, only Arp~151, Mrk\,110 and Mrk~335 have lower \mbh, comparable to what is typically found in LTGs, whereas the remaining twelve ETGs occupy the high-\mbh end of the scaling relation.
\cite{Woo:2013} argued that such massive SMBHs are typically hosted by massive galaxies for which the difference between the methods for measuring \sig should be minimal.
Assuming the \mbh-$M_{\rm dyn}$ relation of AGNs from \cite{Kormendy_Ho:2013} ($\alpha=8.49$, $\beta=1.16$), the average dynamical mass of the ETGs is ${\rm log}(M_{\rm dyn}/M_\odot) \sim 11.1$, a regime where it is likely that the galaxies are slow rotators \citep{Emsellem:2007}. 
Their stellar kinematics have negligible rotational support as reflected in the parameter $\lambda_R <0.1$, where $ \lambda_R \equiv \langle R|v| \rangle /  \langle \sqrt{v^2 + \sigma^2} \rangle$ is a proxy to quantify the observed projected stellar angular momentum per unit mass \citep[][]{Emsellem:2011}. Therefore, contribution from rotational broadening to the kinematics derived from their aperture spectra should be small.
We confirm that for ETGs the difference between $\sigma_{\rm spat}$ and $\sigma_{\rm ap}$ is small: The choice of $\sigma_{\rm ap}^{\rm gal}$ and $\sigma_{\rm spat}^{\rm gal}$ has little effect on the dispersion (see Fig.~\ref{fig:sig_parameters}. 
In the \mbh-\sig plane, ETGs predominantly fall into the high-\mbh regime where their location aligns with the relation of quiescent galaxies.
The observed relation of AGN ETGs is flatter than that of quiescent galaxies, but not as flat as that of LTGs. Since the ETGs sample a broader dynamic range in \mbh, this flattening likely arises from differences in the \mbh distribution (see Sect.~\ref{SubSec:Controlling_selection_effects}). Overall, the slope and intercept are similar to those of AGN LTGs within the uncertainties, suggesting that both follow the same underlying \mbh-\sig relation.

\subsubsection{Intrinsic Scatter}
\label{SubSec:Intrinsic_Scatter_M-sigma-relation}
Constraining the intrinsic scatter of the AGNs' \mbh-\sig relation is complicated by the narrow dynamic range in \mbh.
Furthermore, the \f-factor could only historically only be constrained as sample-average value, which intruded additional scatter to the \mbh-\sig relation relation.
Individual \f factors can vary due to different BLR geometries and viewing angles by more than an order of magnitude, and scatter by 0.41\,dex (\citealt{Villafana:2023}, see also Sect~\ref{SubSec:Consequences_Sample-average_virial_factor_f}).
\cite{Woo:2010}, who were the first to simultaneously constrained slope and intrinsic scatter on the RM AGN sample, report $\epsilon = 0.43$ based on the RM AGN sample.
Since then, calibrations for AGNs seem to have converged around this value, e.g., \cite{Woo:2015} find $\epsilon = 0.41 \pm 0.05$, \cite{Bennert:2021} find $\epsilon = 0.42 \pm 0.08$ and \cite{Caglar:2023} determine $\epsilon = 0.32 \pm 0.06$.
However, previous studies have either suffered from a narrow dynamic range in \mbh covered \citep[e.g.][]{Woo:2015, Bennert:2021}, and/or the use of less precise single-epoch \mbh estimators \citep[e.g.,][]{Grier:2013, Caglar:2023}, which increase $\epsilon$ by about 0.15 dex due to uncertain sample-averaged \f factors \citep{Woo:2015}. 
For individual AGNs, the systematic uncertainties can be as large as $\sim 0.4\,{\rm dex}$ \citep{Pancoast:2014}.
These \mbh measurement uncertainties alone may account for a significant portion of the intrinsic scatter in the \mbh-\sig relation reported in the literature.
Moreover, systematic uncertainties from the host galaxy side may introduce scatter to the \mbh-\sig relation.
Various apertures have been used for measuring \sig in quiescent galaxies, such as $R_{\rm eff}/8$ \citep{Ferrarese&Merritt:2000}, $R_{\rm eff}/2$ \citep[e.g.,][]{Kormendy_Ho:2013}, and $R_{\rm eff}$ \citep[e.g.,][]{Gultekin:2009}.
For AGNs the situation is even worse, as aperture size is often ignored (with the exception of e.g., \citealt{Bennert:2015, Batiste:2017b, Molina:2024}).
While there is no physical motivation which spatial scales \sig should correlate the closest with \mbh, the choice of the right aperture size crucial: In our sample we were able to resolve kinematic substructures, such as fast-rotating disks (see Fig.~\ref{fig:gallery_kin}, NGC\,3227, NGC\,7469), counter-rotating disks (e.g., Mrk\,1310), or circumnuclear spirals (e.g., Mrk\,1044).
Such complex kinematic substructures will affect \sig measurements, depending on what aperture size is used.

\subsubsection*{Optimal aperture for minimizing the intrinsic scatter}
\begin{figure}
   \centering
   \includegraphics[]{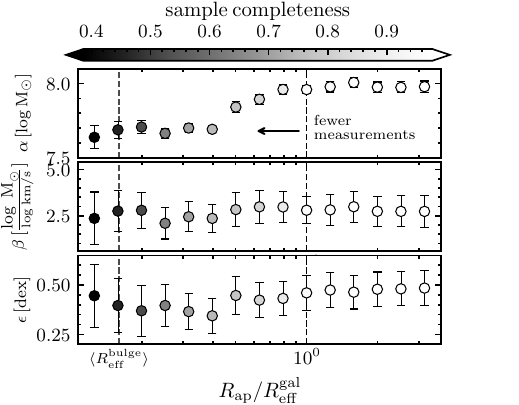} 
   \caption{
   \emph{Parameters of the best-fitting observed \mbh-\sigap relation of AGNs hosted by LTGs, as a function of aperture size $R_{\rm ap}$.
   }
      Data points color-coded by the fraction of AGNs for which we can robustly measure \sigap, which decreases with decreasing $R_{\rm ap}$:
       While \sigap is constrained for the full the sample at $R_{\rm eff}^{\rm gal}$, measuring \sig close to the galaxy center becomes increasingly more challenging for more distant and luminous AGNs. 
   The three panels show the parameters of the best-fitting \mbh-\sigap model parameters to the data, as a function of aperture size. 
  The intrinsic scatter $\epsilon$ reaches its global minimum close to $R_{\rm eff}^{\rm bulge}$, also reflected by lowest $\epsilon$ of the the corresponding scaling relation in row (v) of Table~\ref{tbl:Results}. This \mbh-\sig correlation is on scales of the bulge component.
   }
    \label{fig:parameters_apsize}%
\end{figure}

A generally applicable approach is needed to to define \sigstar consistently across different morphological components. 
We propose using an aperture size in units of the galaxy effective radius, $R_{\rm eff}^{\rm gal}$, to address varying spatial resolution and morphological complexity, thereby enhancing the consistency of the \mbh-\sig relation. 
This straightforward and self-consistent definition aims to minimize scatter in the \mbh-\sig relation.
Fig.~\ref{fig:parameters_apsize} shows the behavior of the best-fitting parameters to the \mbh-\sigap relation of AGNs hosted by LTGs, as a function of varying aperture size $R_{\rm ap}$.
Overall, we observe that the \mbh-\sig relation becomes marginally tighter for smaller aperture sizes below $R_{\rm ap}/R_{\rm eff}^{\rm gal}$.
This may be caused apertures larger than $R_{\rm ap}^{\rm gal}$, the larger-scale outskirts of galaxies are decoupled from the galaxy-intrinsic processes that shape the \mbh-\sig relation.
For instance, galaxy interactions, mergers or stellar accretion streams might affect \sigap at large $R_{\rm ap}>R_{\rm eff}^{\rm gal}$ of individual galaxies, increasing the scatter of the relation.
Within $R_{\rm eff}^{\rm gal}$, the intrinsic scatter $\epsilon$ decreases mildly with decreasing $R_{\rm ap}$. Overall, it stays consistent with $\epsilon = 0.47\,{\rm dex}$, the typical scatter of the relation on galaxy scales (see row (iii) of Table~\ref{tbl:Results}).
For the \mbh-$M_{\rm gal,dyn}$ relation, this behavior is slightly more pronounced: The \mbh-$M_{\rm bulge,dyn}$ relation shows the more significant slope at has a smaller intrinsic scatter compared to the \mbh-$M_{\rm gal,dyn}$ relation (see columns (xii) and (xiii) of Table~\ref{tbl:Results}, suggesting that the bulge represents the spatial scale on which the correlations are the tightest.
However, for many AGNs, stellar kinematics near the galaxy center are often missing, reducing sample completeness from 1 at $R_{\rm eff}^{\rm gal}$ to about 0.5 at $\langle R_{\rm eff}^{\rm bulge} \rangle$. 
As $R_{\rm ap}$ decreases, the intercept $\alpha$ varies significantly, indicating substantial effects from sample down-selection. 
Specifically, smaller apertures preferentially exclude distant galaxies, leading to an over-representation of lower-luminosity AGNs hosted by less massive LTGs. 
We therefore caution to interpret $\alpha$ and $\beta$ on scales of the bulge as the "best" parameters for \mbh-\sig, as this AGN subsample is likely biased.

\subsection{Controlling Selection Effects}
\label{SubSec:Controlling_selection_effects}
\cite{Lauer:2007} pointed out that flux-limited AGN samples are biased towards over-massive BHs compared to local samples of quiescent galaxies. 
This introduces a bias because over-massive BHs are preferentially selected due to the intrinsic scatter of the scaling relations \citep[see also][]{Treu:2007, Peng:2007}.
As such, selection effects can significantly impact black hole mass scaling relations if not properly accounted for. 
In principle, these biases can be corrected if the selection function is well-defined and based solely on AGN parameters \citep{Ding:2020, Ding:2023}, or if it can be statistically modeled using simple assumptions \cite[e.g.][]{Li:2021}.
\begin{figure*}
   \centering
   \includegraphics[]{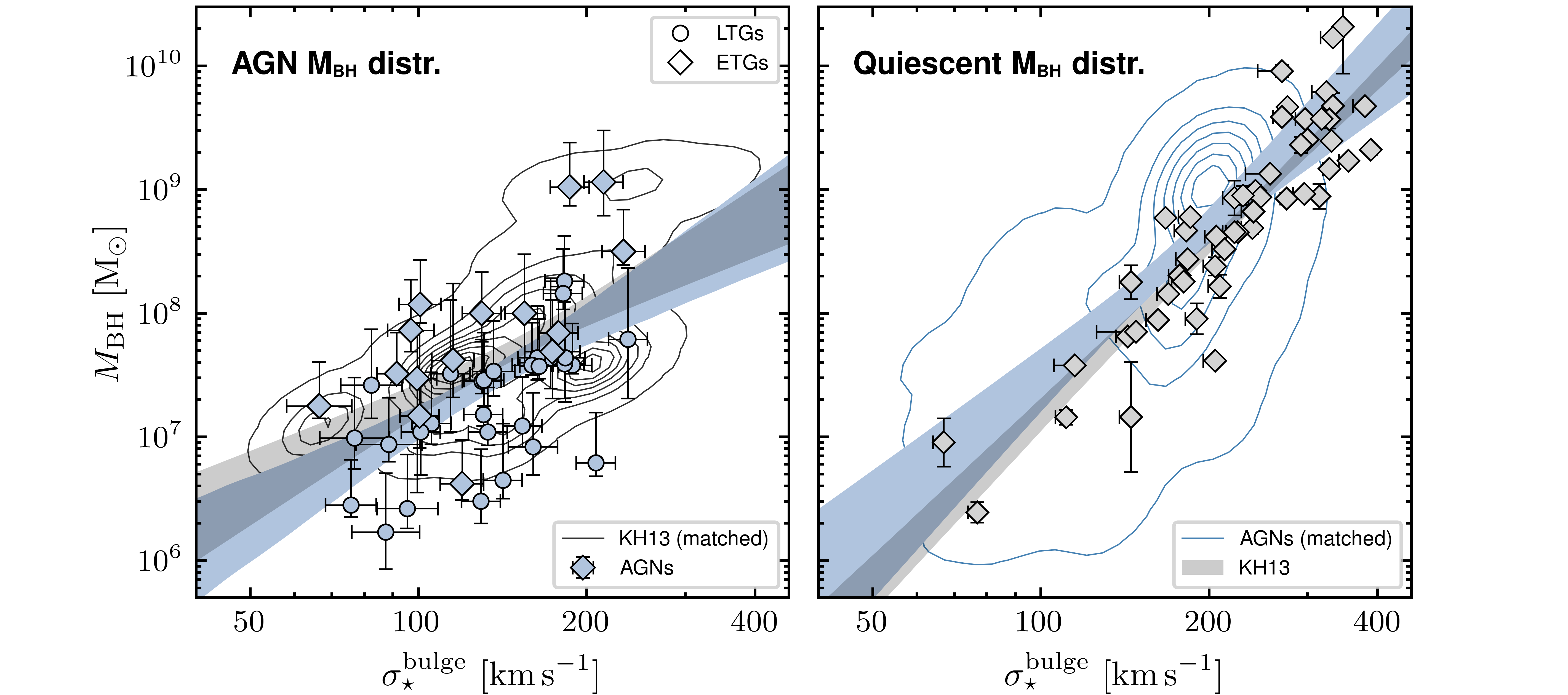}\\ 
   \vspace{.5em} 
   \includegraphics[]{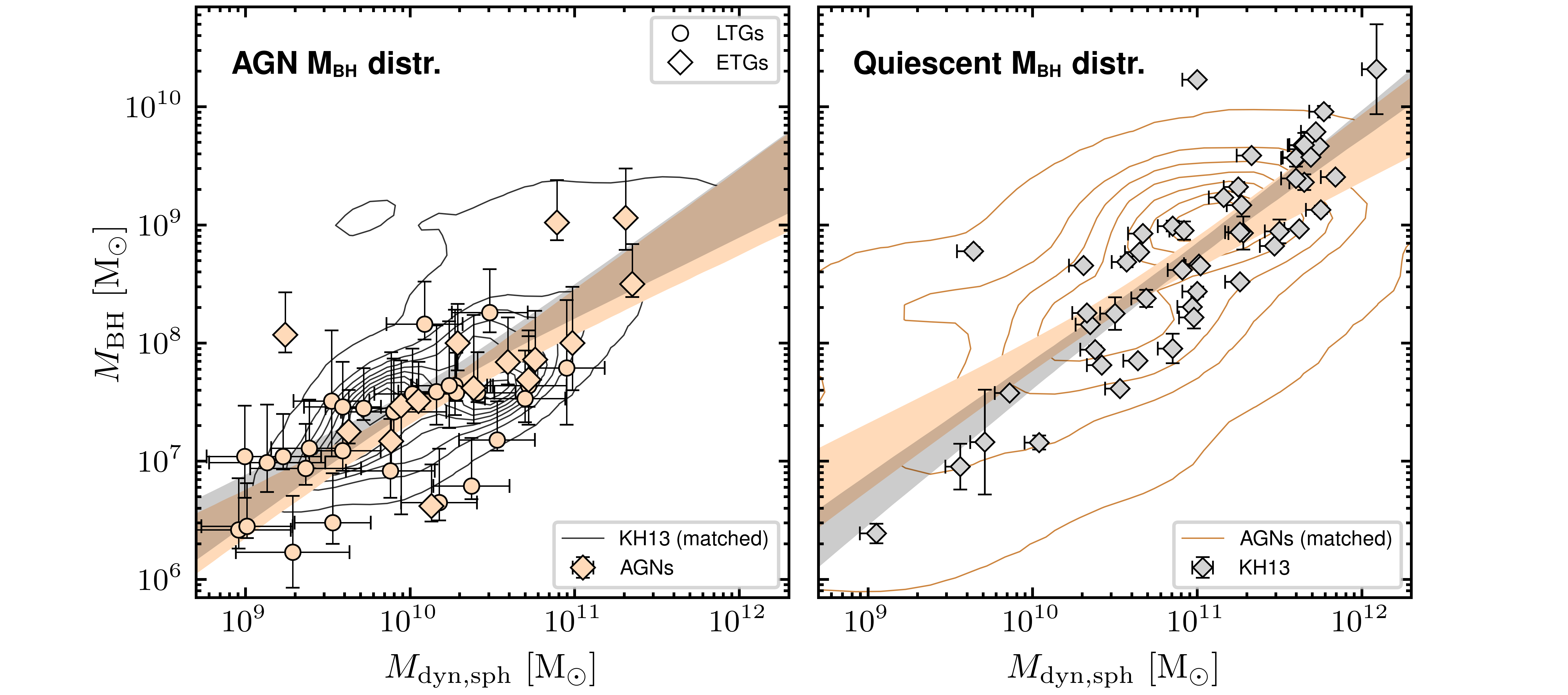} 
   \caption{
   \emph{\mbh-host-galaxy scaling relations of AGNs and quiescent galaxies.}
    (Top Left)
    Colored data points show AGNs hosted by LTGs (circles) and ETGs (squares), with the best-fitting observed relation shown as blue stripe (corresponding to row ii in Table~\ref{tbl:Results}).
    For AGNs, no clear distinction between the relations of ETGs and LTGs is observed.
    The gray contours show the KH13 sample that is resampled in \mbh to match the AGNs' \mbh-distribution (see Sect.~\ref{SubSec:Controlling_selection_effects}). with the fitted relations shown as shaded gray stripe.
    The relation of the \mbh-matched quiescent sample agrees with the AGNs' observed relation, and is significantly flatter than the observed relation of quiescent galaxies shown in the right panel.
    (Top right)
    After empirically matching the \mbh distribution of AGNs (blue contours) to that of quiescent galaxies (KH13 sample, gray data points), both fall onto the same region of the \mbh-\sigstar plane. The best-fitting relations of AGNs are shown as blue and gray stripes, and correspond to the relations in row corresponds to row (v) and (i) respectively, in Table~\ref{tbl:Results}.
    (Bottom left) 
    The same for the \mbh-\mbulgedyn relation, with the observed AGN listed in row (xii) of Table~\ref{tbl:Results}. 
    After matching the \mbh distribution, the relations of AGNs and quiescent galaxies are indistinguishable.
    (Bottom right) Same for matching the AGNs to the distribution of quiescent. The \mbh and \mbulgedyn relation correspond to row (xiv) of Table~\ref{tbl:Results}.
    }
              \label{fig:MBH-scaling-relations}%
\end{figure*}
For AGNs, the biases in the \mbh-\sig relation are dominated by two criteria: 
{\it (i)} measuring a reliable \sigstar which is often drowned by the bright AGN emission, and {\it (ii)} the narrow \mbh range, limited by the detection of low-luminosity AGNs and scarcity of luminous nearby AGNs.
In Sect.~\ref{SubSubSec:two_methods_sigma}, we have directly addressed {\it (i)} by using standardized recipe for consistently measuring \sig in AGN host galaxies.
Regarding {\it (ii)} , we note that the selection of the AGNs in our sample is purely based on \mbh measurement technique, with the vast majority (40/44) having been monitored in RM campaigns.
The selection for such RM campaigns is, to first order, blind to host-galaxy properties and purely based on AGN properties.
As a prerequisite for measuring robust time lags, AGNs must exhibit a broad line that shows sufficient BLR flux variability as well as continuum variability on the relevant timescales.
Compared to higher-mass BHs, lower-mass BHs are more likely to be active and thus included in optically selected type 1 AGN samples (\citealt{Schulze&Wisotzki:2011}, although not in X-ray selected AGN samples, see e.g., \citealt{Zou:2024}). 
This "active fraction bias" is inherent to the RM AGN sample. Additionally, low-luminosity AGNs with weak broad lines \citep[][]{Greene&Ho:2007, Chilingarian:2018} are typically excluded from RM campaigns, introducing an additional luminosity bias. As a result, the \mbh distribution is truncated at both low and high \mbh \citep{Schulze&Wisotzki:2011}, reducing the dynamic range in the \mbh-\sigstar plane and skewing the relation.

\subsubsection*{Matching the \texorpdfstring{\mbh}\ distribution}
These selection effects can be addressed by matching \mbh distributions between AGN and quiescent samples, assuming that differences in \mbh distributions are the primary driver of varying scaling relations.
We correct it by matching the \mbh distribution function of the quiescent population to that of our AGN sample follow the empirical method outlined by \cite{Woo:2013}.
For the implementation, we use a Monte Carlo approach: For each \mbh in the quiescent galaxy sample, we assign a random \sigstar chosen from AGNs that have the same \mbh (within a $\pm 0.15$\,dex bin, the typical uncertainty of \mbh).
By construction, the resulting mock quiescent sample follows the same \mbh distribution as the AGN sample.
We repeat this step for 1000 Monte Carlo samples, and fit the \mbh-\sigstar relation for each, using the method described in Sect.~\ref{SubSubSec:M-sigma-relation_AGNs}.
The left panel of Fig.~\ref{fig:MBH-scaling-relations} shows the results of this experiment. Indeed,when \mbh of quiescent galaxies is resampled to the AGNs' distribution (gray contours), their relation (gray shades)
is flattened, and the best-fitting parameters
$\beta = 8.02\pm\,0.12$ slope of $\beta = 2.38\, \pm \,0.61$, 
are consistent with the relation recovered from directly fitting the AGN (Table~\ref{tbl:Results}, row ii).

\begin{figure}
   \centering
   \includegraphics[width=\columnwidth]{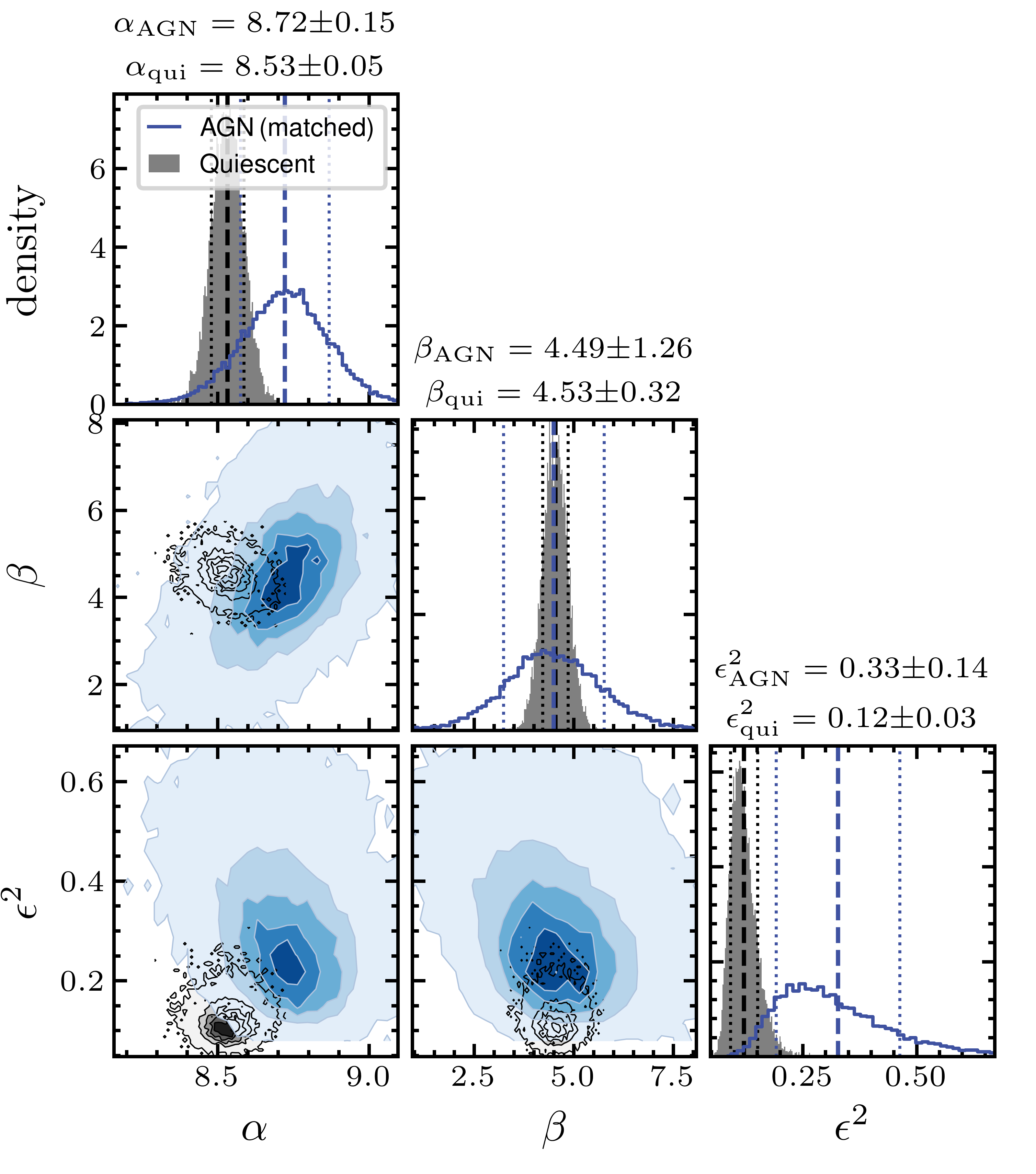} 
   \caption{
   \emph{Corner plot of the free parameters of the linear \mbh-\sig model after matching the \mbh distribution.
   }
    The posterior distributions of the quiescent population from KH13 are shown as black lines/contours, the AGNs from our sample are shown with blue colors. The intercept $\alpha$ and both populations are not significantly different, although associated with larger uncertainties for AGNs. This indicates that quiescent galaxies and AGNs follow similar intrinsic \mbh-\sig relations.
   }
    \label{fig:corner_plot}%
\end{figure}

By construction, the inverse experiment recovers the AGN \mbh-$\sigma_{\rm ap}^{\rm bulge}$ relation for AGNs if they followed the same \mbh distribution as the quiescent KH13 galaxy sample.
We refer to this quiescent-matched relation, highlighted in Table~\ref{tbl:Results}, as the \emph{corrected} scaling relation of AGNs.
The top right hand panel of Fig.~\ref{fig:MBH-scaling-relations} illustrates that after the \mbh-resampling, the AGNs (colored contours) follow the same \mbh-$\sigma_{\rm ap}^{\rm bulge}$ (colored stripes) as quiescent galaxies (gray stripes). 
The posterior distribution of the best-fitting parameters in Fig.~\ref{fig:corner_plot}, confirms that the offsets in $\alpha$ and $\beta$ are statistically insignificant (below the 1$\sigma$ confidence level). 
At our measurement uncertainty, the scaling relations for both populations are indistinguishable. 
Thus, the observed differences in the \mbh-\sigstar relation between AGNs and quiescent galaxies can be attributed to differing \mbh distributions alone.

The relation of \mbh with dynamical bulge mass, shown in the bottom panels of Fig.~\ref{fig:MBH-scaling-relations}, confirms what we find for the \mbh-\sig relation:
AGNs and quiescent galaxies form their own scaling relations that fall in complementary parts of the \mbh-\mbulgedyn plane. 
Fitting the observed \mbh-\mbulgedyn of AGNs returns a relation with shallower slope compared to that of quiescent galaxies.
After matching their \mbh-distributions, the relations are indistinguishable, suggesting that both populations share the same underlying scaling relation.

We note that the RM AGN sample might still contain additional biases which were not considered here.
Their BLR geometry might sample only a fraction of the parameter space \citep{Richards:2011}. From the host-galaxy side, BLR variability may be coupled to gas transport processes on host-galaxy scales, which could entail a secondary correlation with galaxy interactions \citep{Barnes:1996}, or secular processes triggered by e.g., bars \citep[e.g.,][]{Garcia-Burillo:2005}. 
Furthermore, is worth noting that also the quiescent sample suffers from selection biases, as pointed out by e.g., \cite{Bernardi:2007, vandenBosch:2016}: host galaxies with dynamically measured \mbh tend to have higher \sig compared to early type galaxies of the same luminosity, which may artificially increase the normalization \mbh-\sig relation by a factor of $\sim 3$ (see \citealt{Shankar:2016, Shankar:2020}, but \citealt{Kormendy:2020}).

\subsection{The Virial Factor \texorpdfstring{$f$}\ }
\label{SubSec:Consequences_Sample-average_virial_factor_f}
The classical approach for measuring a sample-average virial factor \f involves matching the \mbh-\sig relation of RM AGNs to that of quiescent galaxies.
This is usually done by fitting the VP-\sig relation with fixed slope, and determining the sample average virial factor via the difference of the intercepts ${\rm log}\,f = \alpha_{\rm qui} - \alpha_{\rm AGN}$.
However, this step implicitly assumes that AGNs and quiescent galaxies follow the same \mbh-\sig relation, which so far has little empirical foundation.
Furthermore, the matching is prone to systematic uncertainties introduced by different selection functions between AGN and quiescent galaxy samples; $\langle f \rangle$ can vary by 0.3\,dex depending on what quiescent galaxy sample is used as reference \citep{Ho_Kim:2014}.
In addition, individual \f factors vary by one order of magnitude across the sample, limiting the precision of this approach.

To test the implicit assumptions and reduce systematic uncertainties, we used the independently measured individual \f factors from \texttt{CARAMEL} modeling.
The sample-average $\langle f \rangle$ can be derived from comparing in the intercept between active and inactive galaxies. 
We fit the \mbh-\sig relation for the CARAMEL sample, this time fixing the slope to that of quiescent galaxies due to the limited dynamic range in \mbh ($\alpha_{\rm qui} = 4.53 \pm 0.32$, see Sect~\ref{SubSubSec:M-sigma-relation_quiescent}).
This step is justified, since we demonstrated that both share the same underlying relations (see Sect.~\ref{SubSec:Controlling_selection_effects}).
As opposed to the classical approach, the average of the dynamically measured values, $\langle {\rm log}\,f_{\rm dyn}, \rangle$ must be added to account for the sample average virial factor that is already incorporated in the AGNs' \mbh (i.e., is already included in $\alpha^{\rm AGN}_{\rm qui}$):
\begin{equation}
\label{eq:logf_my_method}
     \langle {\rm log}\,f \rangle = \alpha^{\rm qui} - \alpha^{\rm AGN}_{\rm dyn} + \langle  {\rm log}\,f_{\rm dyn} \rangle
\end{equation}
Fitting the AGNs' observed \mbh-$\sigma_{\rm ap}^{\rm bulge}$ relation with the slope  fixed to the KH13 relation yields $\alpha^{\rm AGN}_{\rm dyn} = 8.20 \pm 0.11$ and $\langle {\rm log}\,f \rangle = 0.65 \pm 0.18$.
This result closely matches $\langle {\rm log}\,f \rangle = 0.65 \pm 0.12$, the value obtained from applying the classical method to the RM AGN sample \citep{Woo:2015}.
It also aligns well with the average of individual \f-factors from dynamically modeling their BLR lags, $\langle {\rm log}\,f_{\rm dyn} \rangle = 0.66 \pm 0.07$ (column 11 of Table~\ref{tbl:bh_parameters}).
We conclude that the classical approach of determining the sample-average virial factor from matching the \mbh-\sig relation of RM AGNs agrees with the independent measurements of \mbh in AGNs.
Importantly, neither do we find significant dependencies of the sample-average \f on host galaxy morphology (as opposed to e.g., \cite{Ho_Kim:2014}), nor do observe such a dependency among the $f_{\rm dyn}$ \citep[see][for more discussion]{Villafana:2023}.

\subsection{Uniform \texorpdfstring{\mbh}\ - Host Galaxy Scaling Relations of Active and Quiescent Galaxies -- Consequences}
AGNs represent a special stage of during BH evolution where the ongoing gas accretion may significantly contribute to grow the SMBH.
However, AGN lifetime and the associated contribution to SMBH growth are only scarcely constrained, so that it is not clear how this should affect the the AGN \mbh scaling relations.
Regardless, there has been no independent and conclusive observational evidence for whether AGNs follow the same \mbh-host-galaxy scaling relations as quiescent galaxies.
While \cite{Woo:2013} demonstrated that selection effects can account for differences in slopes, our independent \mbh measurements reveal for the first time that both the slope $\beta$ and intercept $\alpha$ of the scaling relations for AGNs and quiescent galaxies are the same, indicating that both populations share the same underlying \mbh scaling relation.
This suggests, and we have explicitly tested, that matching the \mbh-\sig relation of AGNs with that of quiescent galaxies is justified for constraining the sample-average virial factor \f. 
In other words, our results reinforce previous calibrations of \f and individual measurements of $f_{\rm dyn}$ from dynamical modeling the BLR lags. 
By covering a larger dynamic range in both host galaxies and BHs, our results also support the use the single-epoch method for estimating \mbh across the explored parameter range, up to  ${\rm log}(M_{\rm BH} / M_\odot) \sim 10^{8.5}$.

\section{Summary}
\label{Sec:Summary}

After more than two decades of study, the \mbh scaling relations have emerged as essential probes of the coevolution between supermassive black holes and their host galaxies.
For AGNs, state-of-the-art observational and computational techniques have enabled more precise measurements of \mbh and \sigstar than were previously possible.
In this work, we used spatially resolved stellar kinematics to calibrate the \mbh-\sigstar relation of the local AGN population.
For a sample of 44 AGNs, the majority of which have precise and independent \mbh measurements from dynamical modeling, we presented IFU data from Keck/KCWI, VLT/MUSE and VLT/VIMOS. We tested different AGN deblending  and analysis techniques that are required to precisely trace the spatially resolved stellar kinematics.
Based on HST imaging data, we spatially resolved \sig across different galaxy morphological components, and studied dependencies of the scaling relation \mbh on morphology and aperture size.
Our key findings can be summarized as follows:

\begin{itemize}

    \item[1.] We find mild evidence that the \mbh-\sigstar correlation of AGNs hosted by LTGs is tightest if the kinematics are measured on scales of the galaxy bulge.
    
    \item[2.] Rotational broadening from the galaxy disk introduces scatter in the \mbh host galaxy relations of AGNs hosted by LTGs. Comparative studies based on higher-redshift AGNs hosted by disk galaxies can use the derived aperture-correction method to statistically infer the underlying \mbh-\sigstar scaling relation. 
    
    \item[3.] After removing the contribution from disk rotation, LTGs follow a \mbh-\sig relation that is similar to that of quiescent galaxies, but offset to lower \sigstar by $0.2\,{\rm dex}$. This suggests, that the dynamically hot disk component of LTGs does coevolve with the SMBH.
    
    \item[4.] The \mbh-\sigstar relation in AGNs is robust, regardless of whether the host galaxies have late-type or early-type morphologies. The intrinsic scatter is primarily driven by galaxy-to-galaxy variations. However, further constraining this scatter is challenging due to the scarcity of AGNs with dynamically measured \mbh$>10^{8},M_\odot$, and the fact that \sig in such luminous AGNs can only be marginally spatially resolved.
    
    \item[5.] The observed flattening of both \mbh-\sig and \mbh-\mdyn relations of AGNs is driven by selection biases that limit the \mbh dynamic range. We demonstrated for the first time that after correcting for this effect, slope \emph{and} intercept of the underlying scaling relations of AGNs match those of quiescent galaxies. This suggests that on average, the current AGN phase does not significantly grow \mbh compared to \mbh in quiescent galaxies.
    
    \item[6.] \mbh of our sample was determined independently of the virial factor. 
    Thus, we present a self-consistent empirical calibration of $\langle f \rangle$ based on spatially resolved kinematics of type-1 AGN.
    The derived value of ${\rm log}\,f= 0.65 \pm 0.18$ matches previous calibrations based on the classical RM AGN sample, as well as the average $\langle f \rangle$ determined from individually measured $f$. A robust understanding of the virial factor is essential for estimating \mbh measurements in the distant Universe via the single epoch method.
    
\end{itemize}

Spatially resolving \sigstar in AGNs is currently feasible only for the local AGN population, which we used in this study to provide the local reference of the \mbh scaling relations. 
It remains an important objective to identify the morphological components and spatial scales across which the \mbh-\sigstar relation of the quiescent population is the tightest. 
This can best be tested on nearby galaxies, for which larger sample size, higher spatial resolution and the lack of a bright AGN PSF make this analysis less sensitive to systematic uncertainties.

\begin{acknowledgements}
    We thank the anonymous referee for their valuable comments that improved the quality of the paper.
    %
    NW is a fellow of the International Max Planck Research School for Astronomy and Cosmic Physics at the University of Heidelberg (IMPRS-HD). This work is co-funded by the DFG under grant HU 1777/3-1.
    %
    VNB gratefully acknowledges assistance from the National Science Foundation (NSF) through grant AST-1909297. 
    This work is based on observations with the NASA/ESA Hubble Space Telescope obtained from the Data Archive at the Space Telescope Science Institute, which is operated by the Association of Universities for Research in Astronomy, Incorporated, under NASA contract NAS5-26555.
    Support for Program number HST-GO 17103 (PI Bennert) and HST-AR 17063 (PI Bennert) was provided through a grant from the STScI under NASA contract NAS5-26555.
    Note that findings and conclusions do not necessarily represent views of the NSF.
    VNB is also grateful for funding provided by a NASA-ADAP grant (80NSSC19K1016).
    TT acknowledges support by the National Science Foundation through grant NSF-AST-1907208.
    VU acknowledges funding support from NSF Astronomy and Astrophysics Research Grant No. AST-2408820, NASA Astrophysics Data Analysis Program (ADAP) grant No. 80NSSC23K0750, and STScI grant Nos. HST-GO-17285.001-A and JWST-GO-01717.001-A. 
    %
    Some of the data presented herein were obtained at Keck Observatory, which is a private 501(c)3 non-profit organization operated as a scientific partnership among the California Institute of Technology, the University of California, and the National Aeronautics and Space Administration. The Observatory was made possible by the generous financial support of the W. M. Keck Foundation. 
    The authors wish to recognize and acknowledge the very significant cultural role and reverence that the summit of Maunakea has always had within the Native Hawaiian community. We are most fortunate to have the opportunity to conduct observations from this mountain. 
\end{acknowledgements}

\facilities{Keck:II (KCWI), 
            VLT:UT4 (MUSE),
            VLT:UT3 (VIMOS),
            HST (ACS, WFCP2, WFC3)
    }
\software{
        \texttt{Astropy} \citep{AstropyCollaboration:2013, AstropyCollaboration:2018},  
         \texttt{SciPy} \citep{SciPy:2020},
         \texttt{Lenstronomy} \citep{Birrer&Amara:2018},
         \texttt{PyParadise} \citep{Husemann:2016a},
         \texttt{pPXF} \citep{Cappellari:2003, Cappellari:2017},
         \texttt{VorBin} \citep{Cappellari:2003},
         \texttt{CubePCA} \citep{Husemann:2022},
         \texttt{BADASS} \citep{Sexton:2021},
         \texttt{QDeblend$^{\rm3D}$} \citep{Husemann:2013, Husemann:2014}, 
         MUSE data processing pipeline \citep{Weilbacher:2020},
         \texttt{LinMix} \citep{Kelly:2007},
          }

\appendix

\section{AGN Spectral Fitting}
\label{Appendix:AGN_spectral_fitting}
\begin{figure}
   \centering
   \includegraphics[width=\columnwidth]{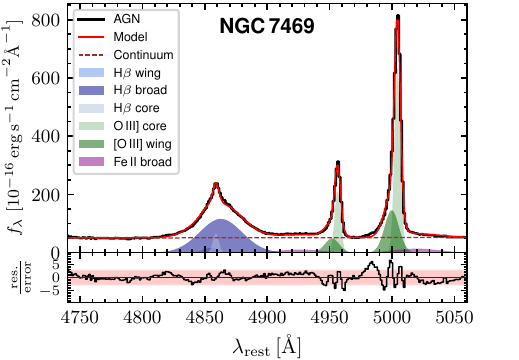} 
   \caption{\emph{Example of the AGN spectral modeling for the case of NGC\,7469.}
   The modeled wavelength range is limited to the rest-frame wavelength range 4750 Å–5100 Å covering the prominent H$\beta$ and [\ion{O}{iii}] emission lines. The spectrum with the full best-fit model and various line components for the BLR and the narrow and core component for H$\beta$ and [\ion{O}{iii}] are individually shown with different line styles and colors. The residual spectrum and the 3$\sigma$ limiting band are shown in the lower panel.}
              \label{fig:QSO_fitting}%
\end{figure}
\begin{deluxetable*}{lcccc}
\tabletypesize{\small} 
\tablecaption{ \emph{AGN parameters estimated from fitting the AGN spectrum.} \label{tbl:agn_parameters}} 
\tablehead{ 
 \colhead{AGN name} & 
  \colhead{$\sigma_{{\rm H}{\beta}}$ } & 
  \colhead{$ f_\lambda (5100 \,{\textrm \AA}$) } &  
 \colhead{$ L_{{\rm bol}}$ } &  
 \colhead{$ {\rm log}\, \lambda_{{\rm Edd}}$}  
 \vspace{-6pt}  \\  
 \colhead{} & 
  \colhead{$[{\rm km \,s}^{-1}]$}  & 
  \colhead{$[10^{-15} \,{\rm erg}\,{\rm s}^{-1} \,{\rm cm}^{-2} \, {\textrm \AA}^{-1}$]} & 
  \colhead{$[10^{43} \,{\rm erg}\,{\rm s}^{-1} ] $} \   
 \vspace{-6pt}  
          } 
\colnumbers  
\startdata 
NGC\,3227          &  1656 $\pm$ 83         &  7.7 $\pm$ 0.4         &  1.23 $\pm$ 0.07       &  $ -2.0 \pm 0.2 $ \\ 
NGC\,6814          &  1466 $\pm$ 73         &  2.3 $\pm$ 0.3         &  0.69 $\pm$ 0.09       &  $ -1.7 \pm 0.3 $ \\ 
NGC\,4593          &  1655 $\pm$ 83         &  2 $\pm$ 0.4           &  1.6 $\pm$ 0.3         &  $ -1.6 \pm 0.4 $ \\ 
NGC\,3783          &  2170 $\pm$ 110        &  10.2 $\pm$ 0.2        &  10.9 $\pm$ 0.2        &  $ -1.5 \pm 0.2 $ \\ 
NGC\,2617          &  2100 $\pm$ 110        &  1.13 $\pm$ 0.08       &  2.6 $\pm$ 0.2         &  $ -2.2 \pm 0.2 $ \\ 
IC\,4329\,A        &  2860 $\pm$ 430        &  2.08 $\pm$ 0.09       &  5.4 $\pm$ 0.2         &  $ -2.0 \pm 0.2 $ \\ 
Mrk\,1044          &  805 $\pm$ 40          &  3.2 $\pm$ 0.2         &  9.6 $\pm$ 0.7         &  $ -0.6 \pm 0.2 $ \\ 
NGC\,5548          &  3500 $\pm$ 170        &  6.8 $\pm$ 0.2         &  20.7 $\pm$ 0.7        &  $ -1.4 \pm 0.2 $ \\ 
NGC\,7469          &  1046 $\pm$ 52         &  5.2 $\pm$ 0.4         &  16.4 $\pm$ 1.2        &  $ -1.1 \pm 0.3 $ \\ 
Mrk\,1310          &  1360 $\pm$ 180        &  0.16 $\pm$ 0.03       &  0.7 $\pm$ 0.1         &  $ -2.7 \pm 0.4 $ \\ 
Mrk\,1239          &  1093 $\pm$ 55         &  3.3 $\pm$ 0.2         &  14.7 $\pm$ 0.7        &  $ -1.4 \pm 0.2 $ \\ 
Arp\,151           &  1170 $\pm$ 350        &  0.46 $\pm$ 0.06       &  2.3 $\pm$ 0.3         &  $ -1.4 \pm 0.3 $ \\ 
Mrk\,50            &  1992 $\pm$ 100        &  1.16 $\pm$ 0.1        &  7.3 $\pm$ 0.6         &  $ -1.7 \pm 0.3 $ \\ 
Mrk\,335           &  1800 $\pm$ 200        &  2 $\pm$ 0.3           &  15.2 $\pm$ 2.4        &  $ -1.2 \pm 0.4 $ \\ 
Mrk\,590           &  3580 $\pm$ 180        &  1.39 $\pm$ 0.07       &  11 $\pm$ 0.5          &  $ -1.6 \pm 0.2 $ \\ 
SBS\,1116+583A     &  1845 $\pm$ 92         &  0.28 $\pm$ 0.03       &  2.6 $\pm$ 0.3         &  $ -1.7 \pm 0.3 $ \\ 
Zw\,229-015        &  1386 $\pm$ 69         &  0.65 $\pm$ 0.02       &  5.9 $\pm$ 0.2         &  $ -1.3 \pm 0.2 $ \\ 
Mrk\,279           &  2010 $\pm$ 100        &  3.24 $\pm$ 0.1        &  35.3 $\pm$ 1.1        &  $ -1.1 \pm 0.2 $ \\ 
Ark\,120           &  1200 $\pm$ 60         &  4.11 $\pm$ 0.06       &  51.8 $\pm$ 0.8        &  $ -1.6 \pm 0.2 $ \\ 
3C\,120            &  1658 $\pm$ 83         &  13.1 $\pm$ 0.1        &  168.1 $\pm$ 1.9       &  $ -0.7 \pm 0.2 $ \\ 
MCG\,+04-22-042    &  1410 $\pm$ 70         &  0.68 $\pm$ 0.03       &  8.9 $\pm$ 0.4         &  $ -1.7 \pm 0.2 $ \\ 
Mrk\,1511          &  1906 $\pm$ 95         &  0.22 $\pm$ 0.01       &  3 $\pm$ 0.2           &  $ -1.7 \pm 0.2 $ \\ 
PG\,1310-108       &  1589 $\pm$ 79         &  1.31 $\pm$ 0.04       &  18.1 $\pm$ 0.6        &  $ -0.3 \pm 0.2 $ \\ 
Mrk\,509           &  2060 $\pm$ 100        &  13 $\pm$ 0.1          &  185 $\pm$ 1.8         &  $ -0.8 \pm 0.2 $ \\ 
Mrk\,110           &  1797 $\pm$ 90         &  1.6 $\pm$ 0.02        &  23.8 $\pm$ 0.3        &  $ -0.9 \pm 0.2 $ \\ 
Mrk\,1392          &  1983 $\pm$ 99         &  0.65 $\pm$ 0.02       &  9.9 $\pm$ 0.3         &  $ -2.3 \pm 0.2 $ \\ 
Mrk\,841           &  2030 $\pm$ 100        &  3.03 $\pm$ 0.05       &  47.6 $\pm$ 0.8        &  $ -1.0 \pm 0.2 $ \\ 
Zw\,535-012        &  1916 $\pm$ 96         &  1.6 $\pm$ 0.03        &  43.7 $\pm$ 0.8        &  $ -1.0 \pm 0.2 $ \\ 
Mrk\,141           &  2780 $\pm$ 140        &  0.89 $\pm$ 0.09       &  18.5 $\pm$ 1.8        &  $ -1.3 \pm 0.3 $ \\ 
RBS\,1303          &  1249 $\pm$ 62         &  2.6 $\pm$ 0.08        &  54.1 $\pm$ 1.7        &  $ -0.2 \pm 0.2 $ \\ 
Mrk\,1048          &  2080 $\pm$ 100        &  8.4 $\pm$ 0.7         &  183 $\pm$ 15          &  $ -0.6 \pm 0.3 $ \\ 
Mrk\,142           &  1291 $\pm$ 65         &  0.7 $\pm$ 0.02        &  16.8 $\pm$ 0.4        &  $ -0.1 \pm 0.2 $ \\ 
RXJ\,2044.0+2833   &  898 $\pm$ 45          &  2.49 $\pm$ 0.04       &  75 $\pm$ 1.3          &  $ -0.3 \pm 0.2 $ \\ 
IRAS\,09149-6206   &  2310 $\pm$ 120        &  27.7 $\pm$ 1.7        &  1110 $\pm$ 67         &  $ -0.1 \pm 0.2 $ \\ 
PG 2130+099        &  1690 $\pm$ 510        &  3.85 $\pm$ 0.08       &  194.1 $\pm$ 4.1       &  $ 0.3 \pm 0.2 $ \\ 
NPM1G\,+27.0587    &  1228 $\pm$ 61         &  1.88 $\pm$ 0.05       &  88.8 $\pm$ 2.3        &  $ -0.8 \pm 0.2 $ \\ 
RBS\,1917          &  1390 $\pm$ 70         &  1.14 $\pm$ 0.08       &  59.6 $\pm$ 4.1        &  $ -0.4 \pm 0.2 $ \\ 
PG\,2209+184       &  1908 $\pm$ 95         &  0.56 $\pm$ 0.02       &  34.2 $\pm$ 1.3        &  $ -1.1 \pm 0.2 $ \\ 
PG\,1211+143       &  1736 $\pm$ 87         &  4.5 $\pm$ 0.7         &  375 $\pm$ 54          &  $ -0.6 \pm 0.4 $ \\ 
PG\,1426+015       &  2630 $\pm$ 140        &  4.29 $\pm$ 0.08       &  402.8 $\pm$ 7.9       &  $ -1.5 \pm 0.2 $ \\ 
Mrk\,1501          &  1870 $\pm$ 170        &  1.08 $\pm$ 0.04       &  103.5 $\pm$ 3.4       &  $ -0.9 \pm 0.2 $ \\ 
PG\,1617+175       &  2030 $\pm$ 140        &  2.67 $\pm$ 0.08       &  439 $\pm$ 13          &  $ -0.1 \pm 0.2 $ \\ 
PG\,0026+129       &  921 $\pm$ 92          &  1.7 $\pm$ 0.2         &  500 $\pm$ 60          &  $ -0.9 \pm 0.3 $ \\ 
3C\,273            &  2120 $\pm$ 110        &  20.9 $\pm$ 1.2        &  7300 $\pm$ 420        &  $ -0.3 \pm 0.2 $ \\ 
\enddata 
\tablecomments{
AGNs are listed in order of increasing redshift (as in Table~\ref{tbl:bh_parameters}). 
(1) {AGN name}.
(2) {Line dispersion of the H$\beta$ BLR component.}
(3) {AGN continuum spectral flux density at 5100\,\AA.}
(4) {Approximate AGN bolometric luminosity from using a bolometric correction factor of 10.}
(5) {Eddington ratio $\lambda_{\rm Edd}$=$L_{\rm bol}$/$L_{\rm Edd}$.}} 
\end{deluxetable*}

\vspace*{-2\baselineskip} 
In order to estimate the AGN parameters \mbh, $L_{\rm bol}$ and $\lambda_{\rm Edd}$ as discussed in Sect.~\ref{SubSec:AGN_parameters}, we model the AGN spectrum in the H$\beta$-\oiii wavelength region.
As a first step, we correct Galactic foreground extinction which can significantly reduce the observed flux and alter the overall shape of the spectra recorded for our extragalactic targets. We correct all KCWI, MUSE, and VIMOS data cubes for Galactic extinction by dividing with the \cite{Cardelli:1989} Milky Way optical extinction curve, before fitting the AGN spectra.
The extinction curve is scaled to the line-of-sight V-band extinction as reported by the NASA/IPAC Extragalactic Database (NED) which is based on 
SDSS stars \citep{Schlafly:2011}.

To get a pure AGN spectrum free from host emission, we collapsed the host-deblended AGN data cube along the spatial axes.
We then subtract the best-fit stellar continuum as determined via \texttt{pPXF} (see Sect.~\ref{SubSec:Spectral_Synthesis_Modeling}.
For a consistent analysis between the datasets that cover different wavelength ranges, we restrict the spectral fitting to the common rest-frame wavelength range 4750\r{A} to 5100\r{A}. To describe the AGN power-law continuum in this narrow wavelength range, we adopt a linear pseudo-continuum.
For the strong emission lines H$\beta$ and \oiii, we use a superposition of broad and narrow Gaussian line profiles: Two broad components for the H$\beta$ line and two broad \ion{Fe}{ii}$\lambda\lambda$4923,\,5018 lines are sufficient to describe the spectral variations across the sample. In addition to the narrow components for each [\ion{O}{iii}] and H$\beta$, we often require a wing component to reproduce the typical asymmetry of those lines in AGN \citep{Greene:2005, Mullaney:2013}. 
We kinematically couple the broad narrow and wing components to each other, tie the [\ion{O}{iii}] doublet line ratio to its theoretical value of 3 \citep{Storey:2000} and that of \ion{Fe}{ii} components to their empirical ratio of 0.81.
With these constraints, we reduce the number of free parameters and increase the robustness of the fit.

An example of the modeling is shown in Fig.~\ref{fig:QSO_fitting}. We list the corresponding line fluxes of the broad H$\beta$ and \ion{Fe}{ii} lines together with their line widths as well as the total [\ion{O}{iii}] flux in Table~\ref{tbl:agn_parameters}. The corresponding errors are estimated from Monte Carlo sampling, plus an addition a 10\% systematic uncertainty introduced from the AGN-host deblending and continuum-subtraction process.

\section{The Importance of AGN-host Deblending}
\label{Appendix:Comparison_AGN-host_Deblending}
\begin{figure}
   \centering
   \includegraphics[]{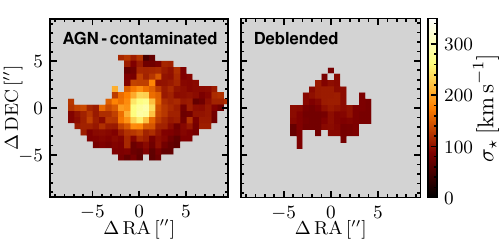} 
   \caption{\emph{AGN deblending impacts the extraction of \sig in Mrk\,110.}
   The left panel shows the spatially resolved \sigspat of Mrk\,110 as extracted from the AGN-contaminated cube (left panel). Its profile shows a steep rise of towards the galaxy nucleus, caused by the prominent \ion{Fe}{ii} broad emission lines blending with important \sig diagnostic lines. After carefully subtracting the bright AGN emission, the measured central \sigspat is smaller by a factor of three (right panel).}
              \label{fig:comparison_deblending}%
\end{figure}
Since close to the galaxy center the AGN is typically orders of magnitude brighter than the host-galaxy stellar continuum, the AGN emission blends the kinematic diagnostic features, making an accurate extraction of host-galaxy parameters challenging.
This can drastically affect the kinematics measured for any extended host-galaxy component.
While this is a well-known problem for tracing ionized gas emission lines of outflows and the extended narrow line region   \citep[e.g.][]{Villar-Martin:2016, Husemann:2016b}, we here demonstrate the impact of the AGN emission on extracting the host-galaxy stellar kinematics.

The AGN spectrum can be described by a power-law spectrum, with an additional contribution from the BLR clouds. 
Depending on observational setup, atmospheric conditions, the AGN/host luminosity ratio, and the AGN spectral classification, the AGN spectrum outshines the stellar continuum within the central 0\farcs4 to $\sim$6\farcs0. Especially for some luminous AGNs, e.g., the narrow-line-Seyfert 1 galaxies Mrk~335, Mrk~1044, with strong \ion{Fe}{ii} blending the \MgIb wavelength region.
In these cases, an accurate extraction of the stellar kinematics close to their nuclei is limited by the contrast between the AGN emission and the host galaxy, together with the accuracy by which the PSF can be modeled and subtracted.

There are two approaches to handle the AGN contamination: 
(i) The AGN spectrum can be included in the spectral synthesis modeling, as e.g., performed by \cite[][in prep.]{Remigio:2024} who use the package \texttt{BADASS} for analyzing a subsample of our AGNs. While, this approach is free of parameters, such as, e.g., the host-galaxy surface brightness profile, it requires a sophisticated treatment of the kinematic coupling between the spectral components, plus a well-considered choice of the starting parameters. 
(ii) We make the well-justified assumption that the broad lines exclusively originate from the spatially unresolved BLR. The package \texttt{QDeblend\textsuperscript{3D}} uses this to extract an empirical PSF, whose subtraction is described in Sect.~\ref{SubSubSec:AGN-host_deblending}.

Fig.~\ref{fig:comparison_deblending} shows the results for \sig from fitting the stellar continuum emission of Mrk\,110 with \texttt{pPXF}, before and after the PSF subtraction with \texttt{QDeblend\textsuperscript{3D}}. In the AGN-contaminated case, i.e., before PSF subtraction, \sig within the central 3\arcsec reaches a central value of $295 \pm 15 \,{\rm km\,s}^{-1}$.
These formal errors drastically underestimate the systematic offset that arises from the AGN contamination: After the AGN-host deblending, the spectral synthesis modeling of the faint host-galaxy signal results in a flat radial profile, where the central spaxel at the AGN location has $\sigma_{\rm spat} = 103 \pm 4 \,{\rm km\,s}^{-1}$. This value is consistent with the $95 \pm 8 \,{\rm km\,s}^{-1}$ reported by \cite{Ferrarese:2001}, which were measured from \CaT in a 2\arcsec$\times$4\arcsec\ long-slit aperture, which is less affected by AGN contamination.
We note that Mrk\,110 represents an extreme case, where the AGN-contamination offsets the central \sig by a factor of three. However, within the bulge effective radii $R_{\rm eff}^{\rm bulge}$, we observe an average increase of 30\% when measuring \sig after not properly subtracting the AGN emission.

\section{Spatially Resolved versus Aperture-integrated \texorpdfstring{\sig}\ }
\label{Appendix:Aperture_vs_Spatially_Resolved}

\begin{figure}
   \centering
   \includegraphics[]{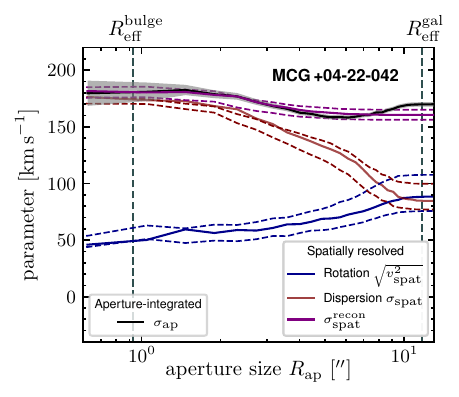} 
   \caption{
   \emph{Comparing different methods to measure \sig in MCG\,+04-22-042.}
   The black line shows the aperture-integrated stellar velocity dispersion  as a function of aperture size $R_{\rm ap}$, with the uncertainty indicated by gray shades.
   By integrating over the intensity-weighted contributions from the spatially resolved measurements of the rotational component $v_{\rm spat}$ and dispersion \sigspat, we can reconstruct the aperture-integrated value across a large dynamic range in aperture sizes ($\sigma_{\rm spat}^{\rm recon}$, purple line). 
   }
              \label{fig:reconstruct_sigma}%
\end{figure}

In Sect.~\ref{SubSubSec:two_methods_sigma} we have defined two methods for measuring the stellar velocity dispersion: The dispersion measured from aperture-integrated $\sigma_{\rm ap}$ and the spatially resolved $\sigma_{\rm spat}$.
As a consistency check, we reconstruct aperture-integrated kinematics from spatially resolved measurements in individual Voronoi cells. 
To achieve this, both contributions from ordered rotation $v_{\rm res}$ and chaotic motion $\sigma_{\rm res}$ must be considered, as it has been done e.g., for quiescent galaxies by \cite{Pinkney:2003} and \cite{Gultekin:2009}, and for AGNs by \cite{Bennert:2015}.
In our case of 2D kinematic fields, the surface area (i.e., number of spaxels) and the associated surface brightness of the host-galaxy stellar emission varies between different Voronoi cells.
To derive consistent flux-weighted kinematics, the spatially resolved values must be weighted by the luminosity of the respective Voronoi cell. 
Thus, we reconstruct the aperture-integrated kinematics from spatially resolved measurements as
\begin{equation}
\label{eq:sigma_spat_re}
\left( \sigma_{\rm ap}^{\rm recon}\right)^2 = \frac{\int_0^{R_{\rm eff}} [ \sigma_{\rm spat}^2 \left( r \right)  +  v_{\rm spat}^2 \left( r \right) ] \cdot I(r) \cdot dr }{\int_0^{R_e} I(r) \cdot dr}    
\end{equation}
with the surface brightness $I(r)$.
In case of reconstructing the kinematics of the bulge, $I(r) = I(R_{e}) \times {\rm exp}( -\kappa_n [(r/R_{e})^{1/n} -1]) $ is described by a Sérsic profile, where the $R_e$ as the bulge effective radius as measured from the photometry presented in \cite[][in prep.]{Bennert:2024}. 
This approach is equivalent to the prescription of the Nuker team (e.g., \citealt{Pinkney:2003, Gultekin:2009}), but for AGNs has only been applied to long-slit spectra \cite[][]{Bennert:2015}.
From the ionized gas kinematics \cite[see][in prep.]{Remigio:2024}, we have noticed that $\sigma_{\rm ap}^{\rm recon}$ and \sigap are not necessarily equal (also see KH13, supplementary material for a discussion), although very complex emission line profiles with high-velocity components are required for the differences to matter. 
Within the context of stellar kinematics where the gradients are small, we detect no significant differences between $\sigma_{\rm ap}^{\rm recon}$ and \sigap;
MCG,+04-22-042 is an arbitrarily selected AGN for which we have coverage of $\sigma_{\rm spat}$ from scales below $R_{\rm eff}^{\rm bulge}$ to beyond $R_{\rm eff}^{\rm gal}$.
In Fig.~\ref{fig:reconstruct_sigma}, we show that with increasing distance from the center, the relative contribution from the bulge component ($\sigma_{\rm spat}$) decreases, while the relative contribution of the ordered disk-like rotation ($v_{\rm spat}$) increases. 
When combined, the reconstructed radial profile of $\sigma_{\rm ap}^{\rm recon}$ matches that of $\sigma_{\rm ap}$ as directly measured from the coadded spectra corresponding to that aperture size.
We have confirmed this behavior for LTGs in the sample, demonstrating the feasibility of disentangling the contributions from random orbital motions versus disk-like galaxy-scale rotation. 
Therefore, we conclude that our approach of inferring the $M_{\rm BH}$ scaling relation from rotation-corrected $\sigma_{\rm spat}$ is self-consistent (see Sect.~\ref{SubSubSec:M-sigma-relation_AGNs}).

However, we note that the spatial resolution in many datasets is too low to spatially resolve the bulge. In addition, the bright AGN emission often prevents measuring robust $\sigma_{\rm spat}^{\rm bulge}$ for 10 out of 38 LTGs. 
In these cases, the finite spatial resolution impacts also our ability to resolve the stellar kinematics on the relevant scales of the disk (few arcseconds), so that rotational broadening likely contributes even to the spatially resolved quantity $\sigma_{\rm spat}^{\rm gal}$.
As a result, disk rotation is poorly spatially resolve, so that the lower spatial resolution might bias $\sigma_{\rm spat}$ of individual AGNs towards higher values if the disk rotation is not resolved.

\section{Stellar Kinematics from Different IFU Datasets}
\label{Appendix:Comparison_IFU_Datasets}

\begin{figure}
   \centering
   \includegraphics[]{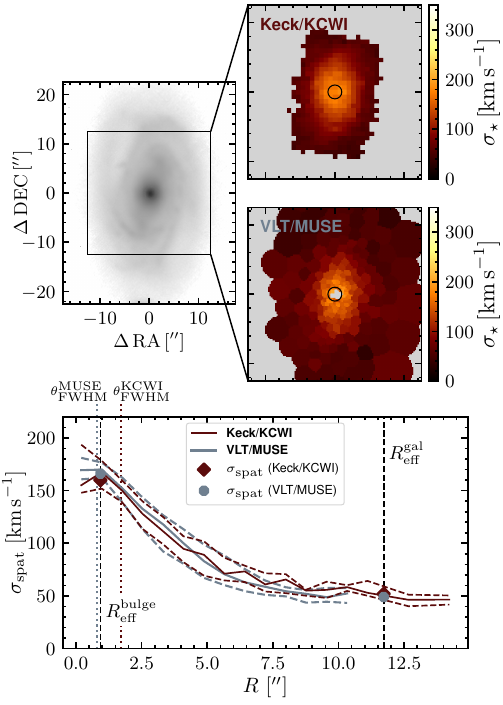} 
   \caption{
   \emph{Comparing \sigspat in MCG\,+04-22-042, extracted from two different datasets.} (Top panels) The top left panel shows the $V$-band AGN-subtracted continuum image of MCG\,+04-22-042, with \sigspat measured from the Keck/KCWI and VLT/MUSE datasets (top right inset panels). While the field coverage of MUSE is superior, we can measure accurate \sigspat at the AGN location only in the KCWI dataset.
   (Bottom panels) The radial profiles of \sigspat measured from the the two datasets are shown, together with the PSF FWHM $\theta_{\rm FWHM}$ of the respective dataset. Across the entire host galaxy, the spatially resolved radial profiles of \sigspat  agree within the uncertainty, which confirms that our method is valid independent of the observational setup.}
              \label{fig:comparison_dataset}%
\end{figure}

A handful of AGNs in our sample (5/44) have been observed with multiple optical IFU instruments, offering different field coverage, depth, spectral and spatial resolution. 
We have demonstrated in Appendix~\ref{Appendix:Comparison_AGN-host_Deblending} that the extraction of the stellar kinematics is limited by the accurate subtraction of the PSF, which is specific to each dataset. 
For the multiply observed objects, observations taken under different conditions with different instruments allow us to obtain independent measurements of the host-galaxy stellar kinematics for consistency checks.
MCG\,+04-22-042 is one of the AGNs that have been observed with both Keck/KCWI and VLT/MUSE.  
We processed each dataset as outlined in Sect.~\ref{Appendix:Comparison_AGN-host_Deblending} and Sect.~\ref{SubSec:Spectral_Synthesis_Modeling}, and here compare the radial profiles of the aperture-integrated \sig, and the spatially resolved \sigspat measurements (described in Sect.~\ref{SubSubSec:two_methods_sigma}).
Fig.~\ref{fig:comparison_dataset} shows the comparison between the radial behavior of the spatially resolved kinematics extracted in MCG\,+04-22-042. 
Compared to the relative small FoV of our KCWI setup(16\arcsec $\times$20\arcsec), the MUSE FoV covers a much larger fraction of the host galaxy. 
Within the overlapping field, the radial profile of \sigspat shows a steep decrease with increasing distance to the center.
The radial profiles extracted from the two datasets agree within the uncertainties, out to the radius where the KCWI coverage stops.
Different observing conditions and instrumental characteristics are reflected in the PSF width, $\theta^{\rm MUSE}_{\rm FWHM} = 1.2$, and $\theta^{\rm KCWI}_{\rm FWHM} = 1.8$. 
However, this difference does not significantly impact \sigspat on scales of the galaxy bulge $R_{\rm eff}^{\rm bulge}$; Analyzing the MUSE and KCWI datasets yields $\sigma_{\rm spat}^{\rm bulge} = 173\pm 5 \,{\rm km/s}$ and $\sigma_{\rm spat}^{\rm bulge} = 169\pm 6 \,{\rm km/s}$, respectively.

For the remaining 4 objects (RBS\,1303, NGC\,5548, NGC\,4593, PG\,1310-108, all observed with VLT/VIMOS), we have carried out the same test for \sigap, which is less sensitive to the differences between instrument characteristics.
While the results generally agree with each other within the error margin, the depth and resolution of the MUSE and KCWI cubes superior to the VIMOS datasets. We therefore adopted the MUSE and KCWI for the analysis in the main part of this work.

\section{Stellar Kinematics from Different Diagnostic Features}
\label{Appendix:Comparison_wavelength_regions}

\begin{figure*}
   \centering
   \includegraphics[]{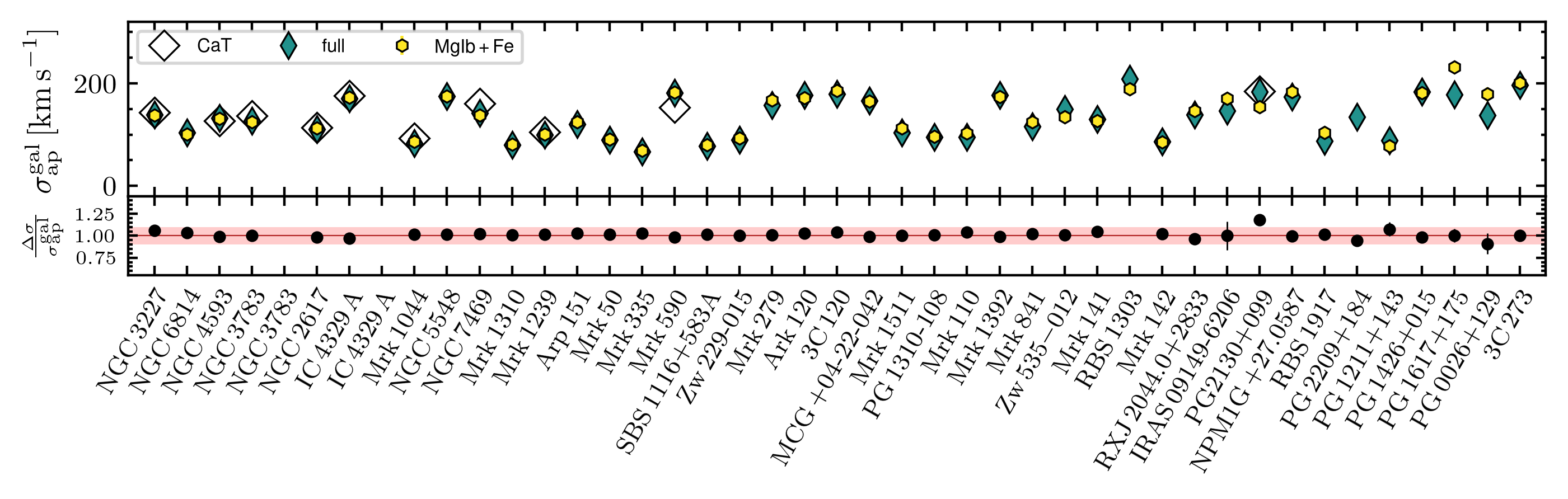} 
   \caption{
   \emph{Comparison of extracting stellar velocity dispersion from different spectral windows.
   }
   The top panel shows $\sigma_{\rm ap}^{\rm gal}$, the aperture-integrated dispersion from coadded spectra within and aperture matched to $R_{\rm eff}^{\rm gal}$, itemized after AGNs in the sample. 
   Colors refer to the different spectral windows within which we measured $\sigma_{\rm ap}^{\rm gal}$.
   The bottom row shows the residual dispersion between measurements from \MgIb versus "full" measurements, with the 5\% uncertainty, a typical uncertainty of \sigap, indicated by the shaded red stripe.
   The differences remain usually below the nominal uncertainties returned by \texttt{pPXF}.
   \MgIb+Fe and full wavelength windows usually provide consistent within the 5\% margin, and we consider both to ben similarly robust and consistent tracers for the stellar kinematics across the sample.
   }
              \label{fig:comparison_fitrange}%
\end{figure*}

In Fig.~\ref{fig:comparison_fitrange}, we compare how the choice of the wavelength range used for fitting the host galaxy emission affects the extracted stellar kinematics (see Sect.~\ref{SubSec:Spectral_Synthesis_Modeling}).
Specifically, we compare the kinematics obtained from the galaxy aperture-integrated spectra in the wavelength ranges 8400\,\AA\-8750\,\AA\,(\CaT), 5100\,\AA-5700\,\AA\,(\MgIb+Fe) and 4700\,\AA\,-5700\,\AA\,(full).
We find that in general, maximizing the wavelength range is favorable to increase the robustness of the parameters inferred through spectral synthesis modeling.
However, PSF subtraction required to remove the AGN emission can severely affect the faint host galaxy stellar emission. As a result, spatially coadding spectra can introduce non-physical artifacts in the spectra, especially near the galaxy nucleus, severely impacting the measured stellar kinematics. This effect is pronounced in two of the brightest AGNs, PG\,1617+175 or PG\,0026+129, where the choice of wavelength range can lead to systematic differences as large as 32\,km/s. This is caused by H$\beta$ AGN residual emission swamping the \MgIb and Fe stellar absorption features, leading to nonphysically high \sig (see Appendix \ref{Appendix:Comparison_AGN-host_Deblending}.
A consistent choice of the wavelength range is therefore a trade-off between narrow wavelength ranges that provide more robust results in bright AGNs, versus larger wavelength ranges providing the more robust results in faint AGNs.
Moreover, the coverage stellar absorption features varies due to varying spectral coverage between the datasets used in this work; While \CaT is available for almost all objects observed with MUSE, KCWI only covers the \MgIb and Fe features.
Overall, choosing the maximum common wavelength range between the datasets provides the best compromise between the three constraints.
We therefore settled with using the wavelength range 4700\,\AA\ -5700\,\AA\ for the spectral synthesis modeling in Sect.~\ref{SubSec:Spectral_Synthesis_Modeling}.

\section{Impact of Aperture Size on \texorpdfstring{\sig}\ }
\label{Appendix:Aperture_Effects}
\begin{figure*}
   \centering
   \includegraphics[]{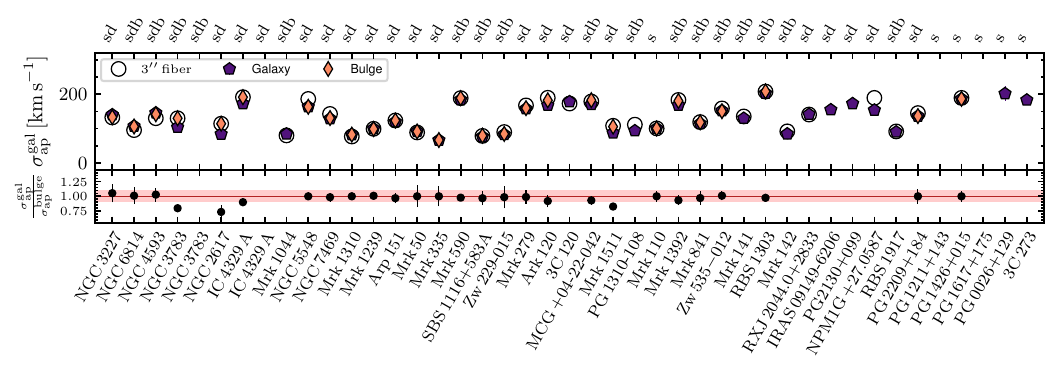} 
   \caption{
   \emph{Comparing integrated \sigap across different host-galaxy morphological components and aperture sizes.} 
   For each object, we show \sigap color-coded by the aperture size over which the host-galaxy emission was integrated prior to spectral synthesis modeling (see Sect.~\ref{SubSec:Surface_photometry}).
   The bottom panel shows the residual between $\sigma_{\rm ap}^{\rm bulge}$ and $\sigma_{\rm ap}^{\rm gal}$, for galaxies where we could robustly measure both quantities. For disk galaxies viewed at low inclination, $\sigma_{\rm ap}^{\rm bulge}$ is systematically larger than $\sigma_{\rm ap}^{\rm gal}$, a difference that can be as large as 40\%.
   }
              \label{fig:sig_ap_list}%
\end{figure*}
As a primary objective of this work, we investigate the dependency of the \mbh-\sigstar relation on aperture size in Sect.~\ref{SubSubSec:M-sigma-relation_AGNs}.
While the crucial role of aperture-size on the spatially resolved kinematics of LTGs is discussed in Sect.~\ref{SubSubSec:two_methods_sigma}, we here examine the effect of aperture size on measuring the aperture-integrated stellar velocity dispersion \sigap.

Fig.~\ref{fig:sig_ap_list} presents $\sigma_{\rm ap}$ for each AGN, measured from aperture-integrated spectra with aperture sizes corresponding to $R_{\rm eff}^{\rm bulge}$, $R_{\rm eff}^{\rm gal}$, or fixed to 3\arcsec (corresponding to the SDSS fiber size). 
The distinction between bulge and galaxy is applicable only for LTGs, which have an identifiable disk component. 
For ETGs, $\sigma_{\rm ap}^{\rm gal}$ is the sole indicator of morphology-matched kinematics since no substructure is detected in these systems. Additionally, small bulge sizes in several galaxies precluded the measurement of $\sigma_{\rm ap}^{\rm bulge}$ on such small scales (e.g. for Mrk\,1044, see also Sect.~\ref{SubSubSec:M-sigma-relation_AGNs}, and discussion in Sect.~\ref{SubSec:Intrinsic_Scatter_M-sigma-relation}).
For the majority of AGNs, changing the aperture size has marginal impact on the stellar kinematics. For ETGs, as long as $R_{\rm eff}^{\rm gal}$ is covered by the aperture, this is to be expected since their bright cores dominate the luminosity and kinematic profiles, which are typically covered by both the 3\arcsec and $R_{\rm eff}^{\rm gal}$-matched aperture.
For LTGs, using bulge- versus galaxy-size apertures makes a significant difference in approximately 50\% of the cases. 
This can be understood from the aperture-size dependent profiles, shown in Fig.~\ref{fig:ap_correction}. For 
LTGs viewed at high inclination, rotational broadening compensates for the drop of \sigspat on galaxy scales, resulting in a flat \sigap profile.
These are the galaxies for which $\sigma_{\rm ap}^{\rm gal} \approx \sigma_{\rm ap}^{\rm bulge}$.
Conversely, if LTGs are observed at low inclination, e.g., NGC\,3783, NGC\,26717 or Mrk\,1511, the high \sigspat in their centers contributes less and less with increasing aperture size, leading to $\sigma_{\rm ap}^{\rm gal} < \sigma_{\rm ap}^{\rm bulge}$.
For individual AGNs in our sample, this effect can be as large as 30\% which is the dominant driver behind differing scaling relation inferred from $\sigma_{\rm spat}^{\rm gal}$ versus $\sigma_{\rm spat}^{\rm bulge}$ (see Sect.~\ref{SubSubSec:M-sigma-relation_AGNs}).

\section{Comparing Stellar and SSP Libraries}
\label{Appendix:Comparison_templates}
\begin{figure*}
   \centering
   \includegraphics[]{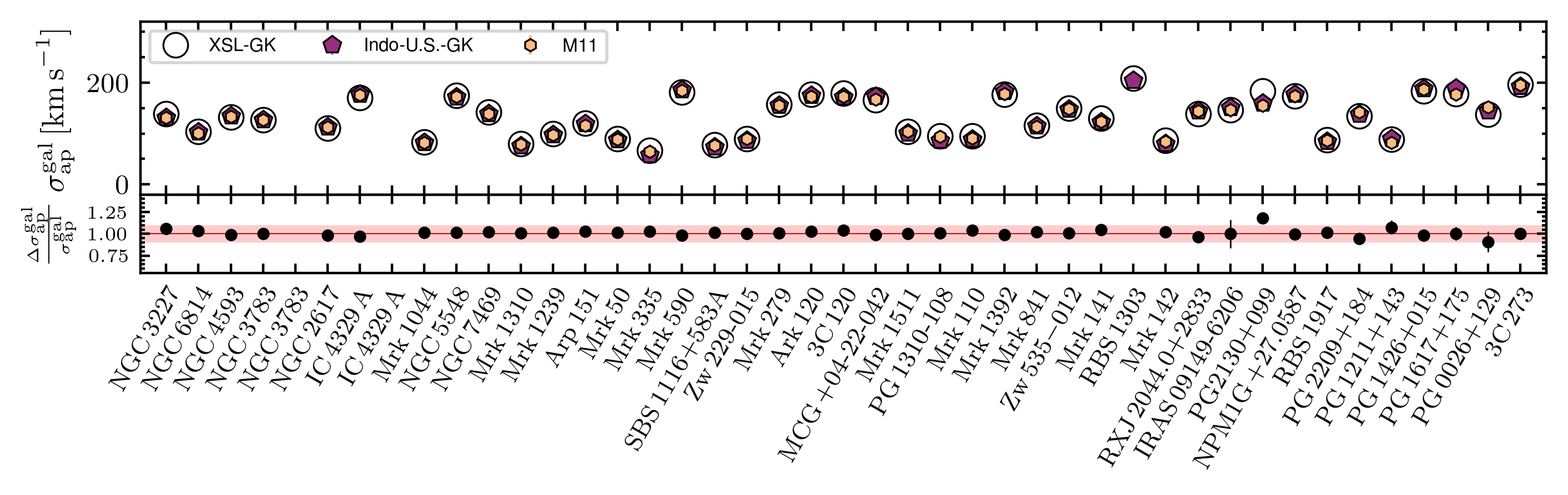} 
   \caption{
   \emph{Comparing stellar and SSP libraries for \sigap of individual AGNs.}
   For each galaxy, we show the aperture-integrated galaxy stellar velocity, color-coded by different template libraries used for the spectral synthesis modeling (see Sect.~\ref{SubSec:Spectral_Synthesis_Modeling}). 
   The differences between the results small, and typically range between  5\,km/s and 10\,km/s.
   }
    \label{fig:comparison_templates}%
\end{figure*}
To understand the robustness of our kinematic measurements, we have tested if the stellar kinematics are sensitive to the choice of stellar or SSP libraries used for fitting the spectra.
In Fig.~\ref{fig:comparison_templates}, we show the velocity dispersion obtained from fitting AGNs' PSF-subtracted aperture-integrated spectra across the restframe 4700\AA-5700\AA\ wavelength range.
We compared the kinematics recovered with templates from M11, Indo-U.S., and XSL. 
Motivated by the assumption that the light from the bulges of late-type galaxies, and early-type galaxies in general, is dominated by old stars, we also selected subsets of G and K giant stars from the XSL and Indo-U.S. libraries. Specifically, we selected temperatures $4400< T_{\rm{eff}} <5000$, surface gravity $0.15 < \rm{log}(g) <3.59$, and metallicity $-2.5 < [\rm{Fe/H}] < 0.34$. 
We refer to these templates as XSL-GK and Indo-U.S.-GK, respectively.

We found that the systematic offsets in \sig are typically $<10\,{\rm km/s}$ and therefore indistinguishable from the nominal uncertainties returned by \texttt{pPXF}.
However, we recognized that, on an individual basis, the best-fitting \sig can differ by up to 30\,km/s.
This is predominantly the case for objects for which the spectra have low S/N due to a strong AGN or a faint host galaxy (e.g., PG\,2130+099).
In these cases, the higher-resolution templates XSL and M11 provide consistent solutions, whereas the lower-resolution template spectra from Indo-U.S. lead to larger uncertainties.
Overall, we do not recognize a significant systematic difference when constraining the library to G and K giants, possibly because the aperture covers the entire galaxy disk.
However, stellar absorption features of \MgIb and Fe are better modeled when choosing the full template library. We therefore prefer the XSL library, which we adopted for the spectral synthesis modeling in Sect.~\ref{SubSec:Spectral_Synthesis_Modeling}.

\section{Comparing Spectral Synthesis Modeling Codes}
\label{Appendix:Comparison_Spectral_Fitting_Codes}
\begin{figure}
   \centering
   \includegraphics[]{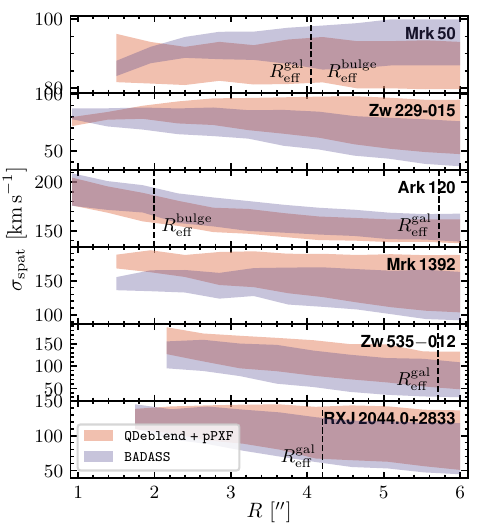} 
   \caption{
   \emph{Comparing kinematics extracted using different spectral synthesis modeling codes.}
   The panels show \sigspat for a subset of AGNs for which the spectra were also fitted by \texttt{BADASS}. 
   Shades indicate the uncertainty range of the kinematic profile, colors indicate the method by which we extracted \sigspat. Dashed lines indicate the bulge and galaxy effective radius.
   For the range across which we can measure \sigspat, the radial profiles extracted from \texttt{QDeblend\textsuperscript{3D}}+\texttt{pPXF} agree within the uncertainties with what we find when using \texttt{BADASS}.
   }
    \label{fig:comparison_badass}%
\end{figure}
We have tested three different codes for fitting the stellar continuum via stellar population synthesis modeling. 
We employed \texttt{pPXF}, \texttt{PyParadise} and \texttt{BADASS} which differ in their fitting methodologies. 
\texttt{pPXF} and \texttt{PyParadise} fit the stellar continuum emission separately from narrow and broad emission lines, which need to be subtracted first.
For the stellar continuum emission, \texttt{pPXF} describes large-scale continuum variations with a polynomial, whereas \texttt{PyParadise} first normalizes the continuum with a running mean before fitting kinematics with the normalized template spectra. This approach effectively removes non-physical continuum variations caused by PSF subtraction with \texttt{QDeblend\textsuperscript{3D}}. Since the continuum shape contains important information about the stellar populations (if AGN contamination is negligible), such a normalization removes information contained in the spectra and thus, effectively, reduces the S/N.
As expected, we observed that the performance of each code depends on the respective dataset. If the dataset covers a large wavelength range, as is the case for the MUSE datasets, \texttt{PyParadise} produces more stable results. 
However, if the analysis is constrained to the wavelength range shared between KCWI, VIMOS, and MUSE, the polynomial used by \texttt{pPXF} provides sufficient accuracy to describe the non-physical continuum variations. 
Moreover, \texttt{pPXF} tends to provide better fits at lower S/N compared to \texttt{PyParadise}, likely due to the S/N loss during continuum normalization in \texttt{PyParadise}.
Since our analysis is constrained to the common wavelength range of 4700\,\AA\ -5700\,\AA\ , we adopted \texttt{pPXF} for our study. We note that for individual AGNs, spurious spectral features near the galaxy need to be masked; otherwise, they would dominate the continuum variation modeled with the polynomial (see Fig.~\ref{fig:Example_Spectral_Fitting}).

The full Bayesian analysis code, \texttt{BADASS}, offers a different approach to fitting AGN spectra. Unlike \texttt{pPXF} and \texttt{PyParadise}, which require the point-like AGN emission to be subtracted first, \texttt{BADASS} fits the AGN spectrum, emission line templates and stellar spectra simultaneously.
An accurate knowledge of the AGN spectrum, combined with sophisticated coupling of the emission line parameters, allows for the robust inference of emission line and stellar kinematic parameters across the FoV of the IFU. With this method, \texttt{BADASS} provides a fundamentally different approach that is independent of the PSF subtraction method.
However, running the full MCMC for \texttt{BADASS} is time-consuming and fine-tuning for individual AGNs is required, depending on the AGN spectral features, absorption line strength, and spectral masking. The details for individual AGNs will be presented in our companion paper \cite[][in prep.]{Remigio:2024}. Here, we focus solely on the quantitative comparison of the inferred stellar kinematics parameters with those obtained using \texttt{pPXF}.
Fig.~\ref{fig:comparison_badass} shows the radial behavior of \sigspat for a subset of AGNs (chosen for good coverage within the effective radius to compare radial trends). We note that the same trends found for this subset also hold for a larger sample which will be presented in \cite{Remigio:2024}.
Within the range where we can robustly measure \sigspat, the radial profiles extracted from \texttt{QDeblend\textsuperscript{3D}}+\texttt{pPXF} are in agreement within the uncertainties. 
This suggests that the two independent methods provide consistent results, regardless of the distance from the AGN.
We conclude that the method used to measure the stellar kinematics in Sect.~\ref{SubSec:Spectral_Synthesis_Modeling} is robust. Furthermore, the nominal uncertainties returned by \texttt{pPXF} do not systematically underestimate the systematic uncertainties induced by the PSF subtraction with \texttt{QDeblend\textsuperscript{3D}}

\bibliographystyle{aasjournal}
\bibliography{references}

\end{document}